\documentclass[12pt,a4paper]{report}

\usepackage[intlimits]{amsmath}
\usepackage{amssymb}
\usepackage{graphicx}
\usepackage[dvips,a4paper,scale={.75,.8}]{geometry}
\usepackage{bbm}
\usepackage{mathrsfs}
\usepackage{axodraw}
\usepackage{wrapfig}
\usepackage{caption}
\usepackage{subfigure}
\usepackage{enumerate}
\usepackage{makeidx}

\newlength{\hoch}
\newlength{\doppelpfeil}
\newlength{\textlength}
\newlength{\overlinelength}
\newcommand{\ol}[2][.625]{%
   \settowidth{\textlength}{\ensuremath{#2}}%
   \setlength{\overlinelength}{3pt}%
   \addtolength{\overlinelength}{0.4\textlength}%
   \makebox[\textlength][s]{\ensuremath{#2}}%
   \hspace{-.5\textlength}\hspace{-\overlinelength}\hspace{#1\overlinelength}
   \overline{%
      \makebox[\overlinelength][s]{%
         \vphantom{\ensuremath{#2}}
      }
   }
   \hspace{-#1\overlinelength}\hspace{.5\textlength}
}

\newcommand{\olra}[2][0]{%
   \settowidth{\doppelpfeil}{\raisebox{\hoch}{$\scriptstyle \leftrightarrow$}}
   \settowidth{\textlength}{\ensuremath{#2}}%
   \settoheight{\hoch}{\ensuremath{#1}}%
   \makebox[\textlength][s]{\ensuremath{#2}}%
   \hspace{-\textlength}%
   \hspace{-#1pt}
   \raisebox{\hoch}{$\scriptstyle\leftrightarrow$}%
   \hspace{-\doppelpfeil}%
   \hspace{\textlength}%
   \hspace{#1pt}%
}

\renewcommand{\slash}[2][4]{\ensuremath{\rlap{\raisebox{1pt}{$\mspace{#1mu}/$}}#2}}

\def\clap#1{\hbox to 0pt{\hss#1\hss}}
\def\mathclap{\mathpalette\mathclapinternal}
\def\mathclapinternal#1#2{\clap{$\mathsurround=0pt#1{#2}$}}

\newcommand{\ls}{{\ensuremath{\scriptscriptstyle L}}}
\newcommand{\rs}{{\ensuremath{\scriptscriptstyle R}}}
\newcommand{\cs}{{\ensuremath{\scriptscriptstyle C}}}
\renewcommand{\L}{\ensuremath{\mathscr{L}}}
\renewcommand{\d}{\text{d}}
\newcommand{\dkbar}{\ensuremath{\overline{\d k}}}
\newcommand{\dpbar}{\ensuremath{\overline{\d p}}}

\newcommand{\ket}[1]{\ensuremath{\!\left| #1 \right\rangle}}
\newcommand{\braket}[2]{\ensuremath{\left.\left\langle #1\vphantom{#2}\,\right|#2\right\rangle}}
\newcommand{\braopket}[3]{\ensuremath{\left\langle#1\left|\vphantom{#1 #3} #2\right|#3 \right\rangle}}
\newcommand{\sun}{\ensuremath{\mathsf{SU(}n\mathsf{)}}}
\newcommand{\sutwo}{\ensuremath{\mathsf{SU(2)}}}
\newcommand{\suthree}{\ensuremath{\mathsf{SU(3)}}}

\DeclareMathOperator{\T}{T}
\DeclareMathOperator{\tr}{tr}
\DeclareMathOperator{\const}{const.}
\DeclareMathOperator{\diag}{diag}

\renewcommand{\i}{\ensuremath{\text{i}}}
\newcommand{\2}{\ensuremath{\sqrt{2}\,}}
\renewcommand{\d}{\ensuremath{\text{d}}}
\renewcommand{\L}{\ensuremath{\mathscr{L}}}
\newcommand{\psib}{\ensuremath{\ol{\psi}}}
\newcommand{\dslash}{\slash[2]{\partial}}
\newcommand{\dyad}{\ensuremath{\olra{\partial}}}
\newcommand{\Dslash}{\slash{D}}

\makeindex

\begin{document}
  \begin{titlepage}
    \hspace*{\fill}DESY-06-151

    \begin{center}
      \vspace{5cm}
      {\Huge \bfseries Field Theory and Standard Model}

      \vspace{2cm}
      {\bfseries\large W. Buchm\"{u}ller, C. L\"{u}deling}

      \vspace{1cm}
      {\it\large  Deutsches Elektronen--Synchrotron DESY, 22607 Hamburg, Germany}

      \vspace{2cm}
      {\bfseries Abstract}\\ 
      This is a short introduction to the Standard Model\\ 
      and the underlying concepts of quantum field theory.

      \vfill
      Lectures given at the European School of High-Energy Physics,\\
      August 2005, Kitzb\"{u}hel, Austria 

    \end{center}
  \end{titlepage}

  \tableofcontents

  \chapter{Introduction}
  
    In these lectures we shall give a short introduction to the standard model of particle
    physics with
    emphasis on the electroweak theory and the Higgs sector, and we shall also attempt
    to explain the underlying concepts of quantum field theory. 

    \medskip

    The standard model of particle physics has the following key features:
    \begin{itemize} 
     \item As a theory of elementary particles, it incorporates relativity and quantum mechanics, 
           and therefore it is based on quantum field theory.
     \item Its predictive power rests on the regularisation of divergent quantum corrections
           and the renormalisation procedure which introduces scale--dependent ``running couplings''.
     \item Electromagnetic, weak, strong and also gravitational interactions are all
           related to local symmetries and described by Abelian and non-Abelian gauge theories.
     \item The masses of all particles are generated by two mechanisms: confinement and spontaneous
           symmetry breaking.
    \end{itemize}
    In the following chapters we shall explain these points one by one. Finally, instead of a summary,
    we will briefly recall the history of ``The making of the Standard Model''\cite{Weinberg:2004kv}. 

    From the theoretical perspective, the standard model has a simple and elegant structure: It 
    is a chiral gauge theory. Spelling out the details reveals a rich phenomenology which can
    account for strong and electroweak interactions, confinement and spontaneous symmetry breaking,
    hadronic and leptonic flavour physics etc. \cite{Nachtmann:1990ta,peskin}.
    The study of all these aspects has kept theorists
    and experimenters busy for three decades. Let us briefly consider these two sides of the
    standard model before we enter the discussion of the details.

    \section{Theoretical Perspective}
    The standard model is a theory of fields with spins 0, $\frac{1}{2}$ and 1. The fermions
    (matter fields) can be arranged in a big vector containing left-handed spinors only:
    \begin{align}
      \Psi^T_L&=\big(\underbrace{q_{\ls \scriptscriptstyle 1}, u_{\rs
          \scriptscriptstyle 1}^\cs, e_{\rs \scriptscriptstyle 1}^\cs,
        d_{\rs \scriptscriptstyle 1}^\cs,l_{\ls \scriptscriptstyle
          1},\left(n_{\rs \scriptscriptstyle 1}^\cs\right)}_\text{1st family},\,\underbrace{q_{\ls
            \scriptscriptstyle 2},\ldots}_\text{2nd},\,\underbrace{\ldots,\left(n_{\rs
              \scriptscriptstyle 3}^\cs\right)}_\text{3rd}\big) \, ,
    \end{align}
    where the fields are the quarks and leptons, all in threefold family replication.
    The quarks come in triplets of colour, i.e., they
    carry an index $\alpha$, $\alpha=1,2,3$, which we suppressed in the above
    expression. The left-handed quarks and leptons come in doublets of weak isospin,
    \begin{align*}
      q_{\ls i}^\alpha&=\begin{pmatrix} u_{\ls i}^\alpha\\ d_{\ls i}^\alpha\end{pmatrix} \quad\text{and}\quad
      l_{\ls i}=\begin{pmatrix} \nu_{\ls i}\\ e_{\ls i}\end{pmatrix} \;,
    \end{align*}
    where $i$ is the family index, $i= 1,2,3$. 
    We have included a right-handed neutrino $n_\rs$ because there is evidence for neutrino masses
    from  neutrino oscillation experiments.

    The subscripts $L$ and $R$ denote left- and right-handed fields, respectively,
    which are eigenstates of the chiral projection operators $P_L$ or $P_R$. The superscript $C$
    indicates the charge conjugate field (the antiparticle).  Note that the charge conjugate of a
    right-handed field is left-handed:
    \begin{align}
      P_L \psi_\ls&\equiv \frac{1-\gamma^5}{2} \psi_\ls = \psi_\ls\;, &
      P_L\psi_\rs^\cs &=\psi_\rs^\cs\;,  & P_L \psi_\rs &=  P_L\psi_\ls^\cs=0\;,\\
      P_R \psi_\rs&\equiv \frac{1+\gamma^5}{2} \psi_\rs = \psi_\rs\;, &
      P_R\psi_\ls^\cs &=\psi_\ls^\cs\;,   & P_R \psi_\ls &= P_R \psi_\rs^\cs=0\;.
    \end{align}
    So all fields in the big column vector of fermions have been chosen left-handed. Altogether there are 48
    chiral fermions. The fact that left- and right-handed fermions carry different
    weak isospin makes the standard model a chiral gauge theory. The threefold replication of
    quark-lepton families is one of the puzzles whose explanation requires physics beyond the
    standard model \cite{ellis}.

    The spin-1 particles are the gauge bosons associated with the fundamental interactions in the
    standard model, 
    \begin{align}
      &G_\mu^A, \, A=1,\dotsc,8: && \text{the gluons of the strong interactions}\;,\\
      &       W^I_\mu \,, I=1,2,3\,,\; B_\mu:
      && \text{the $W$ and $B$ bosons of the electroweak interactions.}
    \end{align}
    These forces are gauge interactions, associated with the symmetry group
    \begin{align}
      G_\text{SM}&=\mathsf{SU(3)}_C\times \mathsf{SU(2)}_W \times \mathsf{U(1)}_Y \; ,
    \end{align}
    where the subscripts $C$, $W$, and $Y$ denote colour, weak isospin and hypercharge, respectively.

    The gauge group acts on the fermions via the covariant derivative $D_\mu$, which is an ordinary
    partial derivative plus a big matrix $A_\mu$ built out of the gauge bosons and the generators of
    the gauge group:
    \begin{align}
      D_\mu \Psi_L &= \left(\partial_\mu \mathbbm{1} + g A_\mu \right)\Psi_L\;.
    \end{align}
    From the covariant derivative we can also construct the field strength tensor,
    \begin{align}
      F_{\mu\nu}&= -\frac{\i}{g} \left[ D_\mu, D_\nu\right]\, ,
    \end{align}
    which is a matrix-valued object as well.

    The last ingredient of the standard model is the Higgs field $\Phi$, the only spin-0 field in
    the theory. It is a complex scalar field and a doublet of weak isospin. It couples 
    left- and right-handed fermions together.

    Written in terms of these fields, the Lagrangean of the theory is rather simple: 
    \begin{align}\label{eq:smlagrangean1}
      \begin{split}
        \L&= -\frac{1}{2} \tr \left[F_{\mu\nu} F^{\mu\nu}\right] +\overline{\Psi}_L \i \gamma^\mu
        D_\mu \Psi_L +\tr \left[\left(D_\mu \Phi\right)^\dagger D^\mu \Phi\right]\\
        &\quad +\mu^2 \,\Phi^\dagger\Phi -\frac{1}{2}\lambda \left(\Phi^\dagger\Phi\right)^2
        +\left(\frac{1}{2}   \Psi^T_L C h \Phi \Psi_L + \text{h.c.}\right)\;.
      \end{split}
    \end{align}
    The matrix $C$ in the last term is the charge conjugation matrix acting on the spinors, $h$ is a
    matrix of Yukawa couplings. All coupling constants are dimensionless, in particular, there is no
    mass term for any quark, lepton or vector boson. All masses are generated via the Higgs mechanism 
    which gives a vacuum
    expectation value to the Higgs field,
    \begin{align}
      \left\langle \Phi \right\rangle &\equiv v = 174\, \text{GeV}\;.
    \end{align}
    The Higgs boson associated with the Higgs mechanism has not yet been found, but its discovery
    is generally expected at the LHC.

    \section{Phenomenological Aspects}
    The standard model Lagrangean~(\ref{eq:smlagrangean1}) has a rich
    structure which has led to different areas of research in particle physics:
    \begin{itemize}
      \item The gauge group is composed of three subgroups with different properties:
        \begin{itemize}
          \item The $\mathsf{SU(3)}$ part leads to quantum chromodynamics, the theory of strong
            interactions \cite{ecker}. Here the most important phenomena are asymptotic
            freedom and confinement: The quarks and gluons appear as free particles only at very
            short distances, probed in deep-inelastic scattering, but are confined into mesons 
            and baryons at large distances.
          \item The $\mathsf{SU(2) \times U(1)}$ subgroup describes the electroweak sector of the
            standard model. It gets broken down to the $\mathsf U(1)_\text{em}$ subgroup of quantum
            electrodynamics by the Higgs mechanism, leading to massive $W$ and $Z$ bosons which are 
            responsible for charged and neutral current weak interactions, respectively.
      \end{itemize}
      \item The Yukawa interaction term can be split into different pieces for quarks and leptons:
        \begin{align}
          \frac{1}{2} \Psi_L^T C h \Phi \Psi_L &= h_{u \, ij} \bar{u}_{\rs i} q_{\ls j} \Phi + h_{d \, ij}
          \bar{d}_{\rs i} q_{\ls j} \widetilde{\Phi} +h_{e \, ij} \bar{e}_{\rs i} l_{\ls j}
          \widetilde{\Phi} +h_{n \, ij} \bar{n}_{\rs i} l_{\ls j} \Phi\;,
        \end{align}
        where $i,j=1,2,3$ label the families and $\widetilde{\Phi}_a =\epsilon_{ab} \Phi_b^*$. 
        When the Higgs field develops a vacuum expectation
        value $\left<\Phi\right>=v$, the Yukawa interactions generate mass terms. The first two terms,
        mass terms for up-type- and down-type-quarks, respectively, cannot be diagonalised 
        simultaneously,
        and this misalignment leads to the CKM matrix and flavour physics \cite{fleischer}. 
        Similarly, the
        last two terms give rise to lepton masses and neutrino mixings \cite{lindner}.
    \end{itemize}


  \chapter{Quantisation of Fields}
    In this chapter we will cover some basics of quantum field theory~(QFT). For a more in-depth
    treatment, there are many excellent books on QFT and its application in particle physics, such as
    \cite{Nachtmann:1990ta,peskin}.

    \section{Why Fields?}
      \subsection{Quantisation in Quantum Mechanics}
      \begin{wrapfigure}[11]{R}[0cm]{4.5cm}
          \begin{picture}(120,80)(0,0)
            \LongArrow(0,10)(120,10)
            \LongArrow(10,0)(10,90)
            \Text(120,5)[t]{$q(t)$}
            \Text(15,90)[l]{$\dot{q}(t)$}
            \Vertex(15,15){1}\Vertex(95,85){1}
            \Curve{(15,15)(30,18)(50,33)(60,50)(70,65)(80,77)(90,83)(95,85)}
          \end{picture}
          \caption{Particle moving in one dimension\label{fig:onedim}}
      \end{wrapfigure}
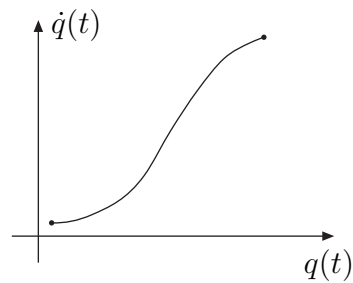
      Quantum mechanics is obtained from classical mechanics by a method called
      quantisation. Consider for example a particle moving in one dimension along a trajectory
      $q(t)$, with velocity $\dot{q}(t)$ (see Fig.~\ref{fig:onedim}). Its motion can be calculated
      in the Lagrangean or the Hamiltonian approach. The Lagrange 
      function $L(q,\dot{q})$\index{Lagrange function} is a function of the position and the
      velocity of the particle, 
      usually just the kinetic minus the potential energy. The equation of motion is obtained by
      requiring that the action, the time integral of the Lagrange function, be extremal, or, in other
      words, that its variation under arbitrary perturbations around the trajectory vanishes:
      \begin{align}
        \delta S &= \delta \int \d t L\left(q(t),\dot{q}(t)\right) =0\;.
      \end{align}
      The Hamiltonian\index{Hamiltonian} of the system, which corresponds to the total energy,
      depends on the coordinate $q$ and its conjugate momentum $p$ rather than $\dot{q}$:
      \begin{align}\index{Conjugate momentum!in quantum mechanics}
        H(p,q)&= p \dot{q} -L\left(q,\dot{q}\right)\, , \qquad p=\frac{\partial L}{\partial \dot{q}}\,.
      \end{align}

      To quantise the system, one replaces the coordinate and the momentum by operators $q$ and $p$
      acting on some Hilbert space of states we will specify later. In the Heisenberg
      picture\index{Heisenberg picture}, the 
      states are time-independent and the operators change with time as 
      \begin{align}
        q(t)&= e^{\i H t} q(0) e^{-\i H t} \;.
      \end{align}
      Since $p$ and $q$ are now operators, they need not commute, and one postulates the commutation
      relation
      \begin{align}\label{eq:commpq}\index{Canonical commutation relations!for $p$ and $q$}
        \left[p(0),q(0)\right]&= -\i \hbar \;,
      \end{align}
      where $h=2\pi\hbar$ is Planck's constant. In the following we shall       
      use units where $\hbar=c=1$. The commutator (\ref{eq:commpq})
      leads to the uncertainty relation\index{Uncertainty relation} 
      \begin{align}\label{heisen}
        \Delta q \cdot \Delta p \geq \frac{1}{2}\,.
      \end{align}
      Note that on Schr\"odinger wave functions the operator $q$ is just the coordinate itself and
      $p$ is $-\i \partial/\partial q$. In this way the commutation relation~(\ref{eq:commpq}) is 
      satisfied.

      As an example example of a quantum mechanical system, consider the harmonic
      oscillator\index{Harmonic oscillator} with the Hamiltonian
      \begin{align}
        H&=\frac{1}{2}\left( p^2 +\omega^2 q^2\right)\,,
      \end{align}
      which corresponds to a particle (with mass 1) moving in a quadratic potential with a strength
      characterised by $\omega^2$. Classically, $H$ is simply the sum of kinetic and potential energy. In
      the quantum system, we can define new operators as linear combinations of $p$ and $q$:
      \begin{subequations}\label{linear}
        \begin{align}
          q&= \frac{1}{\sqrt{2\omega}} \left(a + a^\dagger\right)\;, & p&= -\i
          \sqrt{\frac{\omega}{2}} \left(a - a^\dagger\right)\;, \\
          {\rm i.e.\ ,}\quad a &=\sqrt{\frac{\omega}{2}} q +\i \sqrt{\frac{1}{2\omega}} p\;, &
          a^\dagger&=\sqrt{\frac{\omega}{2}} q -\i \sqrt{\frac{1}{2\omega}} p\;. 
        \end{align}
      \end{subequations}
      $a$ and $a^\dagger$ satisfy the commutation relations
      \begin{align}\label{eq:commaadagger}\index{Canonical commutation relations!for raising and lowering operators}
        \left[ a, a^\dagger\right] &=1\; .
      \end{align}
      In terms of $a$ and $a^\dagger$ the Hamiltonian is given by      
      \begin{align}  
      H=\frac{\omega}{2}\left( a a^\dagger +a^\dagger a\right)\;.
      \end{align} 
      Since Eqs.~(\ref{linear}) are linear transformations, the new operators $a$ and $a^\dagger$ 
      enjoy the same time evolution as $q$ and $p$:
      \begin{align}
        a(t)&= e^{\i Ht} a(0) e^{-\i Ht} = a(0) e^{-\i \omega t}\, ,
      \end{align}
      where the last equality follows from the commutator of $a$ with the Hamiltonian,     
      \begin{align}\label{eq:commHa}
        \left[H,a\right]= -\omega a\,,\quad \left[H,a^\dagger\right]=\omega a^\dagger \; .
      \end{align}
      
      We can now construct the Hilbert space\index{Hilbert space!of the harmonic oscillator}  
      of states the operators act on. We first notice that the
      commutators~(\ref{eq:commHa}) imply that $a$ and $a^\dagger$ decrease and increase the energy of
      a state, respectively. To see this, suppose we have a state $\ket{E}$ with fixed energy, 
      $H\ket{E}=E\ket{E}$. Then 
      \begin{align}
        H a \ket{E}&= \left(a H +\left[H,a\right]\right) \ket{E} = a E \ket{E} - \omega a\ket{E} =
        \left(E-\omega\right) a \ket{E} \;,
      \end{align}
      i.e., the energy of a the state $a\ket{E}$ is $(E-\omega)$. In the same way one can show that
      \mbox{$H a^\dagger \ket{E}=(E+\omega)\ket{E}$}.  From the form of $H$ we can also see that its
      eigenvalues must be positive. This suggests constructing the space of states starting from a
      lowest-energy state $\ket{0}$, the vacuum or no-particle state\index{Vacuum}. This state needs
      to satisfy  
      \begin{align}  
      a\ket{0}=0\;,
      \end{align} 
      so its energy is $\omega/2$. States with more ``particles'', i.e., higher excitations, are
      obtained by successive application of $a^\dagger$: 
      \begin{align}
        \ket{n}&=\left(a^\dagger\right)^n \ket{0}\;, \quad \text{with} \quad
        H\ket{n}=\left(n+\frac{1}{2}\right)\omega \ket{n}\;.
      \end{align}

      \subsection{Special Relativity Requires Antiparticles}%
      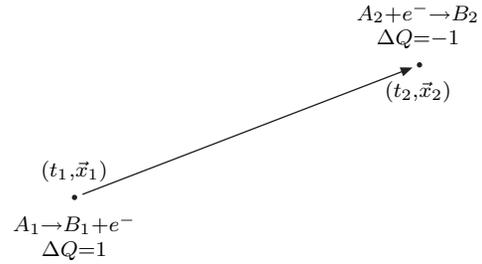
\begin{wrapfigure}[10]{r}[0pt]{6.5cm}
        \vspace{-2ex}
          \begin{picture}(170,100)(0,0)
            \Text(25,15)[c]{$\scriptstyle A_1\to B_1+e^-$}
            \Text(25,5)[c]{$\scriptstyle\Delta Q=1$}
            \Vertex(25,25){1}
            \Text(25,35)[c]{$\scriptstyle (t_1,\vec{x}_1)$}
            \LongArrow(28,26.15)(151,73.45)
            \Text(155,65)[c]{$\scriptstyle (t_2,\vec{x}_2)$}
            \Vertex(155,75){1}
            \Text(155,95)[c]{$\scriptstyle A_2+e^-\to B_2$}
            \Text(155,85)[c]{$\scriptstyle\Delta Q=-1$}            
          \end{picture}
          \caption{Electron moving from $A_1$ to $A_2$\label{fig:electronexchange}}
      \end{wrapfigure}%
      So far, we have considered nonrelativistic quantum mechanics. A theory of elementary particles, 
      however, has to incorporate special relativity. It is very remarkable that quantum mechanics
      together with special relativity implies the existence of antiparticles.  To see this
      (following an argument in~\cite{gravi}), consider
      two system (e.g. atoms) $A_1$ and $A_2$ at positions $\vec{x}_1$ and $\vec{x}_2$. Assume that
      at time 
      $t_1$ atom $A_1$ emits an electron and turns into $B_1$. So the charge of $B_1$ is one unit
      higher than that of $A_1$. At a later time $t_2$ the electron is absorbed by atom $A_2$
      which turns into $B_2$ with charge lower by one unit. This is illustrated in
      Fig.~\ref{fig:electronexchange}. 
      
      According to special relativity, we can also watch the system from a frame moving with relative
      velocity $\vec{v}$. One might now worry whether the process is still causal,
      i.e., whether the emission still precedes the absorption. In the boosted frame (with primed
      coordinates), one has
      \begin{align}
        t_2'- t'_1 &= \gamma \left(t_2-t_1\right) +\gamma \vec{v}
        \left(\vec{x}_2-\vec{x}_1\right)\, ,\quad \gamma=\frac{1}{\sqrt{1-\vec{v}\,^2}}\,.
      \end{align}
      $t_2'- t'_1$ must be positive for the process to remain causal. Since $|\vec{v}|<1$,
      $t_2'-t_1'$ can only be negative for spacelike distances, i.e., $\left(t_2-t_1\right)^2
      -\left(\vec{x}_1 -\vec{x}_2\right)^2<0$. This, however, would mean that the electron travelled
      faster than the speed of light, which is not possible according to special relativity. Hence, 
      within classical physics, causality is not violated.

      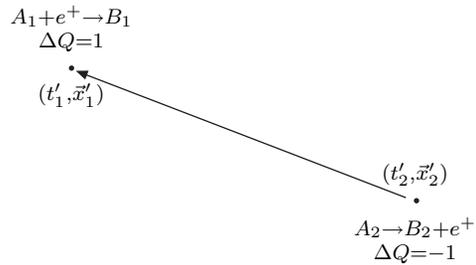
\begin{wrapfigure}[10]{R}[0pt]{6.5cm}
          \begin{picture}(180,100)(0,0)
            \Text(155,15)[c]{$\scriptstyle A_2\to B_2 +e^+$}
            \Text(155,5)[c]{$\scriptstyle\Delta Q=-1$}
            \Vertex(155,25){1}
            \Text(155,35)[c]{$\scriptstyle (t_2',\vec{x}_2')$}
            \LongArrow(151,26.15)(28,73.45)
            \Text(25,65)[c]{$\scriptstyle (t_1',\vec{x}_1')$}
            \Vertex(25,75){1}
            \Text(25,95)[c]{$\scriptstyle A_1 +e^+\to B_1$}
            \Text(25,85)[c]{$\scriptstyle\Delta Q=1$}            
          \end{picture}
          \caption{Positron moving from $A_2$ to $A_1$\label{fig:positronexchange}}
      \end{wrapfigure}
      This is where quantum mechanics comes in. The uncertainty relation leads to a ``fuzzy'' light
      cone, which gives a non-negligible propagation probability for the electron even for slightly
      spacelike distances, as long as
      \begin{align}
        \left(t_2-t_1\right)^2 -\left(\vec{x}_1 -\vec{x}_2\right)^2\gtrsim -\frac{\hbar^2}{m^2}\,.
      \end{align}
      Does this mean causality is violated?

      Fortunately, there is a way out: The antiparticle\index{Antiparticle}. In the moving frame, one can consider the
      whole process as emission of a positron at $t=t_2'$, followed by its absorption at a later time
      $t=t_1'$ (see Fig.~\ref{fig:positronexchange}). So we see that quantum mechanics together
      with special relativity requires the existence of antiparticles for consistency.  In addition,
      particle and antiparticle need to have the same mass.

      In a relativistic theory, the uncertainty relation (\ref{heisen}) also implies that particles
      cannot be localized below their Compton wavelength\index{Compton wavelength}
      \begin{align}
      \Delta x \geq \frac{\hbar}{mc} \,.
      \end{align}  
      For shorter distances the momentum uncertainty $\Delta p > m c$ allows for contributions
      from multiparticle states, and one can no longer talk about a single particle.

      \section{Multiparticle States and Fields}
      In the previous section we saw that the combination of quantum mechanics and special
      relativity has important consequences. First, we need antiparticles, and second, particle
      number is not well-defined. These properties can be conveniently described by means of 
      fields. A field here is a collection of infinitely many harmonic oscillators, corresponding to
      different momenta. For each oscillator, we can construct operators and states just as before
      in the quantum mechanical case. These operators will then be combined into a field operator,
      the quantum analogue of the classical field. These results can be obtained by applying the
      method of canonical quantisation to fields.
      
      \subsection{States, Creation and Annihilation}
      The starting point is a continuous set of harmonic oscillators, which are
      labelled by the spatial momentum $\vec{k}$. We want to construct the quantum fields for
      particles of mass $m$, so we can combine each momentum $\vec{k}$ with the associated energy
      $\omega_k=k^0=\sqrt{\vec{k}^2 +m^2}$ to form the momentum 4-vector $k$. This 4-vector
      satisfies $k^2\equiv k^\mu k_\mu=m^2$. For each $k$ we define creation and annihilation 
      operators, both for particles ($a$, $a^\dagger$) and
      antiparticles ($b$, $b^\dagger$)\index{Creation and annihilation operators}, and construct the
      space of states just as we did for the harmonic oscillator in the previous section. 

      For the states we again postulate the vacuum state, which is annihilated by both particle
      and antiparticle annihilation operators. Each creation operator $a^\dagger(k)$
      ($b^\dagger(k)$) creates a (anti)particle with momentum $k$, so the space of states is:
      \begin{align*}\index{Hilbert space!of the scalar field}  
        \text{vacuum: }&\ket{0}\,,\quad a(k)\ket{0}=b(k)\ket{0}=0\\
        \text{one-particle states: }&a^\dagger(k)\ket{0}\,,\;b^\dagger(k)\ket{0} \\
        \text{two-particle states: } &a^\dagger(k_1)a^\dagger(k_2)\ket{0}\,,\;
        a^\dagger(k_1)b^\dagger(k_2)\ket{0}\,,\;
        b^\dagger(k_1)b^\dagger(k_2)\ket{0}  \\
        &\quad\vdots
      \end{align*}
      Like in the harmonic oscillator case, we
      also have to postulate the commutation relations of these operators, and we choose them in a
      similar way: operators with different momenta correspond to different harmonic oscillators and
      hence they commute. Furthermore, particle and antiparticle operators should commute with each
      other.  Hence, there are only two non-vanishing commutators (``canonical commutation relations''):
      \begin{align}\label{relcom}\index{Canonical commutation relations!for creation and annihilation operators}
        \left[ a(k),a^\dagger(k')\right]&=\left[ b(k),b^\dagger(k')\right] = \left(2\pi\right)^3 2
      \omega_k \,\delta^3\!\!\left(\vec{k}-\vec{k}'\right) \,,
      \end{align}
      which are the counterparts of relation~(\ref{eq:commaadagger}). The expression on the right-hand
      side is the Lorentz-invariant way to say that only operators with the same momentum do not
      commute (the $(2\pi)^3$ is just convention).
      
      Since we now have a continuous label for the creation and annihilation operators, we need a
      Lorentz-invariant way to sum over operators with different momentum. The four components
      of $k$ are not independent, but satisfy $k^2\equiv k_\mu k^\mu=m^2$, and we also require
      positive energy, that is $k^0=\omega_k>0$. Taking these things into account, one is led to the
      integration measure
      \begin{align}
        \begin{split}
          \int \dkbar &\equiv \int \frac{\d^4 k}{\left(2\pi\right)^4}
            \,2\pi\,\delta\!\left(k^2-m^2\right) \,\Theta\!\left(k^0\right)\\
          &= \int \frac{\d^4 k}{\left(2\pi\right)^3}
            \delta\!\left(\left(k^0-\omega_k\right)\left(k^0+\omega_k\right)\right)
            \,\Theta\!\left(k^0\right)\\ 
          &=\int \frac{\d^4 k}{\left(2\pi\right)^3}
            \frac{1}{2\omega_k}\left(\delta\!\left(k^0-\omega_k\right)
              +\delta\!\left(k^0+\omega_k\right) \right) \,\Theta\!\left(k^0\right)\\
          &=\int \frac{\d^3 k}{\left(2\pi\right)^3}\frac{1}{2\omega_k}\,.
        \end{split}
      \end{align}
      The numerical factors are chosen such that they match those in Eq.~({\ref{relcom}) for the 
      commutator of $a(k)$ and $a^\dagger(k)$.

      \subsection{Charge and Momentum}
      Now we have the necessary tools to construct operators which express some properties of fields 
      and states.  The first one is the operator of 4-momentum, i.e., of spatial momentum and
      energy.  Its construction is obvious, since we interpret $a^\dagger(k)$ as a creation operator
      for a state with 4-momentum $k$. That means we just have to count the number of particles with
      each momentum and sum the contributions:\index{Momentum operator}
      \begin{align}
        P^\mu &= \int\dkbar \, k^\mu \left(a^\dagger(k) a(k) + b^\dagger(k)b(k)\right)\,.
      \end{align}
      This gives the correct commutation relations:
      \begin{subequations}
        \begin{align}
          \left[P^\mu,a^\dagger(k)\right]&= k^\mu a^\dagger(k)\,, & \left[P^\mu,b^\dagger(k)\right]&= k^\mu
          b^\dagger(k)\,,\\
          \left[P^\mu, a(k)\vphantom{a^\dagger}\right]&=-k^\mu a(k)\,, & \left[P^\mu,
            b(k)\vphantom{a^\dagger} \right]&=-k^\mu b(k)\,.
        \end{align}
      \end{subequations}  
        
      Another important operator is the charge. Since particles and antiparticles have opposite 
      charges, the
      net charge of a state is proportional to the number of particles minus the number of
      antiparticles: 
      \begin{gather}\index{Charge operator}
        Q=\int\dkbar \left(a^\dagger(k) a(k) - b^\dagger(k)b(k)\right)\,,
      \end{gather}  
       and one easily verifies
       \begin{gather}       
       \left[Q, a^\dagger(k)\right]=a^\dagger(k)\,,\qquad \left[Q,
          b^\dagger(k)\right]=-b^\dagger(k)\,. 
      \end{gather}

      We now have confirmed our intuition that $a^\dagger(k)$ $\left(b^\dagger (k)\right)$
      creates a particle with 4-momentum $k$ and charge +1 (-1). Both momentum and charge are
      conserved: The time derivative of an operator is equal to the commutator of the operator with
      the Hamiltonian, which is the 0-component of $P^\mu$. This obviously commutes with the
      momentum operator, but also with the charge:
      \begin{align}
        \i \frac{\d}{\d t} Q &= \left[Q,H\right] = 0 \,.
      \end{align}

      So far, this construction applied to the case of a complex field. For the special case of 
      neutral particles, one has $a=b$ and $Q=0$, i.e., the field is real.

      \subsection{Field Operator}
      We are now ready to introduce field operators, which can be thought of as Fourier transform
      of creation and annihilation operators:
      \begin{align}\label{eq:fieldoperator}
        \phi (x)&= \int \dkbar \left( e^{-\i kx} a(k) + e^{\i kx} b^\dagger (k)\right)\,.
      \end{align}
      A spacetime translation is generated by the 4-momentum in the following way:
      \begin{align}
        e^{\i y P} \phi(x) e^{-\i y P}&= \phi (x+y).
      \end{align}
      This transformation can be derived from the transformation of the $a$'s:
      \begin{align}
        e^{\i y P} a^\dagger(k) e^{-\i y P} &= a^\dagger(k) +\i y_\mu \left[P^\mu,a^\dagger(k)\right] +
        \mathcal{O}\left(y^2\right) \\
        &= \left(1+ \i y k +\dotsm\right) a^\dagger(k)\\
        &= e^{\i y k} a^\dagger(k)\,.
      \end{align}

      The commutator with the charge operator is
      \begin{align}
        \left[Q,\phi(x)\right] &= -\phi(x)\,, & \left[Q,\phi^\dagger\right]&= \phi^\dagger\,.
      \end{align}

      The field operator obeys the (free) field equation,
      \begin{align}
        \left(\Box +m^2\right)\phi(x)&= \int \dkbar \left( -k^2 +m^2\right)\left(e^{-\i kx}a (k)
          +e^{\i kx} b^\dagger(k)\right)=0,
      \end{align}
      where $\Box = \partial^2/\partial t^2 - \vec{\nabla}^2$ is the d'Alambert operator.

\pagebreak

      \subsection{Propagator}
      \begin{wrapfigure}[10]{r}[0pt]{6.9cm}
          \begin{picture}(180,90)(-8,0)
            \Text(20,55)[c]{$\scriptstyle (t_1,\vec{x}_1)$}
            \Text(20,45)[c]{$\scriptstyle \Delta Q=+1$}
            \Text(160,55)[c]{$\scriptstyle (t_2,\vec{x}_2)$}
            \Text(160,45)[c]{$\scriptstyle \Delta Q=-1$}
            \ArrowArcn(90,-50)(130,120,60)
            \Text(90,90)[c]{$\scriptstyle t_2>t_1 ,\; Q=-1$}
            \ArrowArcn(90,150)(130,-60,-120)
            \Text(90,10)[c]{$\scriptstyle t_1>t_2 ,\; Q=+1$}
          \end{picture}
          \caption{Propagation of a particle or an an\-ti\-par\-tic\-le, depending on the temporal
            order.\label{fig:propagation}} 
      \end{wrapfigure}
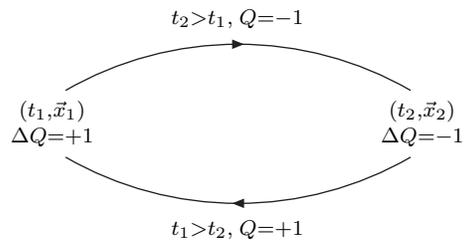

      Now we can tackle the problem of causal propagation that led us to introduce antiparticles. We
      consider the causal propagation of a charged particle between $x_1^\mu=(t_1,\vec{x}_1)$ and
      $x_2^\mu=(t_2,\vec{x}_2)$, see Fig. (\ref{fig:propagation}). The field operator creates
      a state with charge $\pm 1$ ``at position $(t,\vec{x})$'',  
      \begin{align}
        Q\, \phi(t,\vec{x}) \ket{0}&= -\phi(t,\vec{x})\ket{0}\,,\\
        Q\, \phi^\dagger (t,\vec{x}) \ket{0}&= \phi^\dagger (t,\vec{x})\ket{0}\,.
      \end{align}
      
      Depending on the temporal order of $x_1$ and $x_2$, we interpret the propagation of charge
      either as a particle going from $x_1$ to $x_2$ or an antiparticle going the other
      way. Formally, this is expressed as the time-ordered product (using the $\Theta$-function,
      $\Theta(\tau)=1$ for $\tau>0$ and $\Theta(\tau)=0$ for $\tau<0$):
      \begin{align}
        \T \phi(x_2)\phi^\dagger(x_1)&= \Theta(t_2-t_1)\phi(x_2)\phi^\dagger(x_1) +\Theta(t_1-t_2)
        \phi^\dagger(x_1) \phi(x_2)\,.
      \end{align}

      The vacuum expectation value of this expression is the Feynman propagator:\index{Feynman propagator}
      \begin{align}
        \begin{split}
          \i \Delta_\text{F} (x_2-x_1)&=\braopket{0}{\T \phi(x_2)\phi^\dagger(x_1)}{0}\\
          &= \i\int \frac{\d^4 k}{\left(2\pi\right)^4} \frac{e^{\i k (x_2-x_1)}}{k^2-m^2 +\i
            \varepsilon}\;, 
        \end{split}
      \end{align}
      where we used the $\Theta$-function representation\index{Theta@$\Theta$ function}
      \begin{align}
        \Theta(\tau)&= -\frac{1}{2\pi\i} \int_{-\infty}^\infty \!\!\d \omega
        \frac{e^{-\i\omega\tau}}{\omega+\i\epsilon} \,.
      \end{align}
      This Feynman propagator is a Green function for the field equation,
      \begin{align}
          \left(\Box + m^2\right) \Delta_\text{F} (x_2-x_1) &=\int \frac{\d^4
            k}{\left(2\pi\right)^4} \frac{\left(-p^2 +m^2\right)}{p^2-m^2 +\i \varepsilon} e^{-\i
            p\left(x_2-x_1\right)} = -\delta^4\left(x_2-x_1\right)\,.
      \end{align}
      It is causal,  i.e.\ it propagates particles into the future and antiparticles into the
      past. 

      \section{Canonical Quantisation}
      All the results from the previous section can be derived in a more rigorous manner by  
      using the method of canonical quantisation which provides the step from classical to quantum
      mechanics. We now start from classical field theory, where the field at point $\vec{x}$
      corresponds to the position $q$ in classical mechanics, and we again have to construct the 
      conjugate momentum variables and impose commutation relations among them.

      Let us consider the Lagrange density\index{Lagrange density}
      for a complex scalar field $\phi$. Like the Lagrangean in classical mechanics, the
      free Lagrange density is just the kinetic minus the potential energy density,
      \begin{align}\label{eq:Lscalarfree}
        \L&= \partial_\mu \phi^\dagger \partial^\mu \phi- m^2 \phi^\dagger \phi\,.
      \end{align}
      
      The Lagrangean has a $\mathsf{U(1)}$-symmetry, i.e.,
      under the transformation of the field 
      \begin{align}
        \phi\to\phi'= e^{\i \alpha} \phi\,,\quad \alpha=\const\,,
      \end{align}
      it stays invariant.  From Noether's theorem, there is a conserved current $j_\mu$ 
      associated with this symmetry,\index{Noether's theorem} 
      \begin{align}\label{eq:Noetherscalar}
        j^\mu&=\i\phi^\dagger\dyad^\mu\phi =\i\left(\phi^\dagger \partial^\mu
          \phi -\partial^\mu\phi^\dagger \phi\right)\,, & \partial_\mu j^\mu &=0\,.
      \end{align}
      
      The space integral of the time component of this current is conserved in time:
      \begin{align}
        Q&=\int \d^3 x\, \i \phi^\dagger \dyad^{\,0} \phi\,,\quad \partial_0 Q=0\,.
      \end{align}
      The time derivative vanishes because we can interchange derivation and integration and then
      replace $\partial_0 j^0$ by $\partial_i j^i$ since $\partial_\mu j^\mu =\partial_0 j^0 +
      \partial_i j^i=0$. So we are left with an integral of a total derivative which we can
      transform into a surface integral via Gauss' theorem. Since we always assume that all fields
      vanish at spatial infinity, the surface term vanishes.

      Now we need to construct the ``momentum'' $\pi(x)$ conjugate to the field $\phi$. Like in
      classical mechanics, it is given by the derivative of the Lagrangean with respect to the time
      derivative of the field,
      \begin{align}\index{Conjugate momentum!for scalar fields}
        \pi (x)&=\frac{\partial \L}{\partial\dot{\phi}(x)}=\dot{\phi}^\dagger(x)\,, & \pi^\dagger
        (x)&=\frac{\partial \L}{\partial\dot{\phi}^\dagger(x)}=\dot{\phi}\,.
      \end{align}
      
      At this point, we again replace the classical fields by operators which act on some Hilbert
      space of states and which obey certain commutation relations. The commutation relations we
      have to impose are analogous to Eq.~(\ref{eq:commpq}). The only non-vanishing commutators are the
      ones between field and conjugate momentum, at different spatial points but at equal times, 
      \begin{align}\index{Canonical commutation relations!for field operators}
        \left[\pi(t,\vec{x}), \phi(t,\vec{x}')\right]=\left[\pi^\dagger(t,\vec{x}),
          \phi^\dagger(t,\vec{x}')\right] &= -\i \delta^3\left(\vec{x}-\vec{x}'\right)\,,
      \end{align}
      all other commutators vanish.

      These relations are satisfied by the field operator defined in Eq.~(\ref{eq:fieldoperator})
      via the (anti)particle creation and annihilation operators. Its field equation can be derived
      from the Lagrangean,
      \begin{align}
        \partial_\mu \frac{\partial \L}{\partial(\partial_\mu \phi)} -\frac{\partial \L}{\partial
          \phi}&=\left(\Box+m^2\right) \phi^\dagger =0 \,.      
      \end{align}
      
      From the Lagrangean and the momentum, we can also construct the Hamiltonian density, 
      \begin{align}
        \mathscr{H}&= \pi \dot{\phi} +\pi^\dagger \dot{\phi}^\dagger- \L = \pi^\dagger \pi
        +\left(\vec{\nabla} \phi^\dagger\right)\left(\vec{\nabla} \phi\right) +m^2\phi^\dagger
        \phi\,. 
      \end{align}
      Note that canonical quantisation yields Lorentz invariant results,
      although it requires the choice of a particular time direction.

    \section{Fermions}
      Fermions are what makes calculations unpleasant.

      In the previous section we considered a scalar field, which describes particles with spin~0.
      In the standard model, there is just one fundamental scalar field, the Higgs field, which still
      remains to be discovered. There are other bosonic fields, gauge fields which carry spin 1
      (photons, $W^\pm$, $Z^0$ and the gluons). Those are described by vector fields which 
      will be discussed in Chapter~3. Furthermore, there are the matter fields, fermions with spin
      $\frac{1}{2}$, the quarks and leptons. 

      To describe fermionic particles, we need to introduce new quantities, spinor fields.
      These are four-component objects (but not vectors!)\index{Spinors}
      $\psi$, which are defined via a set of $\gamma$-matrices.\index{gamma@$\gamma$-matrices}
      These four-by-four matrices are labelled by a
      vector index and act on spinor indices.  They fulfill the anticommutation relations
      (the Clifford or Dirac algebra), \index{Dirac algebra}
      \begin{align}
        \left\{\gamma_\mu,\gamma_\nu\right\} &= 2 g_{\mu\nu}\mathbbm{1}\,,
      \end{align}
      with the metric $g_{\mu\nu}=\diag(+,-,-,-)$. The numerical form of the $\gamma$-matrices is
      not fixed, rather, one can choose among different possible representations. A common
      representation is the so-called chiral or Weyl representation, which is constructed from the Pauli
      matrices:\index{gamma@$\gamma$-matrices!Weyl representation}
      \begin{align}\label{eq:gammaweyl}
        \gamma^0&=
        \begin{pmatrix}
          0 &\mathbbm{1}_2 \\ \mathbbm{1}_2 &0
        \end{pmatrix}\,,
        & \gamma^i&=
        \begin{pmatrix}
          0 &\sigma^i\\-\sigma^i &0
        \end{pmatrix}\,.
      \end{align}
      This representation is particularly useful when one considers spinors of given
      chiralities. However, for other purposes, other representations are more convenient.
      Various rules and identities related to $\gamma$-matrices are collected in
      Appendix~\ref{sec:algebra-appendix}. 
      
      The Lagrangean for a free fermion contains, just as for a scalar, the kinetic term and the
      mass:
      \begin{align}
        \L&= \psib \i\dslash\psi -m \psib \psi \,.
      \end{align}
      The kinetic term contains only a first-order derivative, the operator $\dslash\equiv
      \gamma^\mu\partial_\mu$. The adjoint spinor $\psib$\index{Adjoint spinor} is defined as $\psib\equiv \psi^\dagger
      \gamma^0$. (The first guess $\psi^\dagger \psi$ is not Lorentz invariant.) 
      To derive the field equation, one has to treat $\psi$ and $\psib$ as independent
      variables. The Euler-Lagrange equation for $\psib$ is the familiar Dirac equation:
      \begin{align}\index{Dirac euation}
        0&=\frac{\partial \L}{\partial \psib} = \left(\i \dslash -m\right)\psi\,,
      \end{align}
      since $\L$ does not depend on derivatives of $\psib$.\footnote{Of course one can shift the
        derivative from $\psi$ to $\psib$ via integration by parts. This slightly modifies the
        computation, but the result is still the same.}

      The Lagrangean again has a $\mathsf{U(1)}$-symmetry, the multiplication of $\psi$ by a constant 
      phase,
      \begin{align}
        \psi\to\psi' =e^{\i\alpha} \psi\,, \qquad \psib\to\psib'=e^{-\i\alpha}\psib\,,
      \end{align}
      which leads to a conserved current and, correspondingly, to a conserved charge,
      \begin{align}      
        j^\mu&= \psib\gamma^\mu\psi\,,\quad \partial^\mu j_\mu=0\,,\quad Q=\int\!\d^3 x\,
        \psib\gamma^0\psi\,.
      \end{align}

      \subsection{Canonical Quantisation of Fermions}
      Quantisation proceeds along similar lines as in the scalar case. One first defines the
      momentum $\pi_\alpha$ conjugate to the field $\psi_\alpha$ ($\alpha=1,\ldots,4$),
      \begin{align}\index{Conjugate momentum!for fermions}
        \pi_\alpha&=\frac{\partial \L}{\partial \dot{\psi}_\alpha} = \i\left(\psib
          \gamma^0\right)_\alpha =\i \psi^\dagger_\alpha\,.
      \end{align}
      Instead of imposing commutation relations, however, for fermions one has to impose
      anticommutation relations. This is a manifestation of the Pauli exclusion principle which
      can be derived from the spin-statistics theorem. The relations are again postulated at 
      equal times (``canonical anticommutation relations''):
      \begin{subequations}\label{eq:fermiquant}
        \begin{align}\index{Canonical anticommutation relations!for spinor fields}
          \left\{\pi_\alpha(t,\vec{x}),\psi_\beta (t,\vec{x}') \right\} &= -\i \delta_{\alpha\beta}
          \delta^3\left(\vec{x}-\vec{x}'\right)\,, \\
          \left\{ \pi_\alpha(t,\vec{x}),\pi_\beta (t,\vec{x}')\right\}=\left\{
            \psi_\alpha(t,\vec{x}),\psi_\beta (t,\vec{x}')\right\}&=0 \,. 
        \end{align}
      \end{subequations}

      In order to obtain creation and annihilation operators, we again expand the field operator
      in terms of plane waves. Because of the four-component nature of the field, 
      now a spinor $u(p)$ occurs, where $p$ is the momentum four-vector of the plane wave: 
      \begin{align}\label{dirac}
        \left(\i \dslash -m\right) u(p) e^{-\i p x}&=0 \,, 
      \end{align}
      which implies
      \begin{align}
        \left(\slash[2]{p}-m\right) u(p)=0 \,.
      \end{align}
      This is an eigenvalue equation for the $4\times 4$-matrix $p_\mu \gamma^\mu$, which has two
      solutions for $p^2=m^2$ and $p^0>0$. They are denoted $u^{(1,2)}(p)$ and represent positive
      energy particles. Taking a positive sign in the exponential in Eq.~(\ref{dirac}), which is 
      equivalent to
        considering $p^0<0$, we obtain two more solutions, $v^{(1,2)}(p)$ that can be interpreted as
      antiparticles. The form of these solutions depends on the representation of the
      $\gamma$-matrices. For the Weyl representation they are given in the appendix.

      The eigenspinors determined from the equations ($\i=1,2$),\index{u@$u$ and $v$ spinors}
      \begin{align}
        \left(\slash[2]{p} -m \right) u^{(i)}(p)&=0\,, & \left(\slash[2]{p} +m \right) v^{(i)}(p)&=0\,,
      \end{align}
      obey the identities:
      \begin{gather}
        \ol{u}^{(i)}(p) u^{(j)}(p) = -\ol{v}^{(i)}(p) v^{(j)}(p)= 2m \delta_{ij}\,,\\
        \sum_i u^{(i)}_\alpha (p) \ol{u}^{(i)}_\beta (p) = \left(\slash[2]{p} +m\right)_{\alpha\beta}\,,
        \qquad \sum_i v^{(i)}_\alpha (p) \ol{v}^{(i)}_\beta (p) = \left(\slash[2]{p}
          -m\right)_{\alpha\beta}  \,.
      \end{gather}
      
      These are the ingredients we need to define creation and annihilation operators
      in terms of the spinor field $\psi(x)$ and its conjugate $\psib(x)$:
      \begin{subequations}
        \begin{align}\label{psiex}
          \psi(x)&=\int \dpbar \sum_i \left(b_i(p) u^{(i)}(p) e^{-\i px} + d^\dagger_i(p) v^{(i)}(p)
            e^{\i px} \right)\,,\\
          \psib(x) &= \int \dpbar \sum_i \left( b_i^\dagger(p) \ol{u}^{(i)}(p) e^{\i px} + d_i(p)
              \ol{v}^{(i)} (p) e^{-\i px}\right)\,. 
        \end{align}
      \end{subequations}
      Here, as before,
      \begin{align}
        \dpbar&=\frac{\d^3 p}{(2\pi)^3}\frac{1}{2 E_p}\,,\quad E_p=\sqrt{\vec{p}^2+m^2}\,.
      \end{align}
      Inverting Eq.~(\ref{psiex}) one obtains
      \begin{align}
        b_i(p)&=\int\! \d^3 x \,\ol{u}^{(i)}(p) e^{\i px} \gamma^0 \psi(x)\,,
      \end{align}
      and similar equations for the other operators.

      The creation and annihilation operators inherit the anticommutator algebra from the field
      operators,
      \begin{subequations}
        \begin{align}\index{Canonical anticommutation relations!for creation and annihilation operators}
          \left\{b_i(\vec{p}\mspace{1.5mu}),b_j^\dagger(\vec{p}\,') \right\} &=
          \left\{d_i(\vec{p}\mspace{1.5mu}),d_j^\dagger(\vec{p}\,') \right\} = (2\pi)^3 2 E_p
          \delta^3\left(\vec{p}-\vec{p}\,'\right)\,, \\
          \left\{b_i(\vec{p}\mspace{2mu}),d_j(\vec{p}\,') \right\} &= \text{all other anticommutators} =0 \,.
        \end{align}
      \end{subequations}

      The momentum and charge operators are again constructed from the creation and
      annihilation operators by ``counting'' the number of particles in each state and summing over
      all states,
      \begin{align}
        P^\mu &= \int\dkbar \, k^\mu \left(b^\dagger(k) b(k) + d^\dagger(k)d(k)\right)\,,\\
        Q&=\int\dkbar \left(b^\dagger(k) b(k) - d^\dagger(k)d(k)\right)\,.
      \end{align}
      These operators have the correct algebraic relations, which involve commutators, since $P^\mu$
      and $Q$ are bosonic operators (not changing the number of fermions in a given state):
      \begin{align}
        \left[P^\mu, b_i^\dagger(p) \right]&= p^\mu b_i^\dagger(p)\,, & \left[P^\mu, d_i^\dagger(p)
        \right]&= p^\mu d_i^\dagger(p)\,,\\
        \left[Q, b_i^\dagger(p) \right]&= b_i^\dagger(p)\,, & \left[Q, d_i^\dagger(p)\right]&=
        -d_i^\dagger(p)\,. 
      \end{align}
      
      An operator we did not encounter in the scalar case is the spin operator $\vec{\Sigma}$
      \index{Spin operator}. It 
      has three components, corresponding to the three components of an angular momentum
      vector\footnote{Actually, $\Sigma$ is constructed as a commutator of $\gamma$-matrices and as
        such has six independent components. But three of these correspond to Lorentz boosts which
        mix time and spatial directions. $\vec{\Sigma}$ is the spin operator in the rest frame.}. Only
      one combination of these components is, however, measurable. This is specified by a choice of
      quantisation axis, i.e., a spatial unit vector $\vec{s}$. The operator that measures the spin
      of a particle is given by the scalar product $\vec{s}\cdot\vec{\Sigma}$. Creation operators
      for particles with definite spin satisfy the commutation relations
      \begin{align}
        \left[\vec{s}\cdot\vec{\Sigma}, d_\pm^\dagger(p) \right]&= \mp\frac{1}{2}
        d_\pm^\dagger(p)\,, & \left[\vec{s}\cdot\vec{\Sigma}, b_\pm^\dagger(p) \right]&= \pm\frac{1}{2}
        b_\pm^\dagger(p)\,.
      \end{align}
      
      In summary, all these commutation relations tell us how to interpret the operators 
      $d_\pm^\dagger(p)$
      ($b_\pm^\dagger(p)$): They create spin-$\frac{1}{2}$ fermions with four-momentum $p^\mu$,
      charge $+1$ ($-1$) and spin orientation $\pm\frac{1}{2}$ ($\mp\frac{1}{2}$) relative to the
      chosen axis $\vec{s}$. Their conjugates $d_\pm(p)$ and $b_\pm(p)$ annihilate those particles.

      This immediately leads to the construction of the Fock space of fermions: We again start from
      a vacuum state $\ket{0}$, which is annihilated by the annihilation operators, and construct
      particle states by successive application of creation operators: 
      \begin{align*}\index{Hilbert space!for the spinor field}
        \text{vacuum: }              &\ket{0}\,,\quad b_i(p)\ket{0}=d_i(p)\ket{0}=0\\ 
        \text{one-particle states: } & b_i^\dagger(p)\ket{0}\,,\;d_i^\dagger(p)\ket{0} \\
        \text{two-particle states: } & b_i^\dagger(p_1)d_j^\dagger(p_2)\ket{0}\,,\ldots\\
        &\quad\vdots
      \end{align*}
      At this point we can verify that the Pauli principle\index{Pauli principle} is indeed
      satisfied, due to the choice of anticommutation relations in Eq.~(\ref{eq:fermiquant}). For a
      state of two fermions with identical quantum numbers, we would get
      \begin{align}
        \underbrace{b_i^\dagger(p) \,b_i^\dagger(p)}_{\text{anticommuting}} \ket{X} =
        -b_i^\dagger(p) \,b_i^\dagger(p) \ket{X}=0 \,,
      \end{align}
      where $\ket{X}$ is an arbitrary state.
      Had we quantised the theory with commutation relations
      instead, the fermions would have the wrong (i.e., Bose) statistics.

      The final expression we need for the further discussion is the propagator. By the same
      reasoning as in the scalar case, it is obtained as the time-ordered product of two field
      operators. The Feynman propagator $S_\text{F}$ for fermions,
      \index{Feynman propagator!for fermions}  which is now a matrix-valued object, is  given by 
      \begin{align}
        \begin{split}
          \i S_\text{F} (x_1-x_2)_{\alpha \beta} &= \braopket{0}{\T \psi_\alpha(x_1)
            \psib_\beta (x_2)}{0}\\
          &=\i \int \frac{\d^4 p}{(2\pi)^4} \frac{\left(\slash[2]{p} + m\right)_{\alpha\beta}}{p^2-
            m^2 +\i\varepsilon} e^{-\i p (x_1-x_2)} \,.
        \end{split}
      \end{align}

      This completes our discussion on the quantisation of free scalar and spinor fields.

   \section{Interactions}
     So far, we have considered free particles and their propagation. A theory of elementary
     particles obviously needs interactions. Unfortunately, they are much more difficult to handle,
     and little is known rigorously (except in two dimensions). Hence, we have to look for
     approximations. 

     By far the most important approximation method is perturbation theory where one treats the
     interaction as a small effect, a perturbation, to the free theory. The interaction strength is
     quantified by a numerical parameter, the coupling constant, and one expresses physical quantities
     as power series in this parameter. This approach has been very successful and has led to
     many celebrated results, like the precise prediction of the anomalous magnetic moment of the
     electron, despite the fact that important conceptual problems still remain to be resolved.

     \pagebreak

     \subsection[$\phi^4$ Theory]{$\boldsymbol \phi^4$ Theory}\index{Phi theory@$\phi^4$ theory}
     
       \begin{wrapfigure}[11]{R}[0pt]{210pt}
         \vspace{-.4cm}
         \begin{picture}(200,100)(0,0)
           \ArrowLine(20,90)(90,60) \ArrowLine(20,70)(80,55) \ArrowLine(20,30)(80,45)
           \ArrowLine(20,10)(90,40)
           \ArrowLine(110,60)(180,90) \ArrowLine(120,55)(180,70) \ArrowLine(120,45)(180,30)
           \ArrowLine(110,40)(180,10)
           \CCirc(100,50){20}{Red}{Red}
           \CTri(100,70)(120,50)(130,80){Red}{Red}
           \CTri(100,70)(80,50)(80,80){Red}{Red}
           \CTri(90,60)(90,40)(50,50){Red}{Red}
           \CTri(80,50)(100,35)(90,20){Red}{Red}
           \CTri(90,35)(115,45)(115,15){Red}{Red}
           \CTri(115,35)(110,60)(135,50){Red}{Red}
           \Text(14,92)[]{$p_1$} 
           \Text(30,52)[]{$\vdots$}
           \Text(14,8)[]{$p_n$}
           \Text(188,92)[]{$p_1'$} 
           \Text(170,52)[]{$\vdots$}
           \Text(188,8)[]{$p_m'$}
         \end{picture}
         \caption{Scattering of $n$ incoming particles, producing $m$ outgoing ones with momenta
           $p_1,\ldots,p_n$ and $p_1',\ldots,p_m'$, respectively. \label{fig:scatter}}
       \end{wrapfigure}
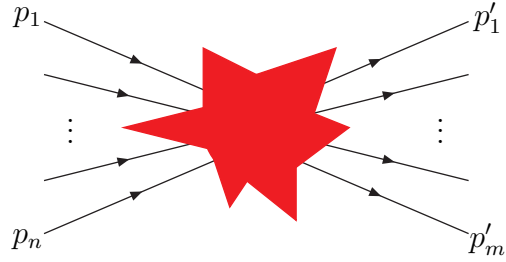
       Let us consider the simplest example of an interacting theory, involving only one real scalar
       field with a quartic self-interaction (a cubic term would look even simpler, but then the
       theory would not have a ground state since the energy would not be bounded from below):
       \begin{align}
         \begin{split}
           \L&=\L_0 +\L_\text{I}\\
           &= \frac{1}{2}\partial_\mu \phi \partial^\mu \phi -\frac{1}{2} m^2\phi^2
           -\frac{\lambda}{4!}\phi^4 \,.  
         \end{split}
       \end{align}
       $\L_0$ is the free Lagrangean, containing kinetic and mass term, while $\L_\text{I}$ is the
       interaction term, whose strength is given by the dimensionless coupling constant $\lambda$. 
        
        In perturbation theory we can calculate various physical quantities, in particular
        scattering cross sections for processes like the one in Fig.~\ref{fig:scatter}: 
        $n$ particles with
        momenta $p_i$ interact, resulting in $m$ particles with momenta $p_j'$. Since the
        interaction is localised in a region of spacetime, particles are free at 
        infinite past and future. In other words, we have free asymptotic states \index{Asymptotic states} 
        \begin{align}
        \ket{p_1,\ldots,p_n\,,\text{in}}\; \text{at}\;, t=-\infty\quad \text{and}\quad
        \ket{p_1',\ldots,p_m'\,,\text{out}}\; \text{at}\; t=+\infty\,. 
        \end{align}
        The transition amplitude for the scattering process is determined by the scalar product of
        incoming and outgoing states, which defines a unitary matrix, the so-called $S$-matrix
        \index{S@$S$ matrix}
        ($S$ for scattering), 
        \begin{align}\label{smatrix}
           \braket{p_1',\ldots,p_m' \,,\text{out}}{p_1,\ldots,p_n \,,\text{in}}=
            \braopket{p_1',\ldots,p_m'}{S}{p_1,\ldots,p_n}\,.
         \end{align}
        
        \begin{wrapfigure}[9]{R}[0pt]{210pt}
          \begin{picture}(200,70)(0,0)
            \ArrowLine(20,50)(80,35) \ArrowLine(20,30)(80,30)
            \ArrowLine(20,10)(80,25)
            \ArrowLine(120,35)(180,50) \ArrowLine(120,30)(180,30)
            \ArrowLine(120,25)(180,10)
            \CCirc(100,30){20}{Red}{Red}
            \CCirc(100,30){17}{Orange}{Orange}
            \CCirc(100,30){10}{Yellow}{Yellow}
            \ArrowLine(20,70)(180,70)
            \Text(14,72)[]{$p_1$} 
            \Text(13,33)[]{$\vdots$}
            \Text(14,8)[]{$p_n$}
            \Text(14,52)[]{$p_2$}            
            \Text(188,72)[]{$p_1'$}
            \Text(188,52)[]{$p_n$}
            \Text(189,33)[]{$\vdots$}
            \Text(188,8)[]{$p_m'$}
          \end{picture}
          \caption{\label{fig:disconnected} A disconnected diagram: One particle does not
            participate in the interaction.}
        \end{wrapfigure}
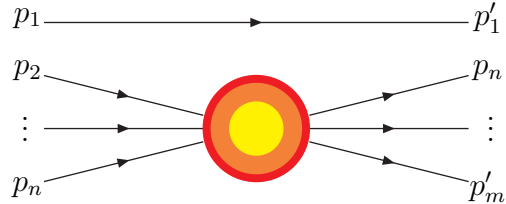

        Detailed techniques have been developed to obtain a perturbative expansion for the $S$-matrix
        from the definition (\ref{smatrix}). The basis are Wick's theorem and the LSZ-formalism. One
        starts from a generalisation of the propagator, the time-ordered product of $k$ fields,
        \begin{align}
          \begin{split}
            \tau(x_1,&\ldots,x_k) \\
            &=  \braopket{0}{\T \phi(x_1),\ldots \phi(x_k)}{0}\;.
          \end{split}
        \end{align}
        First, disconnected pieces involving non-interacting particles have to be subtracted
        (see Fig.~\ref{fig:disconnected}), \index{Disconnected diagrams}
        and the blob in Fig.~\ref{fig:scatter} decomposes into a smaller blob and
        straight lines just passing from the left to the right side.
        From the Fourier transform
        \begin{align}
        \tau(x_1',\ldots, x_m',x_1, \ldots, x_n) \stackrel{\text F.T.}{\longrightarrow} 
        \tilde{\tau}(p_1',\ldots,p_m',p_1,\ldots,p_n)
        \end{align}
        one then obtains the amplitude for the scattering process 
        \begin{align}
          \braopket{p_1',\ldots,p_m'}{S}{p_1,\ldots,p_n} &= \left(2\pi\right)^4
          \delta^4\!\left(\sum_\text{out} p_i'-\sum_\text{in} p_i\right) i\mathcal{M}\,,
        \end{align}
        where the matrix element $\mathcal{M}$ contains all the dynamics of the interaction. Due to
        the translational invariance of the theory, the total momentum is conserved. The matrix
        element can be calculated perturbatively up to the desired order in the coupling $\lambda$ via
        a set of Feynman rules. To calculate the cross section for a particular process, one first
        draws all possible  Feynman diagrams with a given number of vertices and then translates
        them into an analytic expression using the Feynman rules. 

        For the $\phi^4$ theory, the Feynman diagrams are all composed out of three building blocks:
        External lines corresponding to incoming or outgoing particles, propagators and
        4-vertices. The Feynman rules read: 
        \index{Feynman rules! for $\phi^4$ theory}
        
        \begin{enumerate}[i.]
        \item \begin{minipage}{60pt}
            \begin{picture}(50,20)(0,0)\small
              \SetOffset(0,5) \DashArrowLine(3,1)(47,1){2} \Vertex(47,1){2}\Text(25,7)[]{$p$}
            \end{picture}
          \end{minipage}
          \begin{minipage}{75pt}
            \begin{center}
              \raisebox{0pt}{
                $1$
              }
            \end{center}
          \end{minipage}
          \begin{minipage}[t]{282pt}
            External lines: For each external line, multiply by 1 (i.e., external lines
            don't contribute to the matrix element in this theory). However, one needs to keep
            track of the momentum of each particle entering or leaving the interaction. The momentum
            direction is indicated by the arrow.
          \end{minipage}
        \item 
          \begin{minipage}{60pt}
            \begin{picture}(50,20)(0,0)\small
              \SetOffset(0,5)\Vertex(3,1){2}\DashArrowLine(3,1)(47,1){2}\Vertex(47,1){2}\Text(25,7)[]{$p$}
            \end{picture}
          \end{minipage}
          \begin{minipage}{75pt}
            \begin{center}
              \raisebox{-6ex}{
                $\displaystyle \frac{\i}{p^2-m^2 +\i\varepsilon}$
              }
            \end{center}
          \end{minipage}
          \begin{minipage}[t]{282pt}
            Propagators between vertices are free propagators corresponding to the momentum
            of the particle. Note that particles of internal lines need not be on-shell, i.e.,
            $p^2=m^2$ need not hold! 
          \end{minipage}
        \item 
          \begin{minipage}{60pt}
            \begin{picture}(50,20)(0,0)\small
              \DashArrowLine(3,10)(25,-12){2}\DashArrowLine(47,10)(25,-12){2}
              \DashArrowLine(3,-34)(25,-12){2}\DashArrowLine(47,-34)(25,-12){2} \Vertex(25,-12){2} 
            \end{picture}
          \end{minipage}
          \begin{minipage}{75pt}
            \begin{center}
              \raisebox{0pt}{
                $\displaystyle -\i \lambda$
              }
            \end{center}
          \end{minipage}
          \begin{minipage}[t]{282pt}
            Vertices yield a factor of the coupling constant. In this theory, there is only one
            species of particles, and the interaction term does not contain derivatives, so there
            is only one  vertex, and it does not depend on the momenta.
          \end{minipage}
        \item 
          \begin{minipage}{60pt}
            \begin{picture}(50,20)(0,0)\small
              \SetOffset(0,5)\DashArrowArcn(27.5,-29)(39,130,50){2}
              \DashArrowArcn(27.5,31)(39,310,230){2}\Vertex(3,1){2} \Vertex(52,1){2}  \DashLine(3,1)(-2,4){2}
              \DashLine(3,1)(-2,-2){2} \Line(52,1)(57,4) \DashLine(52,1)(57,-2){2}  
            \end{picture}
          \end{minipage}
          \begin{minipage}{75pt}
            \begin{center}  
              \raisebox{-7ex}{
                $\displaystyle \int \frac{\d^4 p}{\left(2\pi\right)^4}$
              }
            \end{center}
          \end{minipage}
          \begin{minipage}[t]{282pt}
            The momenta of internal loops are not fixed by the incoming momenta. For each
            undetermined loop momentum $p$, one integrates over all values of $p$. 
          \end{minipage}
        \end{enumerate}

        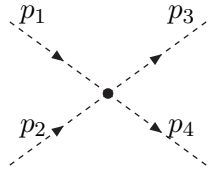
\begin{figure}\small
          \begin{center}
            \subfigure[Tree~graph]{\begin{picture}(82,60)(0,0)
                \DashArrowLine(3,3)(40,30){2} \DashArrowLine(3,57)(40,30){2} \DashArrowLine(40,30)(77,3){2}
                \DashArrowLine(40,30)(77,57){2} \Vertex(40,30){2}
                \Text(12,60)[]{$p_1$} \Text(12,17)[]{$p_2$} \Text(68,60)[]{$p_3$} \Text(68,17)[]{$p_4$}
              \end{picture}}\hspace{9mm}
            \begin{minipage}[b]{220pt}
              \caption{\label{fig:2to2} Feynman graphs for $2\to2$ scattering in $\phi^4$ theory to
                second order. The one-loop graphs all are invariant under the interchange of the internal 
              lines and hence get a symmetry factor of $\frac{1}{2}$.\vspace*{3mm}} 
            \end{minipage}\\
            \subfigure[One-loop graphs]{
              \begin{picture}(100,60)
                \DashArrowLine(3,3)(40,3){2} \DashArrowLine(40,3)(97,3){2} \Vertex(40,3){2}
                \DashArrowLine(3,57)(40,57){2} 
                \DashArrowLine(40,57)(97,57){2} \Vertex(40,57){2} \DashArrowArcn(60,30)(33.8,232,128){2}
                \DashArrowArcn(20,30)(33.8,52,308){2}
                \Text(8,51)[]{$p_1$} \Text(8,9)[]{$p_2$}  \Text(92,51)[]{$p_3$} \Text(92,9)[]{$p_4$}
                \Text(20,30)[]{$p$} \Text(60,30)[l]{$p+p_1 -p_3$}
              \end{picture}
              \hspace{1cm}
              \begin{picture}(110,60)
                \DashArrowLine(3,3)(30,3){2} \Vertex(30,3){2}
                \DashArrowLine(3,57)(30,57){2} \Vertex(30,57){2}
                \DashLine(30,3)(60,23.75){2} \DashArrowLine(60,23.75)(110,57){2}
                \DashLine(30,57)(60,36.75){2} \DashArrowLine(60,36.75)(110,3){2}  
                \DashArrowArcn(50,30)(33.8,232,128){2} \DashArrowArcn(10,30)(33.8,52,308){2}
                \Text(8,51)[]{$p_1$} \Text(8,9)[]{$p_2$}  \Text(90,51)[]{$p_3$} \Text(90,9)[]{$p_4$}
                \Text(10,30)[]{$p$} 
              \end{picture}
              \hspace{1cm} 
              \begin{picture}(100,60)
                \DashArrowLine(3,3)(25,30){2} \DashArrowLine(3,57)(25,30){2} \Vertex(25,30){2} 
                \DashArrowArcn(50,0)(39,130,50){2} \DashArrowArcn(50,60)(39,310,230){2}
                \Vertex(75,30){2} \DashArrowLine(75,30)(97,3){2} \DashArrowLine(75,30)(97,57){2} 
                \Text(3,45)[]{$p_1$} \Text(3,15)[]{$p_2$} \Text(97,45)[]{$p_3$} \Text(97,15)[]{$p_4$} 
                \Text(50,45)[]{$p$} \Text(50,15)[]{$p_1+p_2-p$}
              \end{picture}}            
          \end{center}
        \end{figure}

        As an example, let us calculate the matrix element for the $2\to2$ scattering process to
        second order in $\lambda$. \index{$2\to2$ scattering in $\phi^4$ theory}
        The relevant diagrams are collected in Fig.~(\ref{fig:2to2}). The first-order
        diagram simply contributes a factor of $-\i\lambda$, while the second-order diagrams involve an
        integration:
        \begin{align}
          \begin{split}
            \i \mathcal{M} &= -\i\lambda +\frac{1}{2}\left(-\i\lambda\right)^2\int \frac{\d^4
              p}{\left(2\pi\right)^4} \, \frac{\i}{p^2-m^2}\, \frac{\i}{\left(p+p_1
                -p_3\right)^2-m^2}\\
            &\quad\mspace{35.5mu}+\frac{1}{2}\left(-\i\lambda\right)^2\int \frac{\d^4
              p}{\left(2\pi\right)^4} \, \frac{\i}{p^2-m^2}\, \frac{\i}{\left(p+p_1 
                -p_4\right)^2-m^2}\\
            &\quad\mspace{35.5mu}+\frac{1}{2}\left(-\i\lambda\right)^2\int \frac{\d^4
              p}{\left(2\pi\right)^4} \, 
            \frac{\i}{p^2-m^2}\, \frac{\i}{\left(p_1+p_2 -p\right)^2 -m^2} + 
             \mathcal{O}\left(\lambda^3\right)\,.
          \end{split}
        \end{align}
        The factors of $\frac{1}{2}$ are symmetry factors which arise if a diagram is invariant
        under interchange of internal lines. The expression for $\mathcal{M}$ has a serious problem:
        The integrals do not converge. This can be seen by counting the powers of the integration
        variable $p$. For $p$ much larger that incoming momenta and the mass, the integrand behaves
        like $p^{-4}$. That means that the integral depends logarithmically on the upper integration
        limit,  
        \begin{align}
          \int^\Lambda \frac{\d^4 p}{\left(2\pi\right)^4} \, \frac{\i}{p^2-m^2}\, \frac{\i}{\left(p+p_1
              -p_3\right)^2-m^2}\,  \xrightarrow{\mspace{10mu} \displaystyle p\gg p_i,m
            \mspace{10mu}} \, 
          \int^\Lambda \frac{\d^4 p}{\left(2\pi\right)^4} \frac{-1}{p^4} \, \propto \, \ln \Lambda\,.
        \end{align}
        Divergent loop diagrams are ubiquitous in quantum field theory. They
        can be cured by regularisation, i.e., making the integrals finite by introducing some cutoff
        parameter, and renormalisation, where this additional parameter is removed
        in the end, yielding finite results for observables. This will be discussed in more detail 
        in the chapter on quantum corrections.

        \subsection{Fermions} 
        We can augment the theory by adding a fermionic field $\psi$, with
        a Lagrangean including an interaction with the scalar $\phi$,
        \begin{align}
          \L_\psi&= \underbrace{\psib\left(\i\dslash -m\right) \psi}_\text{free Lagrangean} -
          \underbrace{g \psib \phi \psi}_\text{interaction}\,.
        \end{align}
        
        There are additional Feynman rules for fermions. The lines carry two arrows, one for the
        momentum as for the scalars and one for the fermion number flow, which basically distinguishes
        particles and antiparticles. The additional rules are:\label{sec:fermionrules}
        \index{Feynman rules!for fermions}
        \begin{enumerate}[i.]
        \item \begin{minipage}[t]{60pt}
            \begin{picture}(50,20)(0,0)\small              
              \ArrowLine(3,0)(47,0)\Vertex(47,0){2}\Text(25,10)[]{$p$} \Text(25,4)[]{$\longrightarrow$}
            \end{picture}\\            
            \begin{picture}(50,20)(0,0)\small              
              \ArrowLine(3,0)(47,0)\Vertex(3,0){2}\Text(25,10)[]{$p$} \Text(25,4)[]{$\longrightarrow$}
            \end{picture}
          \end{minipage}
          \begin{minipage}[t]{75pt}
            \begin{center}
              \raisebox{-1pt}{
                $u(p)$
              }\\
              \raisebox{-12pt}{
                $\ol{u}(p)$
              }
            \end{center}
          \end{minipage}
          \begin{minipage}[t]{282pt}
            Incoming or outgoing particles get a factor of $u(p)$ or $\ol{u}(p)$,
            respectively. 
          \end{minipage}
        \item \begin{minipage}[t]{60pt}
            \begin{picture}(50,20)(0,0)\small              
              \ArrowLine(47,0)(3,0) \Vertex(47,0){2}\Text(25,10)[]{$p$} \Text(25,4)[]{$\longrightarrow$}
            \end{picture}\\            
            \begin{picture}(50,20)(0,0)\small              
              \ArrowLine(47,0)(3,0) \Vertex(3,0){2}\Text(25,10)[]{$p$} \Text(25,4)[]{$\longrightarrow$}
            \end{picture}
          \end{minipage}
          \begin{minipage}[t]{75pt}
            \begin{center}
              \raisebox{-1pt}{
                $\ol{v}(p)$
              }\\
              \raisebox{-12pt}{
                ${v}(p)$
              }
            \end{center}
          \end{minipage}
          \begin{minipage}[t]{282pt}
            Incoming or outgoing antiparticles get a factor of $\ol{v}(p)$ or ${v}(p)$,
            respectively. 
          \end{minipage}
        \item \begin{minipage}[t]{60pt}
            \begin{picture}(50,20)(0,0)\small              
              \Vertex(3,0){2} \ArrowLine(3,0)(47,0) \Vertex(47,0){2}\Text(25,10)[]{$p$}
              \Text(25,4)[]{$\longrightarrow$} 
            \end{picture}          
          \end{minipage}
          \begin{minipage}[t]{75pt}
            \begin{center}
              \raisebox{-7pt}{
                $\displaystyle \frac{\i\left(\slash[2]{p} +m\right)}{p^2-m^2 +\i\varepsilon}$
              }
            \end{center}
          \end{minipage}
          \begin{minipage}[t]{282pt}
            Free propagator for fermion with momentum $p$.
          \end{minipage}
        \item \begin{minipage}[t]{60pt}
            \begin{picture}(50,20)(0,0)\small              
              \ArrowLine(3,-10)(25,-10) \Vertex(25,-10){2} \ArrowLine(25,-10)(47,-10)
              \DashLine(25,-10)(25,10){2} 
            \end{picture}          
          \end{minipage}
          \begin{minipage}[t]{75pt}
            \begin{center}
              \raisebox{-1pt}{
                $-\i g$
              }
            \end{center}
          \end{minipage}
          \begin{minipage}[t]{282pt}
            The fermion-fermion-scalar vertex yields a factor of the coupling constant. Again, there
            is no momentum dependence.
          \end{minipage}
        \end{enumerate}


  \chapter{Gauge Theories}
    In addition to spin-0 and spin-$\frac{1}{2}$ particles, the standard model contains spin-1
    particles. They are the quanta of vector fields which can describe strong and electroweak
    interactions. The corresponding theories come with a local (``gauge'') symmetry and are called
    gauge theories.
    
    \section{Global Symmetries v Gauge Symmetries}
      Consider a complex scalar field with the Lagrangean
      \begin{align}\label{eq:Lscalar}
        \L&=\partial_\mu \phi^\dagger \partial^\mu - V\!\left(\phi^\dagger\phi\right)\,,
      \end{align}
      which is a generalisation of the one considered in Eq.~(\ref{eq:Lscalarfree}). This theory has a
      $\mathsf{U(1)}$ symmetry under which $\phi\to\phi' = \exp\!\left\{\i\alpha\right\}\phi$ with
      constant parameter $\alpha$. Usually it is sufficient to consider the variation of the fields
      and the Lagrangean under infinitesimal transformations,
      \begin{align}\label{eq:scalar_u1}
        \delta\phi &= \phi' -\phi = \i \alpha \phi \,,\qquad
        \delta\phi^\dagger= -\i\alpha \phi^\dagger\,,
      \end{align}
      where terms $\mathcal{O}\left(\alpha^2\right)$ have been neglected.
      To derive the Noether current\index{Noether current}, Eq.~(\ref{eq:Noetherscalar}), we
      compute the variation of the Lagrangean under such a transformation:
      \begin{align}\label{eq:derivation}
        \begin{split}
          \delta\L &= \frac{\partial \L}{\partial \phi}\delta \phi + \frac{\partial \L}{\partial
            \phi^\dagger}\delta \phi^\dagger +\frac{\partial \L}{\partial \left(\partial_\mu
              \phi\right)} \underbrace{\delta \left(\partial_\mu \phi\right)}_{=\partial_\mu
            \delta\phi} +\frac{\partial \L}{\partial \left(\partial_\mu \phi^\dagger\right)
          }\delta  \left(\partial_\mu \phi^\dagger\right)\\ 
          &= \underbrace{\left(\frac{\partial \L}{\partial \phi}- \partial_\mu \frac{\partial \L}{\partial
                \left(\partial_\mu \phi\right)}\right)}_{=0\text{ by equation of motion}} \delta
          \phi + \underbrace{\left(\frac{\partial \L}{\partial \phi^\dagger}- \partial_\mu
              \frac{\partial \L}{\partial \left(\partial_\mu \phi^\dagger\right)}\right)}_{=0}
          \delta \phi^\dagger\\ 
          &\quad + \partial_\mu \left(
            \frac{\partial \L}{\partial \left(\partial_\mu \phi\right)} \delta\phi
            +\frac{\partial \L}{\partial \left(\partial_\mu \phi^\dagger\right)}
            \delta\phi^\dagger\right)\\
          &= \alpha \partial_\mu\left(\i \partial^\mu \phi^\dagger \phi -\i
            \phi^\dagger\partial^\mu \phi\right)\\
          &=-\alpha \partial_\mu j^\mu \,.
        \end{split}
      \end{align}
      Since the Lagrangean is invariant, $\delta\L = 0$, we obtain a conserved current for 
      solutions of the equations of motion,
      \begin{align}    
        \partial_\mu j^\mu =0\,.
      \end{align}  
      From the first to the second line we have used that
      \begin{align}
        \frac{\partial \L}{\partial \left(\partial_\mu \phi\right)} \partial_\mu \delta \phi =
        \partial_\mu\left(\frac{\partial \L}{\partial \left(\partial_\mu \phi\right)} \delta
          \phi\right)  - \left(\partial_\mu \frac{\partial \L}{\partial \left(\partial_\mu \phi\right)}\right)
        \delta \phi           
      \end{align}
      by the Leibniz rule. 

      The above procedure can be generalised to more complicated Lagrangeans and
      symmetries. The derivation does not depend on the precise form of $\L$, and up to the second
      line of (\ref{eq:derivation}), it is independent of the form of $\delta \phi$. As a general
      result, a symmetry of the Lagrangean always implies a conserved current, which in turn gives a
      conserved quantity (often referred to as charge, but it can be angular momentum or energy as
      well). 
      
      What is the meaning of such a symmetry? Loosely speaking, it states that ``physics does not
      change'' under such a transformation. This, however, does not mean that the solutions to the
      equations of motion derived from this Lagrangean are invariant under such a
      transformation. Indeed, generically they are not, and only $\phi\equiv 0$ is invariant.

      As an example, consider the Mexican hat potential\index{Mexican hat potential},
      \begin{align}
        V(\phi^\dagger\phi)=-\mu^2 \phi^\dagger\phi +\lambda\left(\phi^\dagger\phi\right)^2\,.
      \end{align}
      This potential has a ring of minima, namely all fields for which
      $\left|\phi\right|^2=\mu^2/(2\lambda)$. This means that any constant $\phi$ with this modulus
      is a solution to the equation of motion,  
      \begin{align}
        \Box \phi +\frac{\partial V}{\partial \phi}\! \left(\phi,\phi^\dagger\right) = \Box \phi -\phi^\dagger
        \left(\mu^2-2\lambda\phi^\dagger\phi \right)=0\,.
      \end{align}
      These solutions are not invariant under $\mathsf{U(1)}$ phase rotations. On the other hand,
      it is obvious that any solution to the equations of motion will be mapped into another
      solution under such a transformation.
      
      This situation is analogous to the Kepler problem: A planet moving
      around a stationary 
      (very massive) star. The setup is invariant under spatial rotations around the star, i.e., the
      symmetries form the group $\mathsf{SO(3)}$. This group is three-dimensional 
      (meaning that any rotation can be built from three independent rotations, e.g.\ around the
      three axes of a Cartesian coordinate system). Thus there are three conserved charges
      which correspond to the three components of angular momentum. The
      solutions of this problem -- the planet's orbits -- are ellipses in a plane, so they are not
      at all invariant under spatial rotations, not even under rotations in the plane of
      motion. Rotated solutions, however, are again solutions.
      
      In particle physics, most experiments are scattering experiments at colliders. For those, the
      statement that ``physics does not change'' translates into ``transformed initial states lead
      to transformed final states'': If one applies the transformation to the initial state and
      performs the experiment, the result will be the same as if one had done the experiment with
      the untransformed state and transformed the result.
    
      \medskip
      
      There is a subtle, but important, difference between this and another type of symmetry,
      gauge symmetry.  A gauge transformation is also a transformation which leaves the Lagrangean
      invariant, but it does relate identical states which describe exactly the same
      physics.
      
      This might be familiar from electrodynamics. One formulation uses electric and magnetic
      fields $\vec{E}$ and $\vec{B}$, together with charge and current densities $\rho$ and
      $\vec{j}$. These fields and sources are related by Maxwell's equations:
      \index{Maxwell's equations}
      \begin{subequations}
        \begin{align}
          \vec{\nabla} \times \vec{E} + \frac{\partial \vec{B}}{\partial t} &=0 \,, & \vec{\nabla}\cdot
          \vec{B}&=0\,,\\ 
          \vec{\nabla} \times \vec{B} - \frac{\partial \vec{E}}{\partial t} &=\vec{j}\,,
          &\vec{\nabla}\cdot\vec{E} &=\rho\,.
        \end{align}
      \end{subequations}
      The first two of these can be identically solved by introducing the potentials $\phi$
      and $\vec{A}$\index{Gauge potential!electromagnetic}, which yield $\vec{E}$ and $\vec{B}$ via
      \begin{align}\label{eq:gt}
        \vec{E}&= -\vec{\nabla} \phi -\frac{\partial\vec{A}}{\partial t}\,, & \vec{B}
        &=\vec{\nabla}\times\vec{A}\,.
      \end{align}
      So we have reduced the six components of $\vec{E}$ and $\vec{B}$ down to the four independent
      ones $\phi$ and $\vec{A}$. However, the correspondence between the physical fields and
      the potentials is not unique. If some potentials $\phi$ and $\vec{A}$ lead to
      certain $\vec{E}$ and $\vec{B}$ fields, the transformed potentials
      \begin{align}\label{eq:maxwellgaugetrafo}
        \vec{A}'&= \vec{A} +\vec{\nabla} \Lambda\,, & \phi'& = \phi- \frac{\partial\Lambda}{\partial t}\,,
      \end{align}
      where $\Lambda$ is a scalar field, give the same electric and magnetic fields.
      
      This transformation~(\ref{eq:gt}) is called gauge transformation.
      \index{Gauge transformation} It is a symmetry of the theory, but it
      is different from the global symmetries we considered before. First, it is a local
      transformation, i.e., the transformation parameter $\Lambda$ varies in space and
      time. Second, it relates physically indistinguishable field configurations, since solutions 
      of the equations of motion for electric and magnetic fields are
      invariant. It is important to note that this gauge transformation is inhomogeneous, i.e.,
      the variation is not multiplicative, but can generate non-vanishing potentials from zero.
      Potentials that are related to $\phi=0$ and $\vec{A}=0$ by a gauge
      transformation are called pure gauge.
      
      Phrased differently, we have expressed the physical fields $\vec{E}$ and $\vec{B}$ in terms 
      of the potentials $\phi$ and $\vec{A}$. These potentials still contain too many degrees of
      freedom for the
      physical fields $\vec{E}$ and $\vec{B}$, since different potentials can lead to the
      same  $\vec{E}$ and $\vec{B}$ fields. So the description in terms of potentials is
      redundant, and the gauge transformation~(\ref{eq:maxwellgaugetrafo}) quantifies just
      this redundancy. Physical states and observables have to be
      invariant under gauge transformations.

    \section{Abelian Gauge Theories}
      The easiest way to come up with a gauge symmetry is to start from a global symmetry and promote
      it to a gauge one, that is, demand invariance of the Lagrangean under local
      transformations (where the transformation parameter is a function of spacetime). To see
      this, recall the Lagrangean with the global $\mathsf{U(1)}$ symmetry from the preceding
      section, 
      \begin{align*}
        \L&=\partial_\mu \phi^\dagger \partial^\mu\phi - V(\phi^\dagger\phi)\,,
      \end{align*}
      and the transformation
      \begin{align*}
        \phi&\to\phi' = e^{\i\alpha} \phi\,,\quad \delta\phi = \phi'-\phi = \i \alpha \phi\,.
      \end{align*}
      
      If we now allow spacetime dependent parameters $\alpha(x)$, the Lagrangean is no longer
      invariant. The potential part still is, but the kinetic term picks up derivatives of
      $\alpha(x)$, so the variation of the Lagrangean is
      \begin{align}
        \delta \L&= \i \partial_\mu \alpha \left(\partial^\mu\phi^\dagger \phi -\phi^\dagger
          \partial^\mu \phi\right) = -\partial_\mu\alpha \,j^\mu\,,
      \end{align}
      the derivative of $\alpha$ times the Noether current of the global symmetry derived
      before. 

      The way to restore invariance of the Lagrangean is to add another field, the gauge
      field, with a gauge transformation just like the electromagnetic potentials in the previous
      section, combined into a four-vector $A^\mu=(\phi,\vec{A})$:
      \begin{align}\label{eq:abeliantrafo}
        A_\mu(x) &\to A'_\mu(x) = A_\mu(x) - \frac{1}{e} \partial_\mu \alpha(x)\,.
      \end{align}
      The factor $\frac{1}{e}$ is included for later convenience. We can now combine the
      inhomogeneous transformation of $A_\mu$ with the inhomogeneous transformation of the
      derivative in a covariant derivative~$D_\mu$\index{Covariant derivative}:
      \begin{align}
        D_\mu \phi &= \left(\partial_\mu +\i e A_\mu\right)\phi\,.
      \end{align}
      This is called covariant derivative because the differentiated object $D_\mu\phi$ transforms in
      the same way as the original field, 
      \begin{align}
        \begin{split}
          D_\mu \phi \longrightarrow \left(D_\mu \phi\right)' &= \left(\partial_\mu +\i e
            A_\mu'\right) \phi'\\ 
          &=\partial_\mu \left(e^{\i\alpha(x)}\phi\right)  +\i e \left(A_\mu(x) - \frac{1}{e}
            \partial_\mu \alpha(x) \right) e^{\i\alpha(x)}\phi\\ 
          &= e^{\i\alpha(x)}D_\mu \phi\,.
        \end{split}
      \end{align}
      So we can construct an invariant Lagrangean from the field and its covariant derivative:
      \begin{align}
        \L&=\left(D_\mu \phi\right)^\dagger \left(D^\mu \phi\right) -
        V\left(\phi^\dagger\phi\right)\,.
      \end{align}
      
      So far this is a theory of a complex scalar with $\mathsf{U(1)}$ gauge invariance. The gauge
      field $A_\mu$, however, is not a dynamical field, i.e., there is
      no kinetic term for it.  This kinetic term should be gauge invariant and contain derivatives
      up to second order. In order to find such a kinetic term, we first construct the field
      strength\index{Field strength} tensor from the commutator of two covariant derivatives: 
      \begin{align}
        \begin{split}
          F_{\mu\nu}&=-\frac{\i}{e}\left[D_\mu, D_\nu\right] = -\frac{\i}{e}\left[\left(\partial_\mu
              +\i e A_\mu\right),\left(\partial_\nu+\i e A_\nu\right) \right]\\
          &= -\frac{\i}{e} \left(\left[\partial_\mu,\partial_\nu\right]+\left[\partial_\nu,\i e
              A_\nu\right] +\left[\i e
              A_\mu,\partial_\nu\right]-e^2\left[A_\mu,A_\nu\right]\right) \\
          &=\partial_\mu A_\nu -\partial_\nu A_\mu\,.
        \end{split}
      \end{align}
      To check that this is a sensible object to construct, we can redecompose $A_\mu$ into the
      scalar and vector potential $\phi$ and $\vec{A}$ and spell out the field strength tensor in
      electric and magnetic fields,
      \begin{align}
        F^{\mu\nu}&=
        \begin{pmatrix}
          0   & -E_1 & -E_2 & -E_3\\
          E_1 &   0  & -B_3 & B_2 \\
          E_2 &  B_3 &   0  & -B_1\\
          E_3 & -B_2 &  B_2 & 0
        \end{pmatrix}\,.
      \end{align}
      This shows that the field strength is gauge invariant, as $\vec{E}$ and $\vec{B}$ are. Of
      course, this can also be shown by straightforward calculation,
      \begin{align}
        \delta F_{\mu\nu} & = \partial_\mu \delta A_\nu -\partial_\nu \delta A_\mu =
        -\frac{1}{e}\left(\partial_\mu\partial_\nu - \partial_\nu\partial_\mu\right)\alpha(x) =0\,,
      \end{align}
      so it is just the antisymmetry in $\mu$ and $\nu$ that ensures gauge invariance.
      
      The desired kinetic term is now just the square of the field strength tensor,
      \begin{subequations}
        \begin{align}
          \L_\text{gaugekin}&= -\frac{1}{4} F_{\mu\nu} F^{\mu\nu}\,,
          \intertext{or, in terms of $\vec{E}$ and $\vec{B}$ fields,}
          \L&=\frac{1}{2}\left(\vec{E}^2-\vec{B}^2\right)
        \end{align}
      \end{subequations}
        
      The coupling to scalar fields via the covariant derivative can also be applied to fermions.
      To couple a fermion $\psi$ to the gauge field, one simply imposes the gauge
      transformation 
      \begin{align}
        \psi&\to\psi' = e^{\i\alpha} \psi\,.
      \end{align}
      In the Lagrangean, one again replaces the ordinary derivative with the covariant one. The
      Lagrangean for a fermion coupled to a $\mathsf{U(1)}$ gauge field is quantum
      electrodynamics~(QED), if we call the fields electron and photon:\index{Lagrangean!QED}
      \begin{align}
        \L_\text{QED}&= -\frac{1}{4} F_{\mu\nu} F^{\mu\nu} +\psib\left(\i \Dslash -m\right) \psi \,.
      \end{align}
      
      Finally, let us note that for a $\mathsf{U(1)}$ gauge theory, different fields may have
      different charges under the gauge group (as e.g.\ quarks and leptons indeed do). For fields
      with charge $q$ (in units of elementary charge), we have to
      replace the gauge transformations and consequently the covariant derivative as follows:
      \begin{align}
        \psi_q&\to \psi_q' = e^{\i q \alpha} \psi_q\,, & D_\mu^{(q)} \psi_q &= \left(\partial_\mu +\i
          q e A_\mu\right) \psi_q\,.
      \end{align}
      
      What have we done so far? We started from a Lagrangean, Eq.~(\ref{eq:Lscalar}) with a
      global $\mathsf{U(1)}$ symmetry (\ref{eq:scalar_u1}). We imposed invariance under local
      transformations, so we had to introduce a new field, the gauge field $A_\mu$. This
      field transformed inhomogeneously under gauge transformations, just in a way to make a
      covariant derivative. This covariant derivative was the object that coupled the gauge field
      to the other fields of the theory.  To make this into
      a dynamical theory, we added a kinetic term for the gauge field, using the field strength
      tensor. Alternatively, we could have started with the gauge field and tried to couple it to
      other fields, and we would have been led to the transformation properties
      (\ref{eq:scalar_u1}).  This is all we need to construct the Lagrangean for QED. For QCD
      and the electroweak theory, however, we need a richer structure: non-Abelian gauge theories.

    \section{Non-Abelian Gauge Theories}
      To construct non-Abelian theories in the same way as before, we first have to discuss
      non-Abelian groups, i.e., groups whose elements do not commute. We will focus on the groups
      $\sun$\index{SUN@$\mathsf{SU(}n\mathsf{)}$}, since they are  
      most relevant for the standard model. $\sun$ is the group of $n\times n$ complex unitary
      matrices with determinant 1. To see how many degrees of freedom there are, we have to count: A
      $n\times n$ complex matrix $U$ has $n^2$ complex entries, equivalent to  $2n^2$ real
      ones. The unitarity constraint, $U^\dagger U=\mathbbm{1}$, is a matrix equation, but not all
      component equations are independent. Actually, $U^\dagger U$  is Hermitean, $\left(U^\dagger
        U\right)^\dagger=U^\dagger U$, so the diagonal entries are real and the 
      lower triangle is the complex conjugate of the upper one. Thus, there are $n +2\cdot\frac{1}{2}
      n (n-1)$ real constraints.  Finally, by taking the determinant of the unitarity constraint,
      $\det \left(U^\dagger U\right)=\left|\det U\right|^2 =1$. Hence, restricting to $\det U=1$ eliminates
      one more real degree of freedom. All in all, we have $2 n^2 - n -2\cdot\frac{1}{2}n (n-1)
      -1=n^2-1$ real degrees of freedom in the elements of $\sun$.
        
      This means that any $U\in\sun$ can be specified by $n^2-1$ real parameters
      $\alpha_a$. The group elements are usually written in terms of these parameters and $n^2-1$
      matrices $T^a$, the generators of the group\index{Group generators}, as an exponential
      \begin{align}
        U & = \exp\left\{\i \alpha_a T^a\right\}= \mathbbm{1} + \i \alpha_a T^a
        +\mathcal{O}\left(\alpha^2\right) \,, 
      \end{align}
      and one often considers only infinitesimal parameters.
      
      The generators are usually chosen as Hermitean matrices\footnote{Actually, the generators
        live in the Lie algebra of the group, and so one can choose any basis one likes, Hermitean
        or not.}.  The product of group elements translates into commutation relations for the
      generators, 
      \begin{align}
        \left[T^a,T^b\right]&=\i f^{abc}  T^c\,,
      \end{align}
      with the antisymmetric structure constants\index{Structure constants} $f^{abc}$, which of
      course also depend on the choice of generators.

      In the standard model, the relevant groups are $\sutwo$ for the electroweak theory and
      $\suthree$ for QCD. $\sutwo$ has three parameters. The generators are usually
      chosen to be the Pauli matrices\index{Group generators!Pauli matrices for $\mathsf{SU(2)}$},
      $T^a=\frac{1}{2}\sigma^a$, whose commutation relations are 
      $\left[\sigma^a,\sigma^b\right]=\i \varepsilon^{abc}\sigma^c$. The common generators of
      $\suthree$ are the eight Gell-Mann matrices,
      $T^a=\frac{1}{2}\lambda^a$\index{Group generators!Gell-Mann matrices for $\mathsf{SU(3)}$}.
      
      To construct a model with a global $\sun$ symmetry, we consider not a single
      field, but an $n$-component vector $\Phi_i$, $i=1,\ldots,n$ (called a multiplet of $\sun$),
      on which the matrices of $\sun$ act by multiplication :
      \begin{align}
        \Phi=
        \begin{pmatrix}
          \Phi_1\\ \vdots \\\Phi_n
        \end{pmatrix}
        &\longrightarrow \Phi' = U \Phi\,, \quad 
        \Phi^\dagger=\left(\Phi_1^\dagger,\cdots,\Phi_n^\dagger\right) \longrightarrow
        \left(\Phi^\dagger\right)'=\Phi^\dagger U^\dagger \,.
      \end{align}
      Now we see why we want unitary matrices $U$: A product $\Phi^\dagger\Phi$ is
      invariant under such a transformation. This means that we can generalise the Lagrangean
      (\ref{eq:Lscalar}) in a straightforward way to include a non-Abelian symmetry:
      \begin{align}
        \L&= \left(\partial_\mu \Phi\right)^\dagger\!\left(\partial^\mu \Phi\right)
        -V\left(\Phi^\dagger\Phi\right) \,.
      \end{align}
      
      If we allow for local transformations $U=U(x)$, we immediately encounter the same problem as
      before: The derivative term is not invariant, because the derivatives act on the matrix $U$
      as well,
      \begin{align}
        \partial_\mu \Phi \to \partial_\mu\Phi' = \partial_\mu\left(U\Phi\right)=U\partial_\mu \Phi
        +\left(\partial_\mu U\right) \Phi\,.
      \end{align}
      To save the day, we again need to introduce a covariant derivative consisting of a partial
      derivative plus a gauge field.  This time, however, the
      vector field needs to be matrix-valued, i.e., $A_\mu = A_\mu^a T^a$, where $T^a$ are the
      generators of the group. We clearly need one vector field per
      generator, as each generator represents an independent transformation in the group. 
        
      The transformation law of $A_\mu$ is chosen such that the covariant derivative is covariant,
      \begin{align}
        \begin{split}
          \left(D_\mu \Phi\right)' &= \left[\left(\partial_\mu + \i g A_\mu\right)\Phi\right]'\\
          &=\left(\partial_\mu + \i g A_\mu'\right)\left(U \Phi\right)\\
          &= U\left(\partial_\mu +U^{-1} \left(\partial_\mu U\right) + \i g U^{-1} A_\mu' U\right) \Phi\\
          &\stackrel{!}{=} U D_\mu \Phi\,.
        \end{split}
      \end{align}
      This requirement fixes the transformation of $A_\mu$ to be \index{Gauge potential!Transformation of}
      \begin{align}\label{eq:NAgaugetrafo}
        A_\mu'&= U A_\mu U^{-1} -\frac{\i}{g} U\partial_\mu U^{-1}\,.
      \end{align}
      In the Abelian case this reduces to the known transformation law, Eq.~(\ref{eq:abeliantrafo}).
      
      For infinitesimal parameters $\alpha=\alpha^a T^a$, the matrix
      $U=\exp\!\left\{\i\alpha\right\}=1+\i \alpha$, and Eq.~(\ref{eq:NAgaugetrafo}) becomes
      \begin{align}
        A_\mu'&= A_\mu -\frac{1}{g} \partial_\mu \alpha +\i\left[\alpha,A_\mu\right]\,,\\
        \intertext{or for each component}
        {A_\mu^a}'&= A_\mu^a- \frac{1}{g} \partial_\mu \alpha^a - f^{abc} \alpha^b A_\mu^c\,.
      \end{align}
      
      Sometimes it is convenient to write down the covariant derivative in component form:
      \begin{align}
        \left(D_\mu \Phi\right)_i &= \left(\partial_\mu\delta_{ij} +\i g T^a_{ij} A_\mu ^a\right)
        \Phi_j \,.
      \end{align}
      
      Next we need a kinetic term, which again involves the field strength, the commutator of
      covariant derivatives:
      \begin{align}
        \begin{split}
          F_{\mu\nu} &= -\frac{\i}{g} \left[ D_\mu, D_\nu\right] = \partial_\mu A_\nu
          -\partial_\nu A_\mu +\i g \left[A_\mu,a_\nu\right]= F_{\mu\nu}^a T^a\,,\\
          F_{\mu\nu}^a&= \partial_\mu A_\nu^a -\partial_\nu A_\mu^a -g f^{abc}
          A_\mu^b A_\nu^c\,.
        \end{split}
      \end{align}
      Now we see  that the field strength is more that just the derivative: There is a quadratic
      term in the potentials.  This leads to a self-interaction of gauge fields, like in QCD,
      where the gluons interact with each other. This is the basic reason for confinement, unlike
      in QED, where the photon is not charged.
      
      Furthermore, when we calculate the transformation of the field strength, we find that it is
      not invariant, but transforms as
      \begin{align}
        F_{\mu\nu}&\to F_{\mu\nu}'= U F_{\mu\nu} U^{-1}\,,
      \end{align}
      i.e., it is covariant. There is an easy way to produce an invariant quantity out of this:
      the trace. Since  $\tr A B =\tr B A$, the  Lagrangean\index{Lagrangean!Non-Abelian gauge field}
      \begin{align}
        \L&= -\frac{1}{2} \tr \left(F_{\mu\nu} F^{\mu\nu}\right) = -\frac{1}{4} F_{\mu\nu}^a F^{a\;\mu\nu}
      \end{align}
      is indeed invariant, as $\tr\left(U F^2 U^{-1}\right)= \tr \left(U^{-1} U F^2\right)=\tr
      F^2$. In the second step we have used a normalisation convention,
      \begin{align}
        \tr \left( T^a T^b\right) &= \frac{1}{2} \delta^{ab}\,,
      \end{align}
      and every generator is necessarily traceless. The factor $\frac{1}{2}$ is arbitrary and
      could be chosen differently, with compensating changes in the coefficient of the kinetic
      term. 
      
      By choosing the gauge group $\suthree$ and coupling the gauge field to fermions, the quarks,
      we can write down the Lagrangean of quantum chromodynamics~(QCD):\index{Lagrangean!QCD} 
      \begin{align}
        \L_\text{QCD}&= -\frac{1}{4} G_{\mu\nu}^a G^{a\;\mu\nu} +\ol{q} \left(\i \Dslash
          -m\right)q \,,
      \end{align}
      where $a=1,\dotsc,8$ counts the gluons and $q$ is a three-component (i.e.\ three-colour) quark.

    \section{Quantisation}
      So far we have only discussed classical gauge theories. If we want to quantise the theory
      and find the Feynman rules for diagrams involving gauge fields, we have a problem: We have
      to make sure we do not count field configurations of $A_\mu$ which are pure gauge, nor that
      we count separately fields which differ only by a gauge transformation, since those are
      meant to be physically identical. On the more technical side, the na\"{\symbol{16}}ve 
      Green function for the free equation of motion does not exist. In the Abelian case, the
      equation is 
      \begin{align}
        \partial_\mu F^{\mu\nu}&=\Box A^\nu -\partial^\nu \partial_\mu A^\mu=\left(\Box g^{\mu\nu}
          -\partial^\nu\partial^\mu\right)A_\mu= 0\,.
      \end{align}
      The Green function should be the inverse of the differential operator in brackets, but the
      operator is not invertible. Indeed, it annihilates every pure gauge mode, as it should,
      \begin{align}
        \left(\Box g^{\mu\nu} -\partial^\nu\partial^\mu\right) \partial_\mu \Lambda &=0\,,
      \end{align}
      so it has zero eigenvalues. Hence, the propagator must be defined in a more clever way.
      
      One way out would be to fix the gauge, i.e., simply demand a certain gauge condition like
      $\vec{\nabla}\cdot \vec{A}=0$ (Coulomb gauge)\index{Gauge conditions} or $n_\mu
      A^\mu=0$ with a fixed 4-vector (axial gauge). It turns out,
      however, that the loss of Lorentz invariance causes many problems in calculations.

      A better way makes use of Faddeev--Popov ghosts\index{Faddeev--Popov ghosts}. In this approach, we add 
      two terms to the Lagrangean, the gauge-fixing term and the ghost term. The gauge-fixing term
      is not gauge invariant, but rather represents a certain gauge condition which can be chosen
      freely. The fact that it is not gauge invariant means that now the propagator is
      well-defined, but the price to pay is that it propagates too many degrees of freedom, namely
      gauge modes. This is compensated by the propagation of ghosts, strange fields which are
      scalars but anticommute and do not show up as physical states but only as internal lines in
      loop calculations.  It turns out that gauge invariance is not lost but rather traded for a
      different symmetry, BRST-symmetry, which ensures that we get physically sensible results.
      
      For external states, we have to restrict to physical states, of which there are two for
      massless bosons. They are labelled by two polarisation vectors
      $\epsilon^\pm_\mu$\index{Polarisation vector} which are
      transverse, i.e., orthogonal to the momentum four-vector and the spatial momentum,
      \mbox{$k_\mu \epsilon^\mu=\vec{k}\vec{\epsilon}=0$}.
      
      The form of the gauge fixing and ghost terms depends on the gauge condition we want to
      take. A common (class of) gauge is the covariant gauge which depends on a parameter $\xi$,
      which becomes Feynman gauge (Landau/Lorenz gauge) for $\xi=1$ ($\xi=0$)%
      
      We now list the Feynman rules for a non-Abelian gauge theory (QCD) coupled to fermions
      (quarks) and ghosts. The fermionic external states and propagators are listed in
      Section~\ref{sec:fermionrules}. \index{Feynman rules!for non-Abelian gauge theories}
        
      \begin{enumerate}[i.]
        \item \begin{minipage}[t]{60pt}
          \begin{picture}(50,20)(0,4)\footnotesize
            \SetOffset(0,0) \Photon(3,1)(40,1){2}{5}
            \Vertex(40,1){2} \Text(20,10)[]{$k$}\Text(20,5.5)[]{$\longrightarrow$} \Text(20,-4)[t]{$\mu$}
          \end{picture}\\
          \begin{picture}(50,20)(0,8)\footnotesize
            \SetOffset(0,0) \Photon(10,1)(47,1){2}{5}
            \Vertex(10,1){2} \Text(30,10)[]{$k$}\Text(30,5.5)[]{$\longrightarrow$} \Text(30,-4)[t]{$\mu$}
          \end{picture}             
        \end{minipage}
        \begin{minipage}[t]{130pt}
          \begin{center}
            \raisebox{-2pt}{
              $\epsilon_\mu (k)$
            }\\
            \raisebox{-16pt}{
              $\epsilon_\mu^* (k)$
            }
          \end{center}
        \end{minipage}
        \begin{minipage}[t]{227pt}
          For each external line one has a polarisation vector.
        \end{minipage}
        \item \begin{minipage}[t]{60pt}
          \begin{picture}(50,20)(0,3)\footnotesize
            \SetOffset(0,0) \Vertex(3,1){2} \Photon(3,1)(47,1){2}{5}
            \Vertex(47,1){2}\Text(25,7)[]{$p$} \Text(1,4)[b]{$\mu$} \Text(49,4)[b]{$\nu$} 
            \Text(1,-3)[t]{$a$} \Text(49,-3)[t]{$b$}
          \end{picture}
        \end{minipage}
        \begin{minipage}[t]{130pt}
          \begin{center}
            \raisebox{-7pt}{
              $\displaystyle \frac{-\i\delta^{ab}}{k^2+\i \varepsilon}\quad\quad$
              \hspace*{50pt}}\\
            \raisebox{-17pt}{$\quad \times\left(g_{\mu\nu} +(1-\xi)\frac{k_\mu
                  k_\nu}{k^2}\right)$
            }
          \end{center}
        \end{minipage}
        \begin{minipage}[t]{227pt}
          The propagator for gauge bosons contains the parameter $\xi$.
        \end{minipage}
        \item \begin{minipage}[t]{60pt}
          \begin{picture}(50,20)(0,5)\footnotesize
            \SetOffset(0,5) \Vertex(3,1){2} \DashArrowLine(3,1)(47,1){3}
            \Vertex(47,1){2}\Text(25,7)[]{$k$} 
            \Text(1,-3)[t]{$a$} \Text(49,-3)[t]{$b$}
          \end{picture}
        \end{minipage}
        \begin{minipage}[t]{130pt}
          \begin{center}
            \raisebox{-6pt}{
              $\displaystyle \frac{-\i\delta^{ab}}{k^2+\i \varepsilon}$
            }
          \end{center}
        \end{minipage}
        \begin{minipage}[t]{227pt}
          The propagator for ghosts is the one of scalar particles. There are no external ghost states.
        \end{minipage}
        \item 
        \begin{minipage}[t]{60pt}
          \begin{picture}(50,20)(0,15)\footnotesize
            \ArrowLine(3,1)(25,1) \Vertex(25,1){2} \ArrowLine(25,1)(47,1)
            \Photon(25,1)(25,20){2}{3} \Text(30,15)[l]{$\mu$}
          \end{picture}
        \end{minipage}
        \begin{minipage}[t]{130pt}
          \begin{center}
            \raisebox{0pt}{
              $\i e \gamma^\mu$
            }
          \end{center}
        \end{minipage}
        \begin{minipage}[t]{227pt}
          In QED, there is just one vertex between photon and electron. 
        \end{minipage}
        \item 
        \begin{minipage}[t]{60pt}
          \begin{picture}(50,20)(0,15)\footnotesize
            \ArrowLine(3,1)(25,1) \Vertex(25,1){2} \ArrowLine(25,1)(47,1)
            \Gluon(25,1)(25,20){2}{3} \Text(30,15)[l]{$\mu$}
          \end{picture}
        \end{minipage}
        \begin{minipage}[t]{130pt}
          \begin{center}
            \raisebox{0pt}{
              $\i \frac{g}{2}\gamma^\mu \lambda^a$
            }
          \end{center}
        \end{minipage}
        \begin{minipage}[t]{227pt}
          In QCD, the basic quark-quark-gluon vertex involves the Gell-Mann matrices.
        \end{minipage}
        \item 
        \begin{minipage}[t]{60pt}
          \begin{picture}(50,30)(0,15)\footnotesize
            \DashArrowLine(3,1)(25,1){3} \Vertex(25,1){2} \DashArrowLine(25,1)(47,1){3}
            \Text(3,-2)[t]{$b$} \Text(47,-2)[t]{$c$} \Text(8,4)[]{$\longrightarrow$} \Text(8,6)[b]{$p$}
            \Gluon(25,1)(25,20){2}{4} \Text(30,15)[l]{$\mu,a$}
          \end{picture}
        \end{minipage}
        \begin{minipage}[t]{130pt}
          \begin{center}
            \raisebox{0pt}{
              $-g f^{abc} p^\mu$
            }
          \end{center}
        \end{minipage}
        \begin{minipage}[t]{227pt}
          The ghosts couple to the gauge field.
        \end{minipage}
        \item 
        \begin{minipage}[t]{60pt}
          \begin{picture}(50,40)(0,30)\footnotesize
            \Gluon(3,3)(25,15){2}{4} \Gluon(25,15)(25,37){2}{4} \Gluon(25,15)(47,3){2}{4} \Vertex(25,15){2}
          \end{picture}
        \end{minipage}
        \begin{minipage}[t]{130pt}
          \begin{center}
            \raisebox{0pt}{
              $g f^{abc} k_\mu +\text{permutations}$ 
            }
          \end{center}
        \end{minipage}
        \begin{minipage}[t]{227pt}
          Three-gluon self-interaction.
        \end{minipage}
        \item 
        \begin{minipage}[t]{60pt}
          \begin{picture}(50,50)(0,40)\footnotesize
            \Gluon(3,3)(25,25){2}{4}  \Gluon(47,47)(25,25){2}{4} \Gluon(3,47)(25,25){2}{4}
            \Gluon(47,3)(25,25){2}{4}  \Vertex(25,25){2}
          \end{picture}
        \end{minipage}
        \begin{minipage}[t]{130pt}
          \begin{center}
            \raisebox{0pt}{
              $-\frac{1}{4}g^2 f^{abc} f^{ade} g^{\mu\nu} g^{\rho\sigma}$
            }\\
            \raisebox{-7pt}{  
              \hspace*{30pt}$+\text{permutations}$
            }
          \end{center}
        \end{minipage}
        \begin{minipage}[t]{227pt}
          Four-gluon self-interaction.
        \end{minipage}
      \end{enumerate}


  \chapter{Quantum Corrections}
    Now that we have the Feynman rules, we are ready to calculate quantum corrections \cite{peskin,
    ecker,thooft}.  As a first
    example we will consider the anomalous magnetic moment of the electron at one-loop order.
    This was historically, and still is today, one of the most important tests of quantum field
    theory.
    The calculation is still quite simple because the one-loop expression is finite. In most
    cases, however, one encounters divergent loop integrals. In the following sections we will 
    study these divergences and show how to remove them by renormalisation. Finally, as an
    application, we will discuss
    the running of coupling constants and asymptotic freedom.

    \section{Anomalous Magnetic Moment}
      The magnetic moment of the electron determines its energy in a magnetic field,
      \begin{align}
        H_\text{mag}&=-\vec{\mu}\cdot\vec{B} \,.\index{Magnetic moment}
      \end{align}
      For a particle with spin $\vec{s}$, the magnetic moment is aligned in the direction of
      $\vec{s}$, and for a classical spinning particle of mass $m$ and charge $e$, its magnitude
      would be the Bohr magneton, $e/(2m)$. In the quantum theory, the magnetic moment is different,
      which is expressed by the Land\'e factor $g_e$\index{Land\'e factor},
      \begin{align}
        \vec{\mu}_e &= g_e\frac{e}{2m} \vec{s}\;.
      \end{align}
      
      We now want to calculate $g_e$ in QED.  To lowest order, this just means solving the Dirac
      equation in an external electromagnetic field $A^\mu=(\phi,\vec{A})$,
      \begin{align}
        \left(\i \Dslash-m\right) \psi =\left[ \gamma^\mu \left(\i \partial_\mu -e A_\mu\right)
          -m\right] \psi=0\,.
      \end{align}
      For a bound nonrelativistic electron a stationary solution takes the form
      \begin{align}
        \psi(x)&= \begin{pmatrix}\varphi(\vec{x})\\ \chi(\vec{x})\end{pmatrix} e^{-\i E t}\,, \quad
          \text{with} \quad \frac{E-m}{m}\ll 1\,.
      \end{align}
      It is convenient to use the following representation of the Dirac matrices:
      \index{gamma@$\gamma$-matrices!Dirac representation}
      \begin{align}\label{eq:gammadirac}
        \gamma^0&=\begin{pmatrix} \mathbbm{1} &0 \\ 0 &-\mathbbm{1}\end{pmatrix}\,,  &
        \gamma^i&=\begin{pmatrix} 0 & \sigma^i \\ -\sigma^i &0\end{pmatrix}\,.
      \end{align}
      One then obtains the two coupled equations
      \begin{subequations}
        \begin{align}
          \left[\left(E- e\phi\right) -m \right]\varphi  -\left(-\i \vec{\nabla} -e
            \vec{A}\right)\cdot \vec{\sigma} \chi &=0\,, \label{eq:prepauli1}\\ 
          \big[\underbrace{-\left(E- e\phi\right) -m}_{\approx -2m}\big] \chi
          +\left(-\i\vec{\nabla} -e \vec{A}\right)\cdot \vec{\sigma} \varphi &=0 \,. \label{eq:prepauli2}
        \end{align}
      \end{subequations}
      The coefficient of $\chi$ in the second equation is approximately independent of $\phi$, so we
      can solve the equation to determine $\chi$ in terms of $\varphi$,
      \begin{align}
        \chi&= \frac{1}{m} \left(-\i\vec{\nabla} -e\vec{A}\right)\cdot \vec{\sigma}\varphi \,.
      \end{align}      
      Inserting this into (\ref{eq:prepauli1}), we get the Pauli equation,\index{Pauli equation}
      \begin{align}
        \left[\frac{1}{2m} \left(-\i \vec{\nabla} -e \vec{A}\right)^2  +e\phi -\frac{e}{2m}
          \vec{B}\cdot \vec{\sigma}\right] \varphi&=\left(E-m\right) \varphi \,.
      \end{align}
      This is a Schr\"{o}dinger-like equation which implies (since $\vec{s}=\frac{1}{2}\vec{\sigma}$),
      \begin{align}
        H_\text{mag}&= -2 \frac{e}{2m} \vec{s} \vec{B}\,.
      \end{align}
      Hence, the Land\'e factor of the electron is $g_e=2$.

      \begin{figure}
        \begin{center}          
          \begin{picture}(400,100)(0,0)\small
            \ArrowLine(5,5)(50,40)  \Text(1,-.5)[]{$p$}
            \ArrowLine(50,40)(95,5) \Text(100,1)[]{$p'$}           
            \Photon(50,95)(50,40){3}{4} \Text(42,78)[r]{$q$} \LongArrow(44,85)(44,70)
            \Text(55,78)[l]{$\mu$}
            \GCirc(50,40){15}{.6}
            \Text(115,50)[]{$\displaystyle =$}
            \SetOffset(-20,0)
            \ArrowLine(155,5)(200,40)  \Text(151,-.5)[]{$p$}
            \ArrowLine(200,40)(245,5) \Text(250,1)[]{$p'$}           
            \Photon(200,95)(200,40){3}{4} \Text(192,78)[r]{$q$} \LongArrow(194,85)(194,70)
            \Text(205,78)[l]{$\mu$}
            \Vertex(200,40){2}
            \SetOffset(0,0)
            \Text(250,50)[]{$\displaystyle +$}
            \SetOffset(-20,0)
            \ArrowLine(285,5)(303,19)  \Text(281,-.5)[]{$p$} 
            \ArrowLine(303,19)(330,40)
            \ArrowLine(330,40)(357,19) \Text(380,1)[]{$p'$}           
            \ArrowLine(357,19)(375,5)
            \Vertex(303,19){2} \Photon(303,19)(357,19){2}{5.5} \Vertex(357,19){2}
            \Text(330,8)[]{$k$} \LongArrow(320,14)(340,14)
            \Photon(330,95)(330,40){3}{4} \Text(322,78)[r]{$q$} \LongArrow(324,85)(324,70)
            \Text(335,78)[l]{$\mu$}
            \Vertex(330,40){2}
            \SetOffset(0,0)
            \Text(375,50)[]{$\displaystyle +\dotsm$}
          \end{picture}
          \caption{Tree level and one-loop diagram for the magnetic moment.\label{fig:mag_mom}}
        \end{center}
      \end{figure}
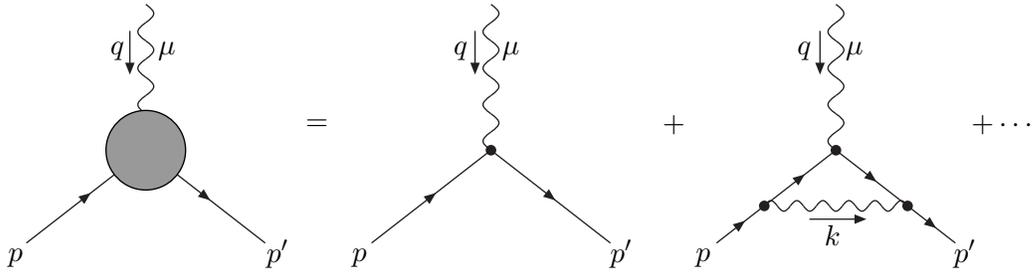

      In QED, the magnetic moment is modified by quantum corrections. 
      The magnetic moment is the spin-dependent
      coupling of the electron to a photon in the limit of vanishing photon momentum.
      Diagrammatically, it is contained 
      in the blob on the left side of Fig.~\ref{fig:mag_mom}, which denotes the complete
      electron-photon coupling. The tree-level diagram is the fundamental electron-photon
      coupling. There are several one-loop corrections to this diagram, but only the so-called vertex
      correction\index{Vertex correction}, where an internal photon connects the two electron lines,
      gives a contribution to 
      the magnetic moment. All other one-loop diagrams concern only external legs, such as an
      electron-positron-bubble on the incoming photon, and will be removed by wave-function
      renormalisation.

      The expression for the tree-level diagram is 
      \begin{align}
        \i \ol{u}(p') e\gamma^\mu u(p)\,.
      \end{align}
      Note that the photon becomes on-shell only for $q\to 0$, so no polarisation vector is included.
      The matrix element of the      
      electromagnetic current can be decomposed via the Gordon identity\index{Gordon identity}
      into convection and spin currents,
      \begin{align}\label{eq:gordon}
        \ol{u}(p')\gamma^\mu u(p) &= \ol{u}(p') \left(\frac{\left(p+p'\right)^\mu}{2m}
          +\frac{\i}{2m} \sigma^{\mu\nu}\left(p'-p\right)_\nu\right) u(p)\,.
      \end{align}
      Here the first term can be viewed as the net flow of charged particles, the second one is the spin
      current. Only this one is relevant for the magnetic moment, since it gives the spin-dependent
      coupling of the electron. 
      
      In order to isolate the magnetic moment from the loop diagram, we first note that the
      corresponding expression will contain the same external states, so it can be written as
      \begin{align}\label{eq:vertexfunction}
        \i \ol{u}(p') e\Gamma^\mu(p,q) u(p) \,, \quad q=p' -p\,,
      \end{align}
      where $\Gamma^\mu(p,q)$ is the correction to the vertex due to the exchange of the photon.
      We can now decompose $\Gamma^\mu$ into
      different parts according to index structure and extract the term $\propto\!
      \sigma^{\mu\nu}$. Using the Feynman rules, we find for $\Gamma^\mu$ in Feynman gauge ($\xi=1$),
      \begin{align}
        \begin{split}\label{eq:vertexcorr}
          \i e\Gamma^\mu (p,q) & = \left(-\i e\right)^3 \int \frac{\d^4 k}{\left(2\pi\right)^4}
          \frac{-\i g_{\rho\sigma}}{k^2+\i \varepsilon} \gamma^\rho  
          \frac{\i \left(\slash[2]{p'} -\slash[1]{k} +m\right)}{\left(p'-k\right)^2 -m^2
            +\i\varepsilon} \gamma^\mu \\
          &\quad \mspace{290mu} \times\frac{\i
            \left(\slash[2]{p}-\slash[1]{k}+m\right)}{\left(p-k\right)^2 -m^2+\i\varepsilon}
          \gamma^\sigma \,.
          \end{split}
      \end{align}
      This integral is logarithmically divergent, as can be seen by power
      counting, since the leading term is $\propto k^2$ in the numerator and $\propto k^6$ in
      the denominator. 

      \medskip

      On the other hand, the part $\propto\! \sigma^{\mu\nu}q_\mu$ is finite and can be extracted
      via some tricks: 
      \begin{itemize}
        \item Consider first the denominator of the integral~(\ref{eq:vertexcorr}). It is the 
          product of three terms 
          of the form $\left(\text{momentum}\right)^2- m^2$, which can be transformed into a  sum at
          the expense of further integrations over the so-called Feynman parameters $x_1$ and $x_2$,
          \index{Feynman parameters}
          \begin{align}
            \frac{1}{A_1 A_2 A_3}&=2 \int_0^1 \d x_1 \int_0^{1-x_1} \d x_2 \frac{1}{\left[A_1 x_1
                +A_2 x_2 +A_3 \left(1-x_1-x_2\right) \right]^3}\,.
          \end{align}
        \item After introducing the Feynman parameters, the next trick is to shift the
          integration momentum $k\to k'$, where 
          \begin{align}
            A_1 x_1 +A_2 x_2 +A_3 \left(1-x_1-x_2\right) &= {\underbrace{\left(k-x_1 p' -x_2
              p\right)}_{k'}}^2 -\left(x_1 p' +x_2 p\right)^2 +\i\varepsilon\;.
          \end{align}
          Note that one must be careful when manipulating divergent integrals. In principle, one should
          first regularise them and then perform the shifts on the regularised integrals, but in this
          case, there is no problem.
        \item For the numerator, the important part is the Dirac algebra of $\gamma$-matrices. A
        standard calculation gives (see appendix) 
          \begin{align}
            \begin{split}
              \gamma^\nu &\left(\slash[2]{p}\,'- \slash[2]{k} +m \right) \gamma^\mu \left(\slash[2]{p}-
                \slash[2]{k} +m \right) \gamma_\nu \\
              &\mspace{50mu} = -2 m^2 \gamma_\mu  -4\i m \sigma^{\mu\nu} \left(p'-p\right)_\nu -
              2\slash[2]{p}\gamma_\mu \slash[2]{p}\,' +\mathcal{O}\!\left(k\right)
              +\mathcal{O}\!\left(k^2\right) \;.
            \end{split}
          \end{align}
          Here we have used again the Gordon formula to trade $\left(p+p'\right)_\nu$ for
          $\sigma_{\nu\rho}q^\rho$, which only is allowed if the expression is
          sandwiched between on-shell spinors $\ol{u}(p')$ and $u(p)$.
        \item Now the numerator is split into pieces independent of $k$, linear and quadratic in
          $k$. The linear term can be dropped under the integral. The quadratic piece leads to a
          divergent contribution which we will discuss later. The integral over the $k$-independent
          part in the limit $q^\mu\to 0$ yields
          \begin{align}
            \int\frac{\d^4 k}{\left(2\pi\right)^4} \,\frac{1}{\left[k^2-\left(x_1+x_2\right)^2 m^2
                +\i\varepsilon\right]^3}  = -\frac{\i}{32\pi^2}\,\frac{1}{\left(x_1+x_2\right)^2 m^2}\,.
          \end{align}
          Now all that is left are the parameter integrals over $x_1$ and $x_2$.
      \end{itemize}
      
      Finally, one obtains the result, usually expressed in terms of the fine structure
      constant $\alpha=e^2/\left(4\pi\right)$,
      \begin{align}
        \i e \ol{u}(p') \Gamma^\mu u(p) = +\i e \ol{u}(p') \left( \frac{\alpha}{2\pi} \frac{\i}{2m}
        \sigma^{\mu\nu} q_\nu +\dotsm \right) u(p)\,,
      \end{align}
      where the dots represent contributions which are not  $\propto \sigma^{\mu\nu} q_\nu$.
      
      Comparison with the Gordon decomposition~(\ref{eq:gordon}) gives the one-loop correction to the
      Land\'e factor,
      \begin{align}
        g= 2\left(1 +\frac{\alpha}{2\pi}\right)\,.\index{Magnetic moment!one-loop correction}
      \end{align}
      This correction was first calculated by Schwinger in 1948. It is often expressed as the
      anomalous magnetic moment $a_e$,  
      \begin{align}
        a_e=\frac{g-2}{2}\,.\index{Magnetic moment!anomalous}
      \end{align}
      Today, $a_e$ is known up to three loops analytically and to four loops numerically
      \cite{kinoshita}. The agreement of theory and experiment is impressive:
      \begin{align}
        \begin{split}
          a_e^\text{exp}&= \left(1159652185.9 \pm 3.8\right) \cdot 10^{-12}\;,\\
          a_e^\text{th}&= \left(1159652175.9 \pm 8.5 \right) \cdot 10^{-12}\;.
        \end{split}
      \end{align}
      This is one of the cornerstones of our confidence in quantum field theory.

    \section{Divergences}
      The anomalous magnetic moment we calculated in the last section was tedious work, but at least
      the result was finite. Most other expressions, however, have divergent  momentum
      integrals. One such example is the vertex function $\Gamma^\mu$ we already considered. It has
      contributions which are logarithmically divergent. We can isolate these by setting $q=0$,
      which yields
      \begin{align}\label{eq:vertex_q=0}
        \Gamma^\mu (p,0)&= -2\i e^2 \int_0^1 \d x_1 \int_0^{1-x_1} \d x_2 \int\frac{\d^4
          k}{\left(2\pi\right)^2} \, \frac{\gamma^\nu\slash[2]{k} \gamma_\mu \slash[2]{k}
          \gamma_\nu}{\left[k^2 -\left(x_1+x_2\right)^2 m^2 +\i \varepsilon \right]^3}\,.
      \end{align}
      
      This expression is treated in two steps:
      \begin{itemize}
        \item First we make the integral finite in a step called
          regularisation\index{Regularisation}. In this step, we 
          have to introduce a new parameter of mass dimension 1. An obvious choice would be a
          cutoff $\Lambda$ which serves as an upper bound for the momentum integration. One might
          even argue that there should be a cutoff at a scale where quantum gravity 
          becomes important, although a
          regularisation parameter has generally no direct physical meaning.
        \item The second step is renormalisation\index{Renormalisation}. The divergences are
          absorbed into the parameters 
          of the theory. The key idea is that the ``bare'' parameters which appear in the Lagrangean
          are not physical, so they can be divergent. Their divergences are chosen such as to cancel
          the divergences coming from the divergent integrals.
        \item Finally, the regulator is removed. Since all divergences have been absorbed into 
          the parameters of the theory,
          the results remain finite for infinite regulator. Of course, one has to make sure the 
          results do not depend on the regularisation method.
      \end{itemize}
      
      The cutoff regularisation, while conceptually simple, is not a convenient method, as it
      breaks Lorentz and gauge invariance. Symmetries, however, are very important for all calculations,
      so a good regularisation scheme should preserve as many symmetries as possible. We will
      restrict ourselves to dimensional regularisation, which is the most common scheme used
      nowadays.

      \subsection{Dimensional Regularisation}\index{Regularisation!dimensional}
      The key idea is to define the theory not in four, but in $d=4-\epsilon$ dimensions \cite{thooft}. 
      If $\epsilon$
      is not an integer, the integrals do converge. 
      Non-integer
      dimensionality might seem weird, but in the end we will take the limit of $\epsilon\to
      0$ and return to four dimensions. This procedure is well-defined and just an intermediate step 
      in the calculation.

      Let us consider some technical issues. In $d$ dimensions, the Lorentz indices
      ``range from 0 to $d$'', in the sense that
      \begin{align}
        g^{\mu\nu} g_{\nu\mu}&=d\;,
      \end{align}
      and there are $d$ $\gamma$-matrices obeying the
      usual algebra,
      \begin{align}
        \left\{ \gamma^\mu, \gamma^\nu\right\} &= 2 g^{\mu\nu} \mathbbm{1}\;,\quad \tr
        \left(\mathbbm{1}\right)=4 \;.
      \end{align}
      $\gamma$-matrix contractions are also modified due to the change in the trace of $g_{\mu\nu}$,
      such as
      \begin{align}
        \gamma^\mu \gamma^\nu \gamma_\mu= -\left(2-\epsilon\right) \gamma^\nu\;, \quad 
        \gamma^\mu\gamma^\nu \gamma^\rho \gamma_\mu= 4 g^{\nu\rho} -\epsilon \gamma^\nu\gamma^\rho\;.
      \end{align}
      The tensor structure of diagrams can be simplified as follows. If a momentum integral over $k$
      contains a factor of $k_\mu k_\nu$, this must be proportional to $g_{\mu\nu} k^2$, since it is
      of second order in $k$ and symmetric in $\left(\mu\nu\right)$. The only symmetric tensor we
      have is the metric (as long as the remaining integrand depends only on the square of $k$ and
      the squares of the external momenta $p_i$), and the coefficient can be obtained by contracting with
      $g^{\mu\nu}$ to yield
      \begin{align}
        \int \frac{\d^4 k}{\left(2\pi\right)^4} k_\mu k_\nu f\!\left(k^2, p^2_i\right) &=
        \frac{1}{d}\, g_{\mu\nu} \int
        \frac{\d^4 k}{\left(2\pi\right)^4} k^2 f\!\left(k^2, p^2_i\right)\;. 
      \end{align}

      The measure of an integral changes from $\d^4 k$ to $\d^d k$. Since $k$ is a
      dimensionful quantity\footnote{In our units where $\hbar=c=1$, the only dimension is mass, so
        everything can be expressed in powers of GeV. The basic quantities have
        $\left[\text{mass}\right]=\left[\text{energy}\right]=\left[\text{momentum}\right]=1$ and
        $\left[\text{length}\right]=\left[\text{time}\right]=-1$, so $\left[\d x^\mu\right]=-1$ and
        $\left[\partial_\mu\right]=1$.} (of mass 
      dimension 1\index{Mass dimension}), we need to  compensate the change in 
      dimensionality by a factor of $\mu^\epsilon$, where $\mu$ is an arbitrary parameter of mass
      dimension 1. The mass dimensions of fields and parameters also change.
      They can be derived from the condition that the action, which is the
      $d$-dimensional integral over the Lagrangean,  be dimensionless. Schematically (i.e.,
      without all numerical factors), a Lagrangean of gauge fields, scalars and fermions reads
      \begin{align}
        \begin{split}
          \L&= \left(\partial_\mu A_\nu\right)^2 +e \partial_\mu A^\mu A_\nu A^\nu + e^2 \left(A_\mu
            A^\mu\right)^2\\
          &\quad + \left(\partial_\mu \phi\right)^2 + \psib\left(\i \dslash- m\right)\psi+ e \psib
          \slash{A}\psi + m^2 \phi^2 +\dotsm\;.  
        \end{split}
      \end{align}
      The condition of dimensionless action, $\left[S\right]=0$, translates into
      $\left[\L\right]= d$, since $\left[\d^d x\right]=-d$. Derivatives have mass dimension 1, and
      so do masses. That implies for the dimensions of the fields (and the limit as $d\to 4$),
      \begin{align}
        \left[A_\mu \right]& = \frac{d-2}{2}\to 1\;,  & \left[\phi \right]& = \frac{d-2}{2} \to 1\;, \\
        \left[ \psi\right]& = \frac{d-1}{2}\to\frac{3}{2}\;, & \left[e \right]& = 2-\frac{d}{2}\to 0\;.
      \end{align}

      How do we evaluate a $d$-dimensional integral? One first
      transforms to Euclidean space
      replacing $k^0$ by $\i k_4$, so that the Lorentzian
      measure $\d^d k$ becomes $\d^d k_\text{E}$.
      In Euclidean space, one can easily convert to 
      spherical coordinates and perform the integral over the angular variables, which
      gives the ``area'' of the $d$-dimensional ``unit sphere'',
      \begin{align}
        \int\frac{\d^d k_\text{E}}{\left(2\pi\right)^d} f\!\left(k^2\right) &= \underbrace{\int
          \frac{\d \Omega_d}{\left(2\pi\right)^d}}_{\textstyle\frac{1}{2^{d-1} \pi^{d/2}}
          \frac{1}{\Gamma(d/2)}} \int_0^\infty \d k_\text{E}\, k_\text{E}^{d-1} 
        f\!\left(k^2\right) \;.
      \end{align}
      The remaining integral can then be evaluated, again often using $\Gamma$-functions. The result
      is finite for $d\neq 4$, but as we let $d\to 4$, the original divergence appears again
      in the form of $\Gamma\!\left(2-d/2\right)$. The $\Gamma$-function has poles at
      negative integers and at zero, so the integral exists for noninteger
      dimension. In the limit $d\to 4$, or equivalently, $\epsilon\to 0$, one has
      \begin{align}
        \Gamma\!\left(2-\frac{d}{2}\right) &=\Gamma \!\left(\frac{\epsilon}{2}\right) =
        \frac{2}{\epsilon} - \gamma_\text{E} +\mathcal{O}\left(\epsilon\right)\,,
      \end{align}
      with the Euler constant $\gamma_\text{E}\simeq 0.58$.
      
      As an example, consider the logarithmically divergent integral (cf.~(\ref{eq:vertex_q=0}))
      \begin{align}
         \int \frac{\d^4 k}{\left(2\pi\right)^4} \frac{1}{\left(k^2 +C\right)^2}\;,
      \end{align}
      where $C=\left(x_1+x_2\right)^2 m^2$. In $d$ Euclidean dimensions, this becomes 
      \begin{align}
         \mu^\epsilon \int \frac{\d^4 k_\text{E}}{\left(2\pi\right)^4} \frac{1}{\left(k_\text{E}^2 +C\right)^2}
        &= \frac{ \mu^\epsilon \Gamma\!\left(2-\frac{d}{2}\right)}{\left(4\pi\right)^{d/2}
          \Gamma(2)}\,\frac{1}{C^{2-d/2}} =\frac{1}{8\pi^2} \, \frac{1}{\epsilon} + \dotsm 
      \end{align}
      For the original expression~(\ref{eq:vertex_q=0}) we thus obtain
      \begin{align}
        \Gamma^\mu\left(p,0\right)&= \frac{\alpha}{2\pi}\,\frac{1}{\epsilon}\, \gamma^\mu
        +\mathcal{O}(1)\;.
      \end{align}

      What have we achieved? In four dimensions, the result is still divergent. However, the
      situation is better than before: We have separated the divergent part from the finite one and
      can take care of the divergence before taking the limit $\epsilon\to 0$. This is
      done in the procedure of renormalisation. 

      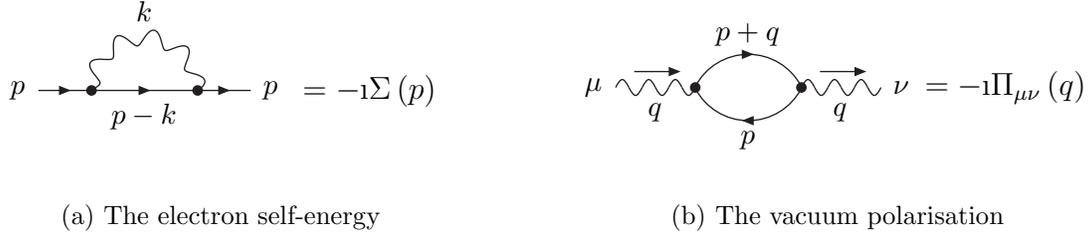
\begin{figure}
        \begin{center}
          \subfigure[The electron self-energy]{\label{fig:self-energy}
            \begin{minipage}{.45\textwidth}
              \begin{picture}(200,80)\small
                \SetOffset(20,22)
                \ArrowLine(10,10)(30,10) \ArrowLine(30,10)(70,10) \ArrowLine(70,10)(90,10)
                \Vertex(30,10){2} \Vertex(70,10){2} \PhotonArc(50,10)(20,0,180){3}{5} 
                \Text(5,10)[r]{$p$} \Text(50,5)[t]{$p-k$} \Text(50,40)[]{$k$} \Text(95,10)[l]{$p$}
                \normalsize
                \Text(110,10)[l]{$\displaystyle=-\i \Sigma\left(p\right)$}
              \end{picture}
            \end{minipage}}
          \hspace{.7cm}
          \subfigure[The vacuum polarisation]{\label{fig:vac-polari}
            \begin{minipage}{.45\textwidth}
              \begin{picture}(200,80) \small
                \SetOffset(15,-17)
                \Photon(0,50)(30,50){3}{3} \ArrowArcn(50,60)(22.5,-28,207) \ArrowArcn(50,40)(22.5,153,28)
                \Photon(70,50)(100,50){3}{3} 
                \Vertex(70,50){2} \Vertex(30,50){2} 
                \Text(-5,50)[r]{$\mu$} \Text(105,50)[l]{$\nu$}
                \Text(85,44)[t]{$q$} \Text(15,44)[t]{$q$} \Text(50,70)[]{$p+q$} \Text(50,30)[]{$p$}
                \LongArrow(77,56)(93,56) \LongArrow(7,56)(23,56)
                \normalsize
                \Text(117,50)[l]{$=-\i \Pi_{\mu\nu}\left(q\right)$}
              \end{picture}
            \end{minipage}}
          
          \caption{One-loop corrections to the propagators of electron and photon. }
          \label{fig:one-loop-propagators}
        \end{center}
      \end{figure}

      There are more divergent one-loop graphs where we can achieve the same: the electron
      self-energy\index{Self-energy!electron} $\Sigma$ in Fig.~\ref{fig:self-energy} (linearly
      divergent) and the photon self-energy or vacuum polarisation\index{Self-energy!photon~(vacuum
        polarisation)} $\Pi_{\mu\nu}$ in Fig.~\ref{fig:vac-polari} (quadratically divergent). The
      self-energy graph has two divergent terms,  
      \begin{align}
        \Sigma(p)&= \frac{3\alpha}{2\pi}\, \frac{1}{\epsilon} m - \frac{\alpha}{2\pi}\,
        \frac{1}{\epsilon}\left(\slash[2]{p} -m\right) +\mathcal{O}(1)\,,
      \end{align}
      which contribute to the mass renormalisation and the wave function renormalisation,
      respectively. The vacuum polarisation seems more complicated since it is a second rank
      tensor. However, the tensor structure is fixed by gauge invariance which requires
      \begin{align}
        q^\mu \Pi_{\mu\nu}\left(q\right) &= 0\;.
      \end{align}
      Therefore, because of Lorentz invariance,
      \begin{align}
        \Pi_{\mu\nu}\left(q\right) &= \left(g_{\mu\nu} q^2 - q_\mu q_\nu\right) \Pi\! \left(q^2\right)
        \,.
      \end{align}
      The remaining scalar quantity $\Pi(q^2)$ has the divergent part
      \begin{align}
        \Pi\!\left(q^2\right) &= \frac{2\alpha}{3\pi} \, \frac{1}{\epsilon}  +\mathcal{O}(1)\,.
      \end{align}

      \subsection{Renormalisation}
      So far we have isolated the divergences, but they are still there. How do we get rid of
      them? The crucial insight is that the parameters of the Lagrangean, the ``bare'' parameters,
      are not observable. Rather, the sum of bare parameters and loop-induced corrections are
      physical. Hence, divergencies of bare parameters  can cancel against divergent loop
      corrections, leaving physical observables finite. 

      To make this more explicit, let us express, as an example, the QED Lagrangean
      in terms of  bare fields $A^\mu_0$ and $\psi_0$ and bare parameters $m_0$ and
      $e_0$, \index{Bare fields} \index{Bare parameters}
      \begin{align}\label{qed}
        \L &= -\frac{1}{4}\left(\partial_\mu A_{0\,\nu} -  \partial_\nu A_{0\,\mu}\right)
        \left(\partial^\mu A^{0\,\nu} -  \partial^\nu A^{0\,\mu}\right) +\psib_0\left(\gamma^\mu
          \left(\i \partial_\mu - e_0 A_{0\,\mu}\right) -m_0\right) \psi_0 \,.
      \end{align}
      The ``renormalised fields'' $A_\mu$ and $\psi$ and the ``renormalised parameters'' $e$ and $m$ are
      then obtained from the bare ones by multiplicative rescaling,\index{Renormalised fields} 
      \begin{align}
        A_{0\,\mu}&= \sqrt{Z_3} A_\mu\;, & \psi_0&= \sqrt{Z_2} \psi\;, \\
        m_0&=\frac{Z_m}{Z_2} m\;,  & e_0 &= \frac{Z_1}{Z_2 \sqrt{Z_3}} \mu^{2-d/2} e \label{eq:ren-e}\;.
      \end{align}
      Note that coupling and electron mass now depend on the mass parameter $\mu$,
      \begin{align}\label{scale}
       e =e(\mu)\;, \quad  m=m(\mu)\;.
      \end{align}
      In terms of the renormalized fields and parameters the Lagrangean (\ref{qed}) reads
      \begin{align}
        \begin{split}
          \L &= -\frac{1}{4}\left(\partial_\mu A_\nu -  \partial_\nu A_\mu\right)
          \left(\partial^\mu A^\nu -  \partial^\nu A^\mu\right) +\psib\left(\gamma^\mu
          \left(\i \partial_\mu - e A_\mu\right) -m\right) \psi  + \Delta \L\;,
        \end{split}
      \end{align}
      where $\Delta \L$ contains the divergent counterterms, \index{Counterterms}
      \begin{align}
        \begin{split}
          \Delta\L&= -\left(Z_3-1\right)\frac{1}{4} F_{\mu\nu} F^{\mu\nu} + \left(Z_2-1\right) \psib
          \i \dslash \psi\\
          &\quad - \left(Z_m-1\right) m \psib\psi - \left(Z_1-1\right)e \psib \slash{A}  \psi\;.        
        \end{split}
      \end{align}
      The counterterms have the same structure as the original Lagrangean and lead to new 
      vertices in the Feynman rules:
      \index{Feynman rules!counterterms}
      \begin{enumerate}[i.]
        \item 
          \begin{minipage}[t]{70pt}
            \begin{picture}(65,20)(0,10)\small 
              \Photon(10,2)(55,2){2}{4.5} \Text(8,2)[r]{$\mu$} \Text(22.25,11)[b]{$q$}
              \Text(57,2)[l]{$\nu$} \LongArrow(12,7)(26,7) \SetWidth{1.5} \Line(29.5,-1)(35.5,5)
              \Line(29.5,5)(35.5,-1) 
            \end{picture}
          \end{minipage}
          \begin{minipage}[t]{105pt}
              \raisebox{-1pt}{
                $-\i \left(Z_3-1\right)$
              }\hspace*{1cm}\\
              \hspace*{.4cm} \raisebox{-1pt}{
                $ \times \left(g_{\mu\nu} q^2 -q_\mu q_\nu\right)$
              }
          \end{minipage}
          \begin{minipage}[t]{245pt}
            Photon wave function counterterm (countertems are generically denoted
            by~\begin{picture}(12,12)(0,0)\SetWidth{1.5} 
              \Line(3,0)(9,6) \Line(3,6)(9,0) \end{picture}). It has
            the same tensor structure as the vacuum polarisation.
          \end{minipage}
        \item 
          \begin{minipage}[t]{70pt}
            \begin{picture}(65,20)(0,5)\small
              \ArrowLine(10,2)(32.5,2) \ArrowLine(32.5,2)(55,2)  \Text(22.25,8)[b]{$p$}
              \SetWidth{1.5} \Line(29.5,-1)(35.5,5) \Line(29.5,5)(35.5,-1) 
            \end{picture}
          \end{minipage}
          \begin{minipage}[t]{105pt}
            \begin{center}
              \raisebox{-1pt}{
                $-\i \left(Z_2-1\right) \slash[2]{p}$
              }
            \end{center}
          \end{minipage}
          \begin{minipage}[t]{245pt}
            Electron wave function counterterm.
          \end{minipage}
          \item 
          \begin{minipage}[t]{70pt}
            \begin{picture}(65,20)(0,5)\small
              \ArrowLine(10,2)(32.5,2) \ArrowLine(32.5,2)(55,2)  \Text(22.25,8)[b]{$p$}
              \CArc(32.5,2)(4.5,0,360)\SetWidth{1.5} \Line(29.5,-1)(35.5,5) \Line(29.5,5)(35.5,-1) 
            \end{picture}
          \end{minipage}
          \begin{minipage}[t]{105pt}
            \begin{center}
              \raisebox{-1pt}{
                $-\i \left(Z_m-1\right) m$
              }
            \end{center}
          \end{minipage}
          \begin{minipage}[t]{245pt}
             Electron mass counterterm.
          \end{minipage}          
          \item 
          \begin{minipage}[t]{70pt}
            \begin{picture}(65,30)(0,10)\small   
              \ArrowLine(10,2)(32.5,2) \ArrowLine(32.5,2)(55,2)  \Photon(32.5,2)(32.5,28){2}{3}
              \SetWidth{1.5} \Line(29.5,-1)(35.5,5) \Line(29.5,5)(35.5,-1) 
            \end{picture}
          \end{minipage}
          \begin{minipage}[t]{105pt}
            \begin{center}
              \raisebox{-1pt}{
                $-\i e\left(Z_1-1\right) \gamma^\mu$
              }
            \end{center}
          \end{minipage}
         \begin{minipage}[t]{245pt}
            Vertex counterterm.
          \end{minipage}
      \end{enumerate}

      \medskip

      The renormalisation constants $Z_i$\index{Renormalisation!constants} are determined by
      requiring that the counterterms cancel the divergences.  They can be determined as power
      series in $\alpha$. The lowest order counterterms are $\mathcal{O}\left(\alpha\right)$ and
      have to be added to the one-loop diagrams.  Calculating e.g.\ the
      $\mathcal{O}\left(\alpha\right)$ correction to the electron-photon vertex, one has
      \begin{align}
        \begin{picture}(70,50)(0,25)
          \ArrowLine(0,20)(15,20) \Vertex(15,20){2} \ArrowLine(15,20)(35,20) \Vertex(35,20){2}
          \ArrowLine(35,20)(55,20) \Vertex(55,20){2} \ArrowLine(55,20)(70,20)
          \Photon(35,20)(35,50){2}{3} \PhotonArc(35,28)(22,198,-18){2}{6} 
        \end{picture}
        \;+\;
        \begin{picture}(50,50)(0,25)
          \ArrowLine(0,20)(25,20) \ArrowLine(25,20)(50,20) \Photon(25,20)(25,50){2}{3}
          \SetWidth{1.5} \Line(22,17)(28,23) \Line(22,23)(28,17)
        \end{picture}
        &= -\i e \gamma^\mu \left(\frac{\alpha}{2\pi}\,\frac{1}{\epsilon} + \left(Z_1-1\right) +
          \mathcal{O}\left(1\right)\right)\;.  
      \end{align}

      \medskip
      \noindent
      Demanding that the whole expression be finite determines the divergent part of $Z_1$,
      \begin{align}
        Z_1 &= 1- \frac{\alpha}{2\pi}\,\frac{1}{\epsilon} +\mathcal{O}\left(1\right)\,.
      \end{align}
      Similarly, the $\mathcal{O}\left(\alpha\right)$ vacuum polarisation now has two contributions,
      \begin{align}
        \begin{split}
          \begin{picture}(70,0)(0,12.1)
            \Photon(0,15)(20,15){2}{3} \Vertex(20,15){2} \ArrowArcn(35,10)(16,160,10)
            \ArrowArcn(35,20)(16,-20,200) \Vertex(50,15){2} \Photon(50,15)(70,15){2}{3}
          \end{picture}
          \;+\;
          \begin{picture}(50,0)(0,12.1)
            \Photon(0,15)(50,15){2}{7} \SetWidth{1.5} \Line(22,12)(28,18) \Line(22,18)(28,12)
          \end{picture}
          &= -\i \left( g_{\mu\nu} q^2 -q_\mu q_\nu \right) \left(\frac{2\alpha}{3\pi} \,
            \frac{1}{\epsilon} +\left(Z_3-1\right) +\mathcal{O}\left(1\right) \right)\;,
         \end{split}
       \end{align}
       which yields
       \begin{align}
           Z_3 = 1- \frac{2\alpha}{3\pi} \,\frac{1}{\epsilon}
          +\mathcal{O}\left(1\right)\;. 
      \end{align}
      
      The other constants $Z_2$ and $Z_m$ are fixed analogously.
      A Ward identity\index{Ward identity}, which follows from gauge invariance, yields the important
      relation  $Z_1=Z_2$. 
      The finite parts of the renormalisation constants are still undetermined. There are different 
      ways to fix them, corresponding to different renormalisation 
      schemes\index{Renormalisation!schemes}. All schemes
      give the same results for physical quantities, but differ at intermediate steps.

      Having absorbed the divergences into the renormalised parameters and fields, we can safely take
      the limit $\epsilon\to 0$. The theory now yields well-defined relations between physical
      observables. Divergencies can be removed to all orders in the loop expansion for renormalisable
      theories \cite{peskin,thooft}. Quantum electrodynamics and the standard model belong to this
      class. The proof is highly non-trivial and has been a major achievement in quantum field theory!

      \subsection{Running Coupling in QED} \index{Running Coupling}
      Contrary to the bare coupling $e_0$, the renormalised coupling $e(\mu)$ depends on the
      renormalisation scale $\mu$ (cf.~(\ref{eq:ren-e})),
      \begin{align*}
        e_0 &= \frac{Z_1}{Z_2 \sqrt{Z_3}} \mu^{-2+d/2} e (\mu) 
           = e(\mu) \mu^{-\epsilon/2} Z_3^{-\frac{1}{2}}\;,
      \end{align*}
      where we have used  
      the Ward identity $Z_1=Z_2$. It is very remarkable that the scale dependence is determined
      by the divergencies. To see this,
      expand Eq.~(\ref{eq:ren-e}) in $\epsilon$ and $e(\mu)$, 
      \begin{align}
        \begin{split}
          e_0 &= e(\mu)\left(1-\frac{\epsilon}{2} \ln \mu + \dotsm \right) \left(1 +
            \frac{1}{\epsilon} \frac{\alpha}{3\pi} + \dotsm\right) \\
          & = e(\mu)\left(\frac{1}{\epsilon} \frac{e^2(\mu)}{12\pi^2} + 1 - \frac{e^2(\mu)}{24\pi^2} \ln\mu
            +\mathcal{O}\left(\epsilon, e^4(\mu)\right)\right)\;,
        \end{split}
      \end{align}
      where we have used $\alpha=e^2/(4\pi)$. Since the bare mass $e_0$ does not depend on $\mu$,
      differentiation with respect to $\mu$ yields
      \begin{align}
        0&=\mu \frac{\partial}{\partial\mu} e_0 = \mu \frac{\partial}{\partial\mu} e -
        \frac{e^3}{24\pi^2} +\mathcal{O}\left(e^5\right) \;,
       \end{align}
      and therefore
      \begin{align}
        \mu \frac{\partial}{\partial\mu} e &= \frac{e^3}{24\pi^2}
        +\mathcal{O}\left(e^5\right) \equiv \beta(e)\;.\label{eq:betafn-qed}
      \end{align}
      This equation is known as the
      renormalisation group equation\index{Renormalisation!group equation}, and the function on the
      right hand side of Eq.~(\ref{eq:betafn-qed}) is the so-called the $\beta$ function
      \index{beta@$\beta$ function},
      \begin{align}
      \beta(e)=\frac{b_0}{(4\pi)^2} e^3 +\mathcal{O}\left(e^5\right)\;,\quad 
      \text{with} \quad b_0=\frac{2}{3}\;.
      \end{align} 
      
      The differential equation~(\ref{eq:betafn-qed}) can easily be integrated. Using a given value
      of $e$ at a scale $\mu_1$, the coupling $\alpha$ at another scale $\mu$ is given by
      \begin{align}
        \alpha(\mu) &= \frac{\alpha\left(\mu_1\right)}{1-\alpha\left(\mu_1\right)
          \frac{b_0}{\left(2\pi\right)} \ln \!\frac{\mu}{\mu_1}}\;.
      \end{align}
      Since $b_0>0$, the coupling increases with $\mu$ until it approaches the so-called Landau
      pole\index{Landau pole} where the denominator vanishes and perturbation theory breaks down.

      What is the meaning of a scale dependent coupling? This becomes clear when
      one calculates physical quantities, such as a scattering amplitude at some momentum
      transfer~$q^2$. In the perturbative expansion one then finds terms 
      $\propto e^2(\mu) \log\!\left(q^2/\mu^2\right)$.
      Such terms make the expansion unreliable unless one chooses $\mu^2\sim q^2$.
      Hence, $e^2\!\left(q^2\right)$ represents the effective interaction strength at a
      momentum (or energy) scale $q^2$ or, alternatively, at a distance of $r\sim 1/q$.

      The positive $\beta$ function in QED implies that the effective coupling strength
      decreases at large distances. Qualitatively, this can be understood as the effect
      of ``vacuum polarisation'': Electron-positron pairs etc.\ from the vacuum screen 
      any bare charge at distances larger than the corresponding Compton wavelength.
      Quantitatively, one finds that the value $\alpha(0)=\frac{1}{137}$, measured in
      Thompson scattering, increases to $\alpha\!\left(M_Z^2\right)=\frac{1}{127}$,
      the value conveniently used in electroweak precision tests.

      \subsection{Running Coupling in QCD}
      Everything we did so far for QED can be extended to non-Abelian gauge theories, in
      particular to QCD \cite{ecker}. It is, however, much
      more complicated, since there are more diagrams to calculate, and we will not be able
      to discuss this in detail.
      The additional diagrams contain gluon self-interactions and ghosts, and they lead to
      similar divergences, which again are absorbed by renormalisation constants. 
      Schematically, these are 
      \begin{align}
        &
        \begin{picture}(70,50)(0,25)
          \ArrowLine(0,20)(15,20) \Vertex(15,20){2} \ArrowLine(15,20)(35,20) \Vertex(35,20){2}
          \ArrowLine(35,20)(55,20) \Vertex(55,20){2} \ArrowLine(55,20)(70,20)
          \Gluon(35,20)(35,50){3}{4} \GlueArc(35,28)(22,198,342){3}{7} 
        \end{picture}
        \;+\;
        \begin{picture}(70,50)(0,25)
          \ArrowLine(0,20)(15,20) \Vertex(15,20){2} \ArrowLine(15,20)(55,20)
          \Vertex(55,20){2} \ArrowLine(55,20)(70,20) \Vertex(35,34){2}
          \Gluon(35,34)(35,55){3}{2} \GlueArc(35,12)(22,18,90){3}{3.5} \GlueArc(35,12)(22,90,162){3}{3.5}
        \end{picture}
        \;+\;
        \begin{picture}(50,50)(0,25)
          \ArrowLine(0,20)(25,20) \ArrowLine(25,20)(50,20) \Gluon(25,20)(25,50){3}{4}
          \SetWidth{1.5} \Line(22,17)(28,23) \Line(22,23)(28,17)
        \end{picture} \quad \leadsto Z_1\;,\\
        &
        \begin{picture}(70,50)(0,25)
          \ArrowLine(0,20)(15,20) \Vertex(15,20){2} \ArrowLine(15,20)(55,20)
          \Vertex(55,20){2} \ArrowLine(55,20)(70,20) \GlueArc(35,20)(20,0,180){3}{7} 
        \end{picture}
        \; +\;
        \begin{picture}(50,50)(0,25)
          \ArrowLine(0,25)(25,25) \ArrowLine(25,25)(50,25) \SetWidth{1.5} \Line(22,22)(28,28)
          \Line(22,28)(28,22) 
        \end{picture}
        \quad\leadsto Z_2\;,\\
        &
        \begin{picture}(70,50)(0,25)
          \Gluon(0,25)(20,25){3}{3} \Vertex(20,25){2} \GlueArc(35,30)(15,198,342){3}{5}
          \GlueArc(35,20)(15,18,162){3}{5} \Vertex(50,25){2} \Gluon(50,25)(70,25){3}{3}  
        \end{picture}
        \; + \;
        \begin{picture}(70,50)(0,25)
          \Gluon(0,25)(20,25){3}{3} \Vertex(20,25){2} \DashCArc(35,30)(15,198,342){3}
          \DashCArc(35,20)(15,18,162){3} \Vertex(50,25){2} \Gluon(50,25)(70,25){3}{3}  
        \end{picture}
        \; + \;
        \begin{picture}(70,50)(0,25)
          \Gluon(0,25)(20,25){3}{3} \Vertex(20,25){2} \ArrowArcn(35,30)(15,342,198)
          \ArrowArcn(35,20)(15,162,18) \Vertex(50,25){2} \Gluon(50,25)(70,25){3}{3}  
        \end{picture}
        \; + \;
        \begin{picture}(50,50)(0,25)
          \Gluon(0,25)(50,25){3}{8}  \SetWidth{1.5} \Line(20,20)(30,30)
          \Line(20,30)(30,20) 
        \end{picture}
        \quad\leadsto Z_3\;.
      \end{align}
      
      \medskip

      The renormalised coupling can again be defined as in QED, Eq.~(\ref{eq:ren-e}),
      \begin{align}
        g_0 &= \frac{Z_1}{Z_2 \sqrt{Z_3}} \mu^{-2+d/2} \, g\,.
      \end{align}
      The
      coefficients of the $1/\epsilon$-divergences depend on the gauge group and on the number of
      different fermions. For a $\mathsf{SU(}N_\text{c}\mathsf{)}$ gauge group with $N_\text{f}$
      flavours of fermions, one obtains the $\beta$ function for the gauge coupling $g$,
      \index{beta@$\beta$ function!for QCD coupling} 
      \begin{align}
        \mu \frac{\partial}{\partial \mu} g &= \frac{b_0}{\left(4\pi\right)^2} g^3
          +\mathcal{O}\left(g^5\right) \,,\quad b_0= -\left(\frac{11}{3}N_\text{c} -\frac{4}{3}
            N_\text{f}\right)\;.
      \end{align}
      Note that for $N_f < 11 N_\text{c}/4$
      the coefficient is negative! Hence, the coupling decreases at high momentum transfers or
      short distances.
      The calculation of this coefficient earned the Nobel Prize in 2004
      for Gross, Politzer and Wilczek. The decrease of the coupling at short distances is the
      famous phenomenon of asymptotic
      freedom.\index{asymptotic freedom} As a consequence, 
      one can treat in deep-inelastic scattering
      quarks inside the proton
      as quasi-free particles, which is the basis of the parton model.

      The coupling at a scale $\mu$ can again be expressed in terms of the coupling at a reference
      scale $\mu_1$,
      \begin{align}
        \alpha(\mu) &= \frac{\alpha\left(\mu_1\right)}{1+\alpha\left(\mu_1\right)
          \frac{\left| b_0 \right|}{\left(2\pi\right)} \ln\! \frac{\mu}{\mu_1}}\;.
      \end{align}
      The analogue of the Landau pole now occurs at small $\mu$ or large distances.
      For QCD with$N_\text{c}=3$ and
      $N_\text{f}=6$, the pole is at the ``QCD scale'' $\Lambda_\text{QCD}\simeq
      300$~MeV. At the QCD scale gluons and quarks are strongly coupled and colour is
      confined \cite{ecker}. Correspondingly,
      the inverse of $\Lambda_\text{QCD}$ gives
      roughly the size of hadrons, $r_\text{had}\sim \Lambda_\text{QCD}^{-1}\sim 0.7$~fm.


  \chapter{Electroweak Theory}
    So far we have studied QED, the simplest gauge theory, and QCD, the prime example of a
    non-Abelian gauge theory. But there also are the weak interactions, which seem rather
    different. They are short-ranged, which requires
    massive messenger particles, seemingly inconsistent with gauge
    invariance. Furthermore, weak interactions come in two types, charged and neutral 
    current-current interactions, which couple quarks and leptons differently. Charged current 
    interactions, mediated by the $W^\pm$ bosons, only involve 
    left-handed fermions and readily change flavour, as in the strange quark decay $s\to u e^-
    \ol{\nu}_e$. Neutral current interactions, on the other hand, couple both left- and
    right-handed fermions, and flavour-changing neutral currents are
    strongly suppressed. 

    Despite these differences from QED and QCD, weak interactions also turn out to be described
    by a non-Abelian gauge theory. Yet the electroweak theory is
    different because of two reasons: It is a chiral gauge theory, and the gauge symmetry is 
    spontaneously broken.

    \section{Quantum Numbers}
      In a chiral gauge theory, the building blocks are massless left- and right-handed
      fermions, 
      \begin{align}
        \psi_L=\frac{1}{2}\left(1-\gamma^5\right) \psi_L\; , \qquad  
           \psi_R=\frac{1}{2}\left(1+\gamma^5\right) \psi_R \; ,
      \end{align}
      with different gauge quantum numbers. For one generation of standard model
      particles, we will have seven chiral spinors: Two each for up- and down-type quark and charged
      lepton, and just one for the neutrino which we will treat as massless in this chapter, i.e.,
      we omit the right-handed one. The electroweak gauge group is a product of two groups,
      $\mathsf{G}_{EW}=\mathsf{SU(2)}_W \times \mathsf{U(1)}_Y$
      \index{Electroweak theory!gauge group}. Here the subscript $W$ stands for 
      ``weak isospin'', which is the quantum number associated with the $\mathsf{SU(2)}_W$ factor, and
      the $\mathsf{U(1)}$ charge is the hypercharge $Y$.  
      
      The assignment of quantum numbers, which corresponds to the grouping into representations of 
      the gauge group, is obtained as follows: The non-Abelian group $\mathsf{SU(2)}_W$ has a chargeless 
      one-dimensional singlet ($\mathbf{1}$) representation and charged multidimensional representations,
      starting with the two-dimensional doublet ($\mathbf{2}$) representation\footnote{Here we use
        ``representation'' as meaning ``irreducible representation''. Of course we can build
        reducible representations of any dimension.}. We are not allowed to mix
      quarks and leptons, since weak interactions do not change colour, nor
      left- and right-handed fields, which would violate Lorentz symmetry. 
      The $\mathsf{U(1)}_Y$ factor is Abelian, so it only has
      one-dimensional representations. This means we can assign different hypercharges we to the
      various singlets and doublets of $\mathsf{SU(2)}_W$.

      Furthermore, we know that charged currents connect up- with down-type
      quarks and charged leptons with neutrinos, and that the $W^\pm$ bosons couple only to
      left-handed fermions. This suggests to form doublets  from $u_\ls$ and $d_\ls$, and
      from $e_\ls$ and $\nu_\ls$, and to keep the right-handed fields as singlets. So we 
      obtain the $\mathsf{SU(2)}_W$ multiplets
       \begin{align}
        q_\ls&=\begin{pmatrix} u_\ls \\ d_\ls\end{pmatrix}\,, &&u_\rs\,, && d_\rs\,,  &
        l_\ls&=\begin{pmatrix} \nu_\ls \\  e_\ls\end{pmatrix}\,, &&  e_\rs\,,
      \end{align}
      with the hypercharges (which we will justify later) \index{Electroweak theory!quantum numbers}
      \begin{align}\label{eq:hypercharges}
        \begin{array}{rccccc}
          \text{field:} & q_\ls & u_\rs & d_\rs & l_\ls & e_\rs\\
          \text{hypercharge:\rule[12pt]{0pt}{1pt}} & \frac{1}{6} & \frac{2}{3} & -\frac{1}{3} &
          -\frac{1}{2} & -1 
        \end{array}\;.
      \end{align}

      With these representations, we can write down the covariant derivatives. The
      $\mathsf{SU(2)}_W$ has three generators, which we choose to be the Pauli matrices, and 
      therefore three gauge fields $W_\mu^I$, $I=1,2,3$. The $\mathsf{U(1)}_Y$ gauge field is
      $B_\mu$\index{Electroweak theory!$W^I$ and $B$ bosons}, and the coupling constants are
      $g$ and $g'$, respectively. The covariant derivatives acting on the left-handed fields are
      \begin{align}
        D_\mu \psi_\ls &= \left(\partial_\mu +\i g W_\mu + \i g' Y B_\mu \right)\psi_\ls\,,\quad
        \text{where } W_\mu=\tfrac{1}{2} \sigma^I W_\mu^I\,,\\
        \intertext{while the right-handed fields are singlets under $\mathsf{SU(2)}_W$, and hence do
        not couple to the $W$ bosons,}
        D_\mu \psi_\rs&= \left(\partial_\mu + \i g' Y B_\mu\right) \psi_\rs\;.
      \end{align}
      From the explicit form of the Pauli matrices,
      \begin{align}
        \sigma^1&=\begin{pmatrix} 0&1\\1&0\end{pmatrix}\,, & \sigma^2&= \begin{pmatrix}
          0&-\i \\ \i&0\end{pmatrix}\,, &\sigma^3&=\begin{pmatrix} 1&0\\0&-1\end{pmatrix} \,,
      \end{align}
      we see that $W^1_\mu$ and $W_\mu^2$ mix up- and down-type quarks, while
      $W_\mu^3$ does not, like the $\mathsf{U(1)}$ boson $B_\mu$.
      
      \medskip
      
      It is often convenient to split the Lagrangean into the free (kinetic) part and the
      interaction Lagrangean, which takes the form $\left(\text{current}\right)\cdot
      \left(\text{vector field}\right)$. In the electroweak theory, one has
      \begin{align}\label{eq:LintWY} \index{Electroweak theory!Isospin and hypercharge currents}
        \L_\text{int} &= - g J_{W,\,\mu}^I W^{I\,\mu} - g' J_{Y,\,\mu} B^\mu\;,\\
        \intertext{with the currents}
        J_{W,\,\mu}^I&= \ol{q}_\ls \gamma_\mu \tfrac{1}{2}\sigma^I q_\ls + \ol{l}_\ls \gamma_\mu
        \tfrac{1}{2}\sigma^I l_\ls \label{eq:Wcurrent}\;,\\
        J_{Y,\,\mu} &= \frac{1}{6} \ol{q}_\ls \gamma_\mu q_\ls - \frac{1}{2} \ol{l}_\ls \gamma_\mu
          l_\ls +\frac{2}{3} \ol{u}_\rs \gamma_\mu u_\rs - \frac{1}{3} \ol{d}_\rs \gamma_\mu
          d_\rs - \ol{e}_\rs \gamma_\mu e_\rs \;.\label{eq:Ycurrent}
      \end{align}
      These currents have to be conserved, $\partial_\mu J^\mu=0$, to allow a consistent coupling
      to gauge bosons.

      \subsection{Anomalies}
        Before considering the Higgs mechanism which will lead to the identification of the 
        physical $W^\pm$, $Z$ and
        $\gamma$ bosons of the standard model, let us briefly discuss anomalies. We will see that
        the choice of hypercharges in~(\ref{eq:hypercharges}) is severely constrained
        by the consistency of the theory.

        Suppose we have a classical field theory with a certain symmetry and associated conserved 
        current. After quantising the theory, the resulting quantum field theory might not have that
        symmetry anymore, which means the current is no longer conserved. This is called an
        anomaly. Anomalies are not a problem for global symmetries, where the quantised theory just
        lacks that particular symmetry. For gauge symmetries, however, the currents have to be
        conserved, otherwise the theory is inconsistent.

        \begin{figure}[htbp]
          \begin{center}
            \begin{picture}(400,100)(0,0)
              \Text(10,50)[]{\Large $\displaystyle\mathscr{A} \propto$}
              \Photon(65,50)(100,50){2}{5} \Photon(140,80)(175,80){2}{5}
              \Photon(140,20)(175,20){2}{5}  \ArrowLine(100,50)(140,80) \ArrowLine(140,80)(140,20)
              \ArrowLine(140,20)(100,50) \Vertex(100,50){2} \Vertex(140,80){2} \Vertex(140,20){2}
              \Text(50,50)[]{$J^A$} \Text(190,80)[]{$J^B$} \Text(190,20)[]{$J^C$}
              \Text(116,76)[]{$\psi_\ls$}
              \Text(215,50)[]{\Large $\displaystyle -$}
              \Photon(260,50)(295,50){2}{5} \Photon(335,80)(370,80){2}{5}
              \Photon(335,20)(370,20){2}{5}  \ArrowLine(295,50)(335,80) \ArrowLine(335,80)(335,20)
              \ArrowLine(335,20)(295,50) \Vertex(295,50){2} \Vertex(335,80){2} \Vertex(335,20){2}
              \Text(245,50)[]{$J^A$} \Text(385,80)[]{$J^B$} \Text(385,20)[]{$J^C$}
              \Text(311,76)[]{$\psi_\rs$}
            \end{picture}
          \end{center}
          \caption{\label{fig:triangle}The gauge anomaly is given by triangle diagrams with chiral
            fermions in the loop.}
        \end{figure}
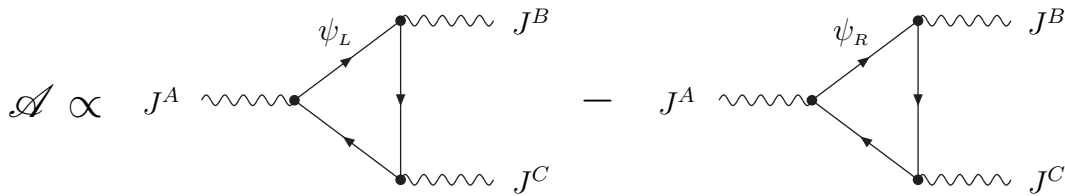

        Anomalies are caused by certain one-loop diagrams, the so-called triangle diagrams
        \index{Triangle diagrams} (see Fig~\ref{fig:triangle}). The left- and right-handed fermions
        contribute with different sign, so if they have the same quantum numbers, the anomaly
        vanishes. This is the case in QED and QCD, which thus are automatically anomaly free. In
        general, for currents $J^A$, $J^B$ and $J^C$, the anomaly $\mathscr{A}$ is 
        the difference of the traces of the generators $T^A$, $T^B$ and $T^C$ in the
        left- and right-handed sectors,
        \begin{align}
          \mathscr{A} &= \tr \left[\left\{T^A, T^B\right\} T^C\right]_L - \tr
          \left[\left\{T^A, T^B\right\} T^C\right]_R \stackrel{!}{=} 0\,.
        \end{align}
        Here the trace is taken over all fermions. 
        For the electroweak theory, in principle there are four combinations of currents, containing
        three, two, one or no $\mathsf{SU(2)}_W$ current. However, the trace of any odd number of
        $\sigma^I$ matrices vanishes, so we only have to check the $\mathsf{SU(2)}_W^2
        \mathsf{U(1)}_Y$ and $\mathsf{U(1)}_Y^3$ anomalies.
        
        The $\mathsf{SU(2)}_W$ generators are $\frac{1}{2} \sigma^I$, whose anticommutator is
        $\left\{\frac{1}{2} \sigma^I, \frac{1}{2} \sigma^J\right\}
        =\frac{1}{2}\delta^{IJ}$. Furthermore, only the left-handed fields contribute, since the
        right-handed ones are $\mathsf{SU(2)}_W$ singlets. Hence the
        $\mathsf{SU(2)}_W^2\mathsf{U(1)}_Y$ anomaly is 
        \begin{align}
          \mathscr{A} &= \tr \left[\left\{\frac{1}{2} \sigma^I, \frac{1}{2} \sigma^J\right\}
            Y\right]_L = \frac{1}{2}\delta^{IJ}\, \tr\left[Y\right]_L =\frac{1}{2}\delta^{IJ}\,
          \biggl(\underbrace{3}_{\mathclap{N_c}}\cdot \frac{1}{6} -\frac{1}{2}\biggr)=0\;.
        \end{align}
        We see that it only vanishes if quarks come in three colours!

        The $\mathsf{U(1)}_Y^3$ anomaly also vanishes:
        \begin{align}
          \begin{split}
            \mathscr{A} &= \tr \left[\left\{Y,Y\right\}Y\right]_L - \tr
            \left[\left\{Y,Y\right\}Y\right]_R = 2 \Bigl(\tr \left[Y^3\right]_L -\tr
              \left[Y^3\right]_R \Bigr)\\
            &= 2\left(3\cdot 2\left(\frac{1}{6}\right)^3 + 2
              \left(-\frac{1}{2}\right)^3 -3\left(\frac{2}{3}\right)^3 -3\left(-\frac{1}{3}\right)^3
              -\left(-1\right)^3 \right)\\
            &=0 \;.
          \end{split}
        \end{align}
        This vanishing of the anomaly is again related to the number of colours. It does not vanish in
        either the left- or right-handed sector, nor in the quark and lepton sector
        individually. Hence, the vanishing of anomalies provides a deep connection between quarks and 
        leptons in the
        standard model, which is a hint to grand unified theories where anomaly cancellation is often
        automatic.
        
        Anomaly cancellation is not restricted to the electroweak gauge currents, but applies to the
        strong force and gravity as well: Mixed $\mathsf{SU(3)}_C$-$\mathsf{U(1)}_Y$ anomalies
        vanish by the same argument as above: 
        Only the $\mathsf{SU(3)}_C^2\mathsf{U(1)}_Y$ triangle contributes, but it is $\tr
        \left[Y\right]_L -\tr \left[Y\right]_R =0$. The same is true for the last possible anomaly,
        the gravitational one, where two non-Abelian gauge currents are replaced by the 
        energy-momentum tensor $T_{\mu\nu}$.

        Hence, the standard model is anomaly free, as it should be. For this, all
        particles of one generation with their strange hypercharges have to  conspire
        to cancel the different anomalies. A ``standard model'' without quarks, for instance, 
        would not
        be a consistent theory, nor a ``standard model'' with four colours of quarks. Note that
        a right-handed neutrino, suggested by neutrino masses, does not pose any problem,
        since it is a complete singlet, without any charge, and thus it does not
        contribute to any gauge anomaly.
    
    \section{Higgs Mechanism}
      The electroweak model discussed so far bears little resemblance to the physics of weak
      interactions. The gauge bosons $W_\mu^I$ and $B_\mu$ are
      massless, implying long-range forces, because a mass term $ m^2
      W_\mu W^\mu$ would violate gauge invariance. Furthermore, the fermions are massless as well,
      again because of gauge invariance: A mass term mixes left- and right-handed fermions,
      \begin{align}
        m\psib\psi &= m\left(\psib_\ls \psi_\rs +\psib_\rs \psi_\ls \right)\,,
      \end{align}
      and since these have different gauge quantum numbers, such a term is not gauge invariant. The
      way out is the celebrated Higgs mechanism\index{Higgs!mechanism}: Spontaneous symmetry breaking
      generates masses for 
      the gauge bosons and fermions without destroying gauge invariance. A simpler version of this
      effect is what happens in superconductors: The condensate of Cooper pairs induces an effective
      mass for the photon, so that electromagnetic interactions become short-ranged, leading to the
      Meissner--Ochsenfeld effect where external magnetic fields are expelled from the
      superconductor, levitating it.

      The key ingredient for the Higgs mechanism is a complex scalar field $\Phi$, which is a
      doublet under $\mathsf{SU(2)}_W$ with hypercharge $-\frac{1}{2}$, which
      has four real degrees of freedom. The crucial feature of the Higgs field is its
      potential ,which is of the  Mexican hat form:
      \begin{align}\label{eq:LHiggs}\index{Higgs!potential}
        \L &= \left(D_\mu \Phi\right)^\dagger \left(D^\mu \Phi\right)
        -V\!\left(\Phi^\dagger\Phi\right) \,,
        \intertext{with}
        D_\mu \Phi &= \left(\partial_\mu +\i g W_\mu -\frac{\i}{2} g' B_\mu \right) \Phi\notag\;,\\
        V\!\left(\Phi^\dagger\Phi\right) &= -\mu^2 \,\Phi^\dagger\Phi +\frac{1}{2}\lambda
        \left(\Phi^\dagger\Phi\right)^2 \,,\quad \mu^2>0 \;.
      \end{align}
      This potential has a minimum away from the origin, at $\Phi^\dagger\Phi=v^2\equiv
      \mu^2/\lambda$. In the vacuum, the Higgs field settles in this minimum. At first sight, the
      minimisation of the potential only fixes the modulus $\Phi^\dagger\Phi$, i.e., one of the four
      degrees of freedom. The other three, however, can be eliminated by a gauge transformation, and
      we can choose the following form of $\Phi$, which is often referred to as unitary
      gauge:\index{Unitary gauge}
      \begin{align}\label{eq:Hunitarygauge}
        \Phi &=\begin{pmatrix} v +\frac{1}{\2} 
          H\!\left(x\right)\\0 \end{pmatrix}\,,\quad H = H^*  
      \end{align}
      Here we have eliminated the upper component and the imaginary part of the lower one. We
      have also shifted the lower component to the vacuum value, so that the dynamical field
      $H\!\left(x\right)$ vanishes in the vacuum.

      In unitary gauge, the Higgs Lagrangean~(\ref{eq:LHiggs}) becomes
      \begin{align}
        \begin{split}\label{eq:Lunitarygauge}
          \L &= \frac{\lambda}{2} v^4 \\
          &\quad + \frac{1}{2} \partial_\mu H\, \partial^\mu H -\lambda v^2 H^2
          +\frac{\lambda}{\2} v H^3 +\frac{\lambda}{8} H^4 \\
          &\quad +\frac{1}{4}\left(v+\frac{1}{\2} H \right)^2 \left(W_\mu^1,W_\mu^2,W_\mu^3,
            B_\mu\right)   
          \begin{pmatrix} 
            \begin{matrix}
              g^2 & 0\\ 0 & g^2
            \end{matrix} & \text{\Large 0}\\
            \text{\rule[20pt]{0pt}{1pt}\Large 0} &\begin{matrix} g^2 & g g'  \\ g g' & g'^2\end{matrix}
          \end{pmatrix}
          \begin{pmatrix} W^{1\,\mu} \\W^{2\,\mu} \\W^{3\,\mu} \\  B^\mu\end{pmatrix}\,. 
        \end{split}
      \end{align}
      The first line could be interpreted as vacuum energy density, i.e., a cosmological
      constant. However, such an interpretation is on shaky grounds in quantum field theory, so
      we will ignore this term\footnote{Generally, nothing prevents us from adding an arbitrary
        constant to the Lagrangean, obtaining any desired ``vacuum energy''. For example, the Higgs
        potential is often written as $\left(\Phi^\dagger\Phi-v^2\right)^2$, so that
        its expectation value vanishes in the vacuum. These potentials just differ by the a shift
        $\sim v^4$, and are indistinguishable within QFT.}. The second line describes a real scalar
      field $H$ of mass $m_H^2=2\lambda v^2$ with cubic and quartic self-interactions. The most
      important line, however, is the last one: It contains mass terms for the vector bosons! A
      closer look at the mass matrix reveals that it only is of rank three, so it has one zero
      eigenvalue, and the three remaining ones are $g^2$, $g^2$, and $\left(g^2 +g'^2\right)$. In
      other words, it describes one massless particle, two of equal nonzero mass and one which is
      even heavier, i.e., we have identified the physical $\gamma$, $W^\pm$ and $Z$ bosons.

      The massless eigenstate of the mass matrix, i.e., the photon,  is the linear combination \mbox{$A_\mu
      =-\sin\theta_\text{W} W^3_\mu +\cos \theta_\text{W} B_\mu$}, the orthogonal combination is the
      $Z$ boson, \mbox{$Z_\mu = \cos \theta_\text{W}W^3_\mu + \sin\theta_\text{W} B_\mu$}. Here we
      have introduced the Weinberg angle $\theta_\text{W}$\index{weak mixing angle}, which is defined by
      \begin{align}
        \sin \theta_\text{W} &= \frac{g'}{\sqrt{g^2+g'^2}}\;, \quad\cos \theta_\text{W}
        = \frac{g}{\sqrt{g^2+g'^2}}\;.
      \end{align}
      To summarise, the theory contains the following mass eigenstates:
      \begin{itemize}
        \item Two charged vector bosons $W^\pm$ with mass $M_W^2=\frac{1}{2} g^2 v^2$,
        \item two neutral vector bosons with masses $M_Z=\frac{1}{2}\left(g^2+g'^2\right) v^2
          =M_W^2\cos^{-2}\theta_\text{W}$ and \mbox{$M_\gamma=0$}, 
        \item and one neutral Higgs boson with mass $m_H^2=2\lambda v^2$.
      \end{itemize}

      The Higgs mechanism and the diagonalisation of the vector boson mass matrix allow us to 
      rewrite the interaction
      Lagrangean~(\ref{eq:LintWY}), which was given in terms of the old fields $W^I_\mu$ and $B_\mu$
      and their currents~(\ref{eq:Wcurrent}) and~(\ref{eq:Ycurrent}), in terms of the physical field. 
      The
      associated currents are separated into a charged current (for $W^\pm_\mu$) and neutral
      currents (for $A_\mu$ and $Z_\mu$):\index{Electroweak theory!charged current}
      \begin{align}
        \L_\text{CC} &= -\frac{g}{\2} \sum_{i=1,2,3\mspace{-10mu}} \left(\ol{u}_{Li} \gamma^\mu
       d_{Li} +\ol{\nu}_{Li}\gamma^\mu
       e_{Li} \right) W_\mu^+ +\text{h.c.}\;,\\
        \begin{split}
          \L_\text{NC} &= - g J^3_\mu W^{3\,\mu} - g' J_{Y\,\mu} B^\mu \\
          &= - e J_{\text{em}\,\mu} A^\mu -\frac{e}{\sin2\theta_\text{W}} J_{Z\,\mu}  Z^\mu\;,
        \end{split}
        \intertext{with the electromagnetic and $Z$ currents}\index{Electroweak theory!neutral current}
        J_{\text{em}\,\mu}&= \sum_{\substack{i=u,d,c,\\s,t,b,e,\mu,\tau}}\!\!\! \psib_i \gamma_\mu Q_i
        \psi_i\,,\quad \text{with the electric charge }\quad Q_i= T^3_i+Y_i\;,\\
        J_{Z\,\mu} &=
        \sum_{\substack{i=u,d,c,s,t,b\\e,\mu,\tau,\nu_e,\nu_\mu,\nu_\tau}}\!\!\!\psib_i \gamma_\mu
        \left(v_i -a_i\gamma^5\right)\psi_i\;.\label{eq:Zcurrent}
      \end{align}
      Here the fermions $\psi_i$ are the sum of left- and right-handed fields,
      \begin{align}
      \psi_i &= \psi_{Li} + \psi_{Ri}\;.
      \end{align}

      The coupling to the photon, the electric charge $Q$, is given by the sum of the third
      component of weak isospin $T^3$ ($\pm\frac{1}{2}$ for doublets, zero for singlets) and the
      hypercharge~$Y$. This reproduces the known electric charges of quarks and leptons, which 
      justifies the hypercharge assignments in (\ref{eq:hypercharges}). The coupling 
      constant $e$ is related to the original couplings and the weak mixing angle:
      \begin{align}
        e&= g \sin\theta_\text{W} = g' \cos \theta_\text{W}\;.
      \end{align}
      The photon couples only vector-like, i.e., it does not distinguish between different
      chiralities.  The $Z$ boson, on the other hand, couples to the  
      vector- and axial-vector currents of different fermions $\psi_i$ (i.e., their left-and
      right-handed components) with different strengths. They are given by the
      respective couplings $v_i$ and $a_i$, which are universal for all families. In particular, the
      $Z$ couples in the same way to all leptons, a fact known as lepton universality.

      \medskip

      The Higgs mechanism described above is also called spontaneous symmetry
      breaking\index{Spontaneous symmetry breaking}. This term, however, is somewhat misleading:
      Gauge symmetries are never broken, but only hidden. The Lagrangean~(\ref{eq:Lunitarygauge})
      only has a manifest $\mathsf{U(1)}$ symmetry associated with the massless vector field, so it
      seems we have lost three gauge symmetries. This, however, is just a consequence of
      choosing the unitary gauge. The Higgs mechanism can also be described in a manifestly
      gauge invariant way, and all currents remain conserved.

      The ``spontaneous breaking of gauge invariance'' reshuffles the degrees of freedom of
      the theory: Before symmetry breaking, we have the complex Higgs
      doublet (four real degrees of freedom) and four massless vector fields with two degrees of
      freedom each, so twelve in total. After symmetry breaking (and going to unitary gauge), three
      Higgs degrees of freedom are gone (one remaining), but they have resurfaced as extra
      components of three  massive vector fields\footnote{Remember that a massless vector only has
        two (transverse) degrees of freedom, while a massive one has a third, longitudinal, mode.}
      (nine), and one vector field stays massless (another two). So there still are twelve degrees
      of freedom.

    \section{Fermion Masses and Mixings}
      The Higgs mechanism  generates masses not only for the gauge bosons, but also for the
      fermions. As already emphasized, direct mass terms are not allowed in the standard
      model. There are, however, allowed Yukawa couplings of the Higgs doublet to two fermions.
      They come in three classes, couplings to quark
      doublets and either up- or 
      down-type quark singlets, and to lepton doublet and charged lepton singlets. Each term is
      parametrised by a $3\times 3$-matrix in generation space,
      \begin{align}
        \label{eq:LYukawa}
        \L_\text{Y}&= \left(h_u\right)_{ij} \ol{q}_{\ls\,i} u_{\rs\,j} \Phi +\left(h_d\right)_{ij}
        \ol{q}_{\ls\,i} d_{\rs\,j} \widetilde{\Phi} + \left(h_e\right)_{ij} \ol{l}_{\ls\,i} e_{\rs\,j}
        \widetilde{\Phi} + \text{h.c.}\;,
      \end{align}
      where $\widetilde{\Phi}$ is  given by
      $\widetilde{\Phi}_a =\epsilon_{ab} \Phi_b^*$. 

      These Yukawa couplings effectively turn into mass terms once the electroweak symmetry is
      spontaneously broken: A vacuum expectation value $\left<\Phi_X\right>=v$ inserted in the
      Lagrangean~(\ref{eq:LYukawa}) yields
      \begin{align}
        \L_m &= \left(m_u\right)_{ij} \ol{u}_{\ls\,i}  u_{\rs\,j} +\left(m_d\right)_{ij}
        \ol{d}_{\ls\,i}  d_{\rs\,j} +\left(m_e\right)_{ij} \ol{e}_{\ls\,i}  e_{\rs\,j} + \text{h.c.}\;.
      \end{align}
      Here the mass matrices are $m_u=h_u v$ etc., and $u_\ls$, $d_\ls$ and $e_\ls$ denote the
      respective components of the quark and lepton doublets $q_\ls$ and $l_\ls$.

      The mass matrices thus obtained are in general not diagonal in the basis where the charged
      current is diagonal. They can be diagonalised by bi-unitary
      transformations, 
      \begin{subequations}
        \begin{align}
          {V^{(u)}}^\dagger m_u \widetilde{V}^{(u)} 
          &= \diag\!\left(m_u,m_c,m_t\right)\;, \\
          {V^{(d)}}^\dagger m_d \widetilde{V}^{(d)} 
          &= \diag\!\left(m_d,m_s,m_b\right)\;, \\
          {V^{(e)}}^\dagger m_e \widetilde{V}^{(e)} 
          &= \diag\!\left(m_e,m_\mu,m_\tau\right)\;, 
          \intertext{with unitary matrices $V$,}
          {V^{(u)}}^\dagger V^{(u)} &= \mathbbm{1}\,, \quad \text{etc.} \notag
        \end{align}
      \end{subequations}
      This amounts to a change of basis from the weak eigenstates (indices $i,j,\ldots$) to mass
      eigenstates (with indices $\alpha,\beta,\ldots$):
      \begin{align}
        u_{\ls\,i} &= V^{(u)}_{i\alpha} u_{\ls\,\alpha}\,, & d_{\ls\,i} &= V^{(d)}_{i\alpha }
        d_{\ls,\alpha}\,, & u_{\rs\,i} &= \widetilde{V}^{(u)}_{i\alpha } u_{\rs\,\alpha}\,, & d_{\rs\,i}
        &= \widetilde{V}^{(d)}_{i\alpha } d_{\rs\,\alpha} \,.
      \end{align}
      The up- and down-type matrices $V^{(u)}$ and $V^{(d)}$ are not identical, which has an
      important consequence: The charged current couplings are now no longer diagonal, but rather
      \begin{align}
        \L_\text{CC} &= -\frac{g}{\2} V_{\alpha\beta} \ol{u}_{\ls\,\alpha} \gamma^\mu d_{\ls\,\beta}
        W^+_\mu +\text{h.c.}\;,
        \intertext{with the CKM matrix\index{CKM matrix}}
        V_{\alpha\beta}&= {V^{(u)}}^\dagger_{\alpha i} {V^{(d)}}_{i \beta}\;, 
      \end{align}
      which carries the information about
      flavour mixing in charged current interactions. Because of the unitarity of the transformations,
      there is no flavour mixing in the neutral current.

      We saw that the Higgs mechanism generates fermion masses since direct mass terms are not
      allowed due to gauge invariance. There is one possible exception: a right-handed
      neutrino, which one may add to the standard model to have also neutrino masses.  
      It is a singlet of the standard model gauge group and can therefore have a
      Majorana mass term\index{Majorana mass} which involves
      the charge conjugate fermion
      \begin{align}
        \psi^\cs &= C \psib^T \;, 
      \end{align}
      where $C=\i\gamma^2\gamma^0$ is the charge conjugation matrix. 
      As the name suggests, the charge conjugate spinor has charges opposite to the original one. It
      also has opposite chirality, $P_\ls \psi_\rs^\cs =\psi_\rs$. Thus we can produce a mass term
      $\psib^\cs\psi$ (remember that a mass term always requires both chiralities), which only
      is gauge invariant for singlet fields.

      So a right-handed neutrino $\nu_\rs$ can have the usual Higgs coupling
      and a Majorana mass term, 
      \begin{align}
        \L_\text{$\nu$,mass} &= h_{\nu\, ij} \ol{l}_{\ls\,i} \nu_{\rs\,j} \Phi + \frac{1}{2} M_{ij}
        \ol{\nu}_{\rs\,i} \nu_{\rs\,j} +\text{h.c.}\;,
      \end{align}
      where $i,j$ again are family indices.

      The Higgs vacuum expectation value $v$ turns the coupling matrix $h_\nu$ into the Dirac mass
      matrix $m_D=h_\nu v$. The eigenvalues of the Majorana mass matrix $M$ can be much
      larger than the Dirac masses, and a diagonalisation of the $\left(\nu_\ls,\nu_\rs\right)$
      system leads to three light modes $\nu_i$ with the mass matrix 
      \begin{align}
        m_\nu &= - m_D M^{-1} m_D^T\;.
      \end{align}
      Large Majorana masses naturally appear in grand unified theories. For $M\sim
      10^{15}$~GeV, and $m_D \sim m_t \sim 100$~GeV for the largest Dirac mass,
      one finds $m_\nu\sim 10^{-2}$~eV, which is consistent with results from neutrino
      oscillation experiments. 
      This ``seesaw mechanism'', which explains the smallness of neutrino masses 
      masses as a consequence of large Majorana mass terms, successfully relates neutrino
      physics to grand unified theories\index{See-saw mechanism}.

    \section{Predictions}
      The electroweak theory contains four parameters, the two
      gauge couplings and the two parameters of the Higgs potential: $g$, $g'$, $\mu^2$ and 
      $\lambda$. They can be traded
      for four other parameters, which are more easily measured: The fine-structure constant
      $\alpha$, the Fermi constant $G_\text{F}$ and  the $Z$ boson mass $M_Z$, which are 
      known to great
      accuracy, and the Higgs mass $m_H$ which is not yet known. 

      \begin{figure}[htbp]
        \begin{center}
          \subfigure[]{
          \begin{picture}(140,80)
            \small
            \Photon(20,40)(60,40){2}{5}
            \Text(10,40)[]{$W$}
            \ArrowLine(78,46)(110,70)
            \Text(120,70)[]{$f$}
            \ArrowLine(110,10)(78,34)
            \GCirc(70,40){10}{.8}
            \Text(120,10)[]{$\ol{f}'$}
          \end{picture}}
          \subfigure[]{
          \begin{picture}(140,80)
            \small
            \Photon(20,40)(60,40){2}{5}
            \Text(10,40)[]{$Z$}
            \ArrowLine(78,46)(110,70)
            \Text(120,70)[]{$f$}
            \ArrowLine(110,10)(78,34)
            \GCirc(70,40){10}{.8}
            \Text(120,10)[]{$\ol{f}$}
          \end{picture}}
          \begin{minipage}[b]{152pt}
            \caption{Decays of the $W$ and $Z$ bosons into two fermions. In $W$ decays, the fermion
              and antifermion can have different flavour. The grey blobs indicate
              higher order corrections which must be included to match the experimental
              precision.\label{fig:WZdecay}}    
            \vspace*{-.5cm}
          \end{minipage}
        \end{center}        
      \end{figure}
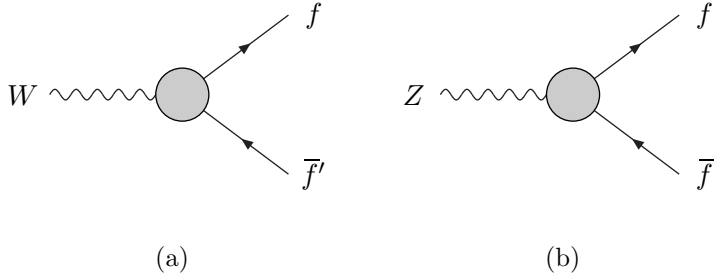

      At LEP, $W$ and $Z$ bosons were produced in huge numbers. There are many observables
      related to their production and decay (Fig.~\ref{fig:WZdecay}). These include:
      \begin{itemize}
        \item The  $W$ mass $M_W$ and the decay widths $\Gamma_W$ and $\Gamma_Z$.
        \item Ratios\index{R ratios@$R$ ratios} of partial decay widths, for
          example, the ratio of the partial $Z$ width into bottom quarks to that into all hadrons,
          \begin{align}
            R_b&= \frac{1}{\Gamma\!\left(Z\to \text{hadrons}\right)} \Gamma\!\left(Z\to
              b\ol{b}\right)\;.
          \end{align}
        \item Forward-backward asymmetries\index{Forward-backward asymmetries}: In $e^+ e^-\to
          Z/\gamma \to f \ol{f}$ reactions, the 
          direction of the outgoing fermion is correlated with the incoming electron. This
          is quantified by the asymmetries $A_\text{fb}^f$, 
          \begin{align}
            A_\text{fb}^f &= \frac{\sigma_\text{f}^f -\sigma_\text{b}^f}{\sigma_\text{f}^f
              +\sigma_\text{b}^f}\,,\quad \text{for }  f=\mu,\tau,b,c\,,
          \end{align}
          where $\sigma_\text{f}^f$ is the cross section for an outgoing fermion in
          the forward direction, i.e., $\theta\in\left[0,\pi/2\right]$ in Fig.~\ref{fig:Afb}, while
          $\sigma_\text{b}^f$ is the cross section for backward scattering. 

          \begin{figure}[htbp]
            \begin{center}
              \subfigure[]{
                \begin{picture}(160,100)
                  \small
                  \ArrowLine(15,90)(50,50) \ArrowLine(50,50)(15,10) \Vertex(50,50){2}
                  \Photon(50,50)(100,50){2}{7} \Vertex(100,50){2} \ArrowLine(100,50)(145,90)
                  \ArrowLine(145,10)(100,50)  \Text(5,90)[]{$e^-$} \Text(5,10)[]{$e^+$}
                  \Text(75,60)[]{$Z,\gamma$} \Text(155,90)[]{$f$} \Text(155,10)[]{$\ol{f}$}
                \end{picture}
              }
              \hfill
              \subfigure[]{
                \begin{picture}(110,100)
                  \small
                  \LongArrow(15,50)(57,50) \LongArrow(105,50)(63,50)  \LongArrow(60,50)(95,85)
                  \LongArrow(60,50)(25,15) \CArc(60,50)(30,0,45) \Text(25,60)[]{$e^-$} \Text(98,42)[]{$e^+$}
                  \Text(102,85)[]{$f$} \Text(18,15)[]{$\ol{f}$} \Text(78,57.5)[]{$\theta$} 
                  {\SetColor{Red} \Vertex(60,50){2}}
                \end{picture}
              }
              \hfill
              \begin{minipage}[b]{150pt}
                \caption{The forward-backward asymmetry $A_\text{fb}$: In the process \mbox{$e^+ e^-\to Z/\gamma\to
                  f\ol{f}$}, there is a correlation between the directions of the outgoing fermion and
                  the incoming electron. This asymmetry has been measured for several types of final
                  state fermions, mostly at LEP with center of mass energy $\sqrt{s} =M_Z$.}
                \vspace*{-.8cm}
              \end{minipage}
            \end{center}
            \label{fig:Afb}
          \end{figure}
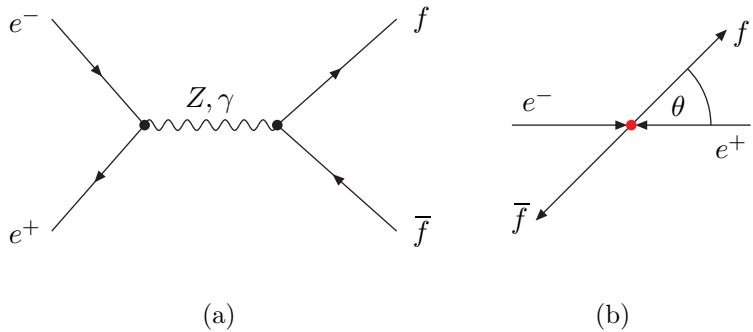
          Also important are double, left-right and forward-backward asymmetries,
          \begin{align}
             A^\text{fb}_\text{LR} &= 
            \frac{\sigma_\text{Lf}^f -\sigma_\text{Lb}^f-\sigma_\text{Rf}^f +\sigma_\text{Rb}^f}
                 {\sigma_\text{Lf}^f +\sigma_\text{Lb}^f+\sigma_\text{Rf}^f +\sigma_\text{Rb}^f}
                 \equiv \frac{3}{4} A_f\;.
          \end{align}   
          The reason for these asymmetries is the presence of the axial couplings $a_i$ in the $Z$ boson
          current~(\ref{eq:Zcurrent}), which lead to different cross sections for the processes $Z\to
          f_\ls \ol{f}_\rs$ and $Z\to f_\rs \ol{f}_\ls$. Thus, one can deduce the $a_i$ and $v_i$
          couplings for fermions from the forward-backward asymmetries, and finally the weak mixing
          angle, on which the vector- and axial-vector couplings of the $Z$ boson depend,
          \begin{align}
            \sin^2\theta_\text{eff}^\text{lept} &=\frac{1}{4}\left(1-\frac{v_l}{a_l}\right)\;.
          \end{align}

        \item Electroweak measurements by now are very precise, and require the inclusion of
          $W$ boson loops in theoretical calculations, so that they test the non-Abelian nature
          of the electroweak theory. The theoretical predictions critically depend on the
          the electromagnetic coupling at the electroweak scale, $\alpha(m_Z)$, which differs
          from the low energy value $\alpha(0)$ in particular by hadronic corrections,
          $\Delta\alpha_\text{had}(m_Z)$.
          
          An important observable is the $\rho$
           parameter, defined by\index{rho pa@$\rho$ parameter}
          \begin{align}
            \rho&= \frac{M_W^2}{M_Z^2 \cos^2\theta_\text{W}}\,.
          \end{align}
          At tree level, $\rho=1$. Loop corrections to the masses of the gauge
          bosons, and therefore to $\rho$, due to quark or Higgs boson loops as in Fig.~\ref{fig:WZsunset},
          are an important prediction of the electroweak theory.

          \begin{figure}[htbp]
            \begin{center}
              \subfigure[Heavy quark corrections\label{fig:Htoploop}]{
                \begin{picture}(140,80)\small
                  \Photon(5,40)(50,40){2}{6} \ArrowArc(70,40)(20,0,180) \ArrowArc(70,40)(20,180,360)
                  \Photon(90,40)(135,40){2}{6} \Text(20,50)[]{$W^+$} \Text(120,50)[]{$W^+$} 
                  \Text(70,70)[]{$\ol{b}$} \Text(70,10)[]{t}  \Vertex(50,40){2} \Vertex(90,40){2}
                \end{picture}
                \hspace{.2cm}
                \begin{picture}(140,80)\small
                  \Photon(5,40)(50,40){2}{6} \ArrowArc(70,40)(20,0,180) \ArrowArc(70,40)(20,180,360)
                  \Photon(90,40)(135,40){2}{6} \Text(20,50)[]{$Z$} \Text(120,50)[]{$Z$} 
                  \Text(70,70)[]{$\ol{t}$} \Text(70,10)[]{t}  \Vertex(50,40){2} \Vertex(90,40){2}
                \end{picture}
              }
              \subfigure[Higgs corrections\label{fig:WZsunsetH}]{
                \begin{picture}(140,80)\small
                  \Photon(5,20)(135,20){2}{12} \Vertex(40,20){2} \Vertex(100,20){2}
                  \DashCArc(70,20)(30,0,180){3} \Text(20,10)[]{$W^\pm,Z$} \Text(120,10)[]{$W^\pm,Z$}
                  \Text(70,60)[]{$H$}  
                \end{picture}
                \hspace{.2cm}
                \begin{picture}(140,80)\small
                  \Photon(5,20)(135,20){2}{12.5} \Vertex(70,22){2} \DashCArc(70,52)(30,0,360){3}
                  \Text(20,10)[]{$W^\pm,Z$} \Text(120,10)[]{$W^\pm,Z$}  \Text(110,50)[]{$H$}
                \end{picture}
              }
              \caption{Radiative corrections to the masses of the $W$ and $Z$ bosons, 
                which depend on the masses of the particles in the loop.
                Diagrams with gauge boson self-interactions have been omitted.}
            \end{center}
            \label{fig:WZsunset}
          \end{figure}
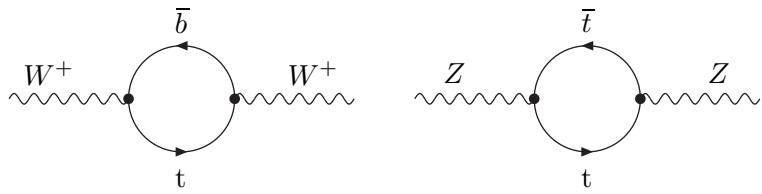
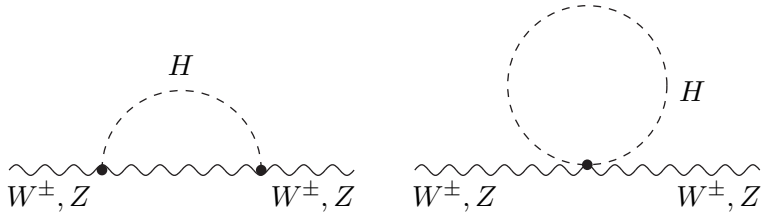 

          The tree level value $\rho=1$ is protected by an approximate $\mathsf{SU(2)}$ symmetry,
          called custodial symmetry, which is only broken by the $\mathsf{U(1)_Y}$ gauge
          interaction and by Yukawa couplings. Thus the corrections
          depend on the fermion masses, and are dominated by the top quark, as in
          Fig~\ref{fig:Htoploop}. The leading correction is 
          \begin{align}
            \Delta \rho^{(t)} &= \frac{3 G_F m_t^2}{8\pi^2 \2} \propto \frac{m_t^2}{M_W^2}\;.
          \end{align}
          This led to the correct prediction of the top mass from electroweak precision data before
          the top quark was discovered at the TeVatron.
          
          The correction due to the Higgs boson diagrams in
          Fig.~\ref{fig:WZsunsetH} again depends on the Higgs mass, but this time
          the effect is only logarithmic:
          \begin{align}
            \Delta \rho^{(H)} &= -C \ln\!\frac{m_H^2}{M_W^2}\;.
          \end{align}
          From this relation, one can obtain a prediction for the mass of the Higgs boson. Clearly,
          the accuracy of this prediction strongly depends on the experimental error on the top mass,
          which affects $\rho$ quadratically.
          
          However, the Higgs mass (weakly) influences many other quantities, and from precision
          measurements one can obtain a fit for the Higgs mass. This is shown in the famous
          blue-band plot, Fig.~\ref{fig:blueband}. 
        
          \item A characteristic prediction of any non-Abelian gauge theory is the self-interaction
          of  the gauge bosons. In the electroweak theory, this can be seen in
          the process $e^+ e^-\to W^+ W^-$. \index{Gauge boson self-interaction}
                
          \begin{figure}
            \begin{center}
              \subfigure[\label{fig:eeWWt}]{
                \begin{picture}(120,90)
                  \small
                  \ArrowLine(20,75)(55,60) \Vertex(55,60){2} \ArrowLine(55,60)(55,20) \Vertex(55,20){2}
                  \ArrowLine(55,20)(20,5) \Photon(55,60)(90,75){2}{5} \Photon(55,20)(90,5){2}{5}
                  \Text(10,75)[]{$e^-$} \Text(10,5)[]{$e^+$} \Text(65,40)[]{$\nu_e$} \Text(105,75)[]{$W^-$}
                  \Text(105,5)[]{$W^+$}  
                \end{picture}
              }
              \hfill
              \subfigure[]{
                \begin{picture}(160,90)
                  \small
                  \ArrowLine(20,75)(50,40) \Vertex(50,40){2} \ArrowLine(50,40)(20,5)
                  \Photon(50,40)(100,40){2}{6} \Vertex(100,40){2} \Photon(100,40)(130,75){2}{5}
                  \Photon(100,40)(130,5){2}{5} \Text(10,75)[]{$e^-$} \Text(10,5)[]{$e^+$}
                  \Text(75,50)[]{$Z/\gamma$} \Text(145,75)[]{$W^-$} \Text(145,5)[]{$W^+$} 
                \end{picture}
              }
              \hfill
              \begin{minipage}[b]{145pt}
                \caption{The process $e^+ e^-\to W^+ W^-$.  The diagrams of panel (b) contain triple gauge boson
                  vertices, $\gamma WW$ and $ZWW$.\label{fig:eeWW}}
                \vspace*{1cm}  
              \end{minipage}
            \end{center}        
          \end{figure}
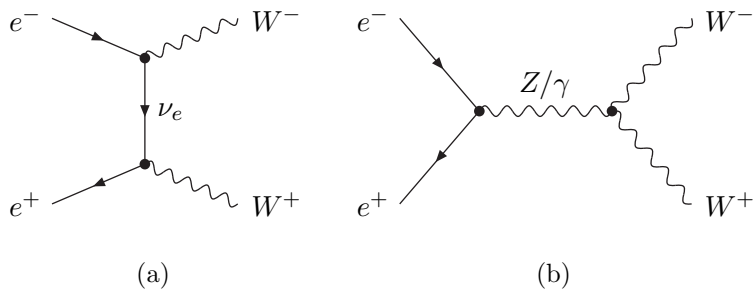
          
          \begin{figure}[b]
            \begin{center}
              \subfigure[$W$ boson pair production cross section at LEP2. Predictions which ignore the
              $WWW$ vertex deviate
              substantially.\label{fig:eeWWplot}]{\includegraphics*[width=.46\textwidth]{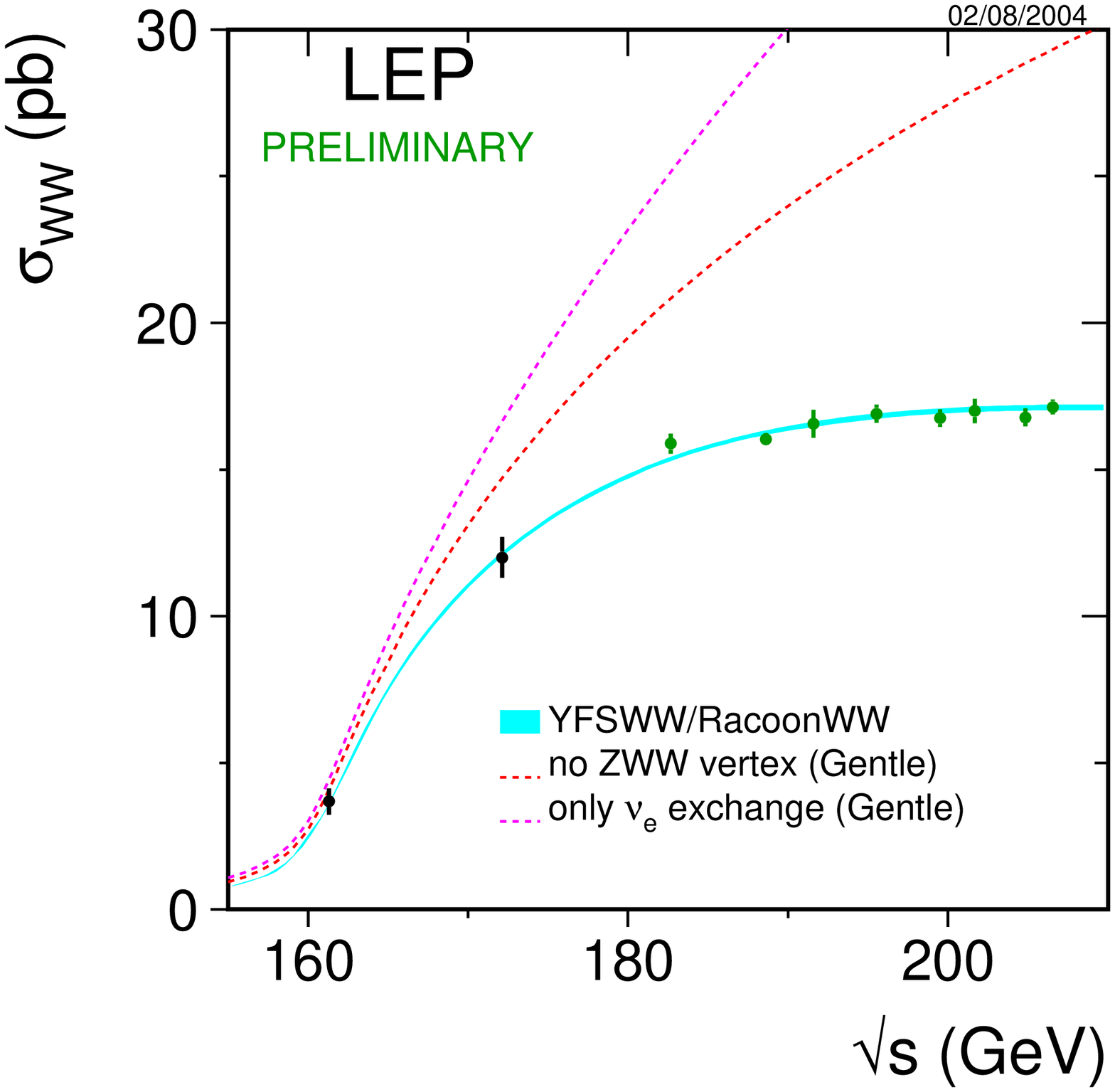}}  
              \hspace{.02\textwidth}
              \subfigure[For $Z$ pair production, there is no triple $Z$ vertex, which agrees well with
              the experimental
              result.\label{fig:eeZZplot}]{\raisebox{-4pt}{\includegraphics*[width=.488\textwidth]{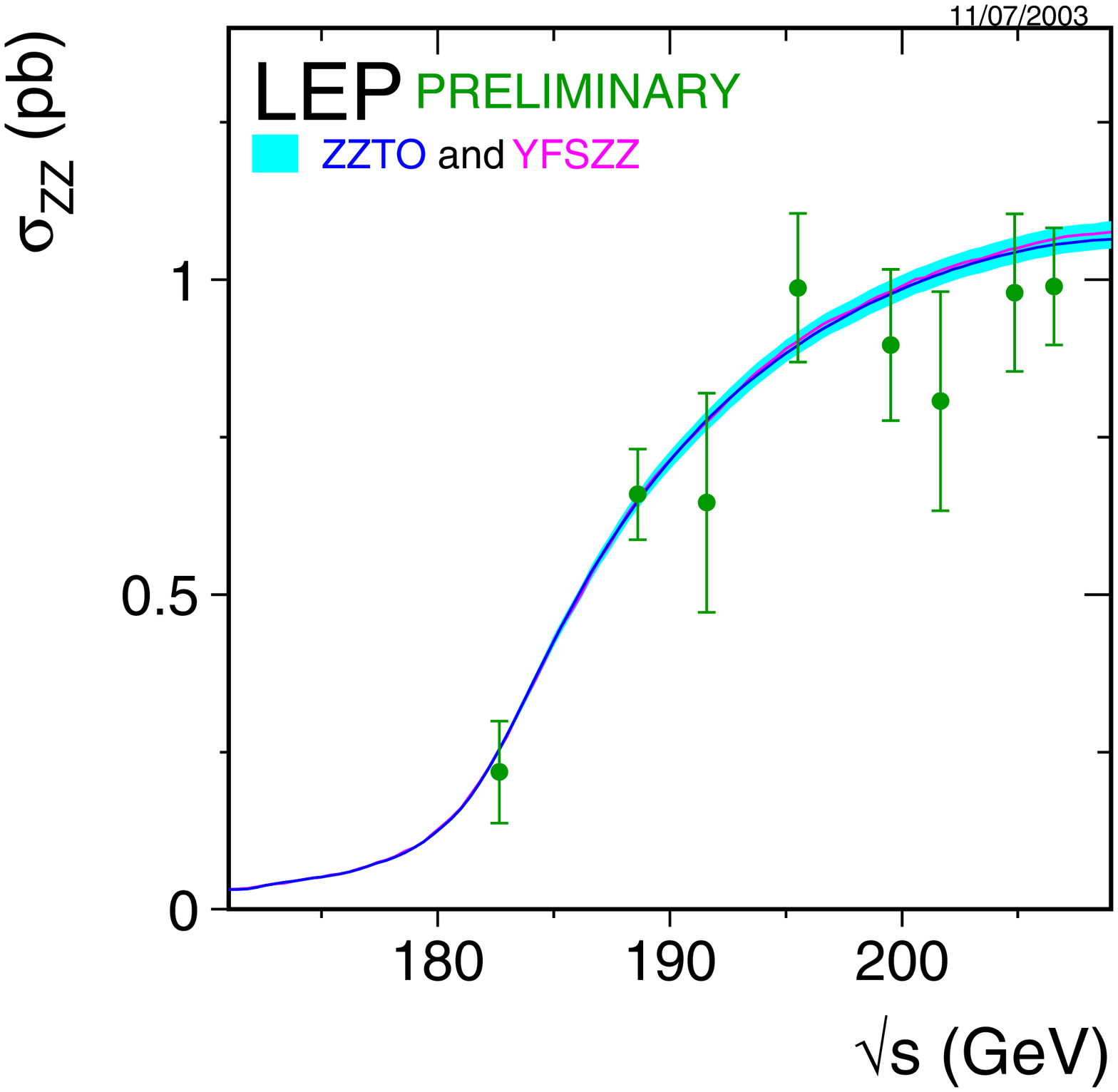}}}  
            \end{center}
            \vspace{-.7cm}
            \caption{\label{fig:pairproduction}Gauge boson pair production cross sections at LEP2
              energies. From~\cite{ewwg}.}
          \end{figure}

          The tree-level diagrams are given in
          Fig.~\ref{fig:eeWW}, and Fig.~\ref{fig:eeWWplot} shows the measured cross section from
          LEP, compared with theoretical predictions. Clearly, the full calculation including all
          diagrams agrees well with data, while the omission of the $\gamma WW$ and $ZWW$
          vertices leads to large discrepancies.  For the process \mbox{$e^+ e^-\to ZZ$}, on the other hand,
          there is no triple gauge boson ($ZZZ$ or $\gamma ZZ$) vertex, so at tree level
          one only has the $t$-channel diagram which is similar to the diagram in
          Fig.~\ref{fig:eeWWt}, but with an electron instead of the neutrino. The agreement
          between theory and data is evident from Fig.~\ref{fig:eeZZplot}. 
        \end{itemize}

      \subsection{Fermi Theory}

        \begin{wrapfigure}[8]{R}[0pt]{90pt}
          \begin{picture}(90,60)\small
            \ArrowLine(5,40)(30,40) \Vertex(30,40){1.5} \ArrowLine(30,40)(70,60)
            \Photon(30,40)(50,20){2}{4} \Vertex(50,20){1.5} \ArrowLine(50,20)(70,30)
            \ArrowLine(70,10)(50,20) \Text(10,30)[]{$\mu^-$} \Text(80,60)[]{$\nu_\mu$}
            \Text(80,30)[]{$e^-$} \Text(80,10)[]{$\ol{\nu}_e$} 
          \end{picture}
          \caption{\label{fig:mudecay}$\mu$ decay}
        \end{wrapfigure}
        The exchange of a $W$ boson with momentum $q$ in a Feynman diagram contributes 
        a factor of $\left(M_W^2-q^2\right)^{-2}$ to the amplitude. For low-energy processes like muon
        decay (see Fig.~\ref{fig:mudecay}), the momentum transfer is much smaller than the mass of the $W$
        boson. Hence, to good approximation one can ignore $q^2$ and replace the
        propagator by $M_W^{-2}$. This amounts to introducing an effective
        four-fermion vertex (see Fig.~\ref{fig:4Fermivertex}),
        \begin{align}
        \L_\text{CC}^\text{eff} = - \frac{G_\text{F}}{\sqrt{2}} J_\text{CC}^\mu J_{\text{CC}\mu}^\dagger\;,
        \end{align}        
        where $G_\text{F}$ is Fermi's constant, \index{Fermi constant $G_\text{F}$}
        \begin{align}\label{eq:GFermi}
          G_\text{F} &= \frac{g^2}{4\2 M_W^2}= \frac{1}{2\2 v^2}\,,
        \end{align}
        which is inversely proportional to the Higgs vacuum expectation value $v^2$. A 
        four-fermion theory for the weak interactions was first introduced by Fermi in
        1934. Since it is not renormalisable, it cannot be considered a fundamental
        theory. However, one can use it
        as an effective theory at energies small
        compared to the $W$ mass. This is sufficient for many applications in
        flavour physics, where the energy scale is set by the masses of leptons, kaons and $B$ mesons.
        
        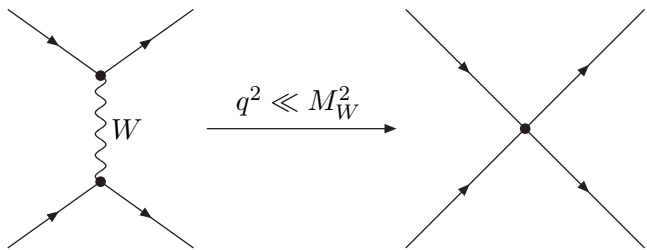
\begin{figure}[htbp]
          \begin{center}
            \begin{picture}(250,100)\small
              \ArrowLine(5,95)(40,70) \ArrowLine(40,70)(75,95) \Vertex(40,70){2}
              \Photon(40,70)(40,30){2}{5} \Text(50,50)[]{$W$}
              \Vertex(40,30){2} \ArrowLine(5,5)(40,30) \ArrowLine(40,30)(75,5) \LongArrow(80,50)(150,50)
              \Text(115,60)[]{$q^2\ll M_W^2$}  \ArrowLine(155,95)(200,50) \ArrowLine(200,50)(245,95)
              \Vertex(200,50){2} \ArrowLine(155,5)(200,50) \ArrowLine(200,50)(245,5) 
            \end{picture}
            \hspace{.2cm}
            \begin{minipage}[b]{180pt}
              \caption{$W$ boson exchange can be described in terms of the Fermi theory, an
                effective theory for momentum transfers  small compared to  the $W$
                mass. The $W$ propagator is replaced by a four-fermion vertex $\propto
                G_\text{F}$.\label{fig:4Fermivertex}} 
              \vspace*{.6cm}            
            \end{minipage}
          \end{center}                
        \end{figure}

    \section{Summary}
    
      \begin{figure}
        \begin{center}
          \includegraphics[width=\textwidth]{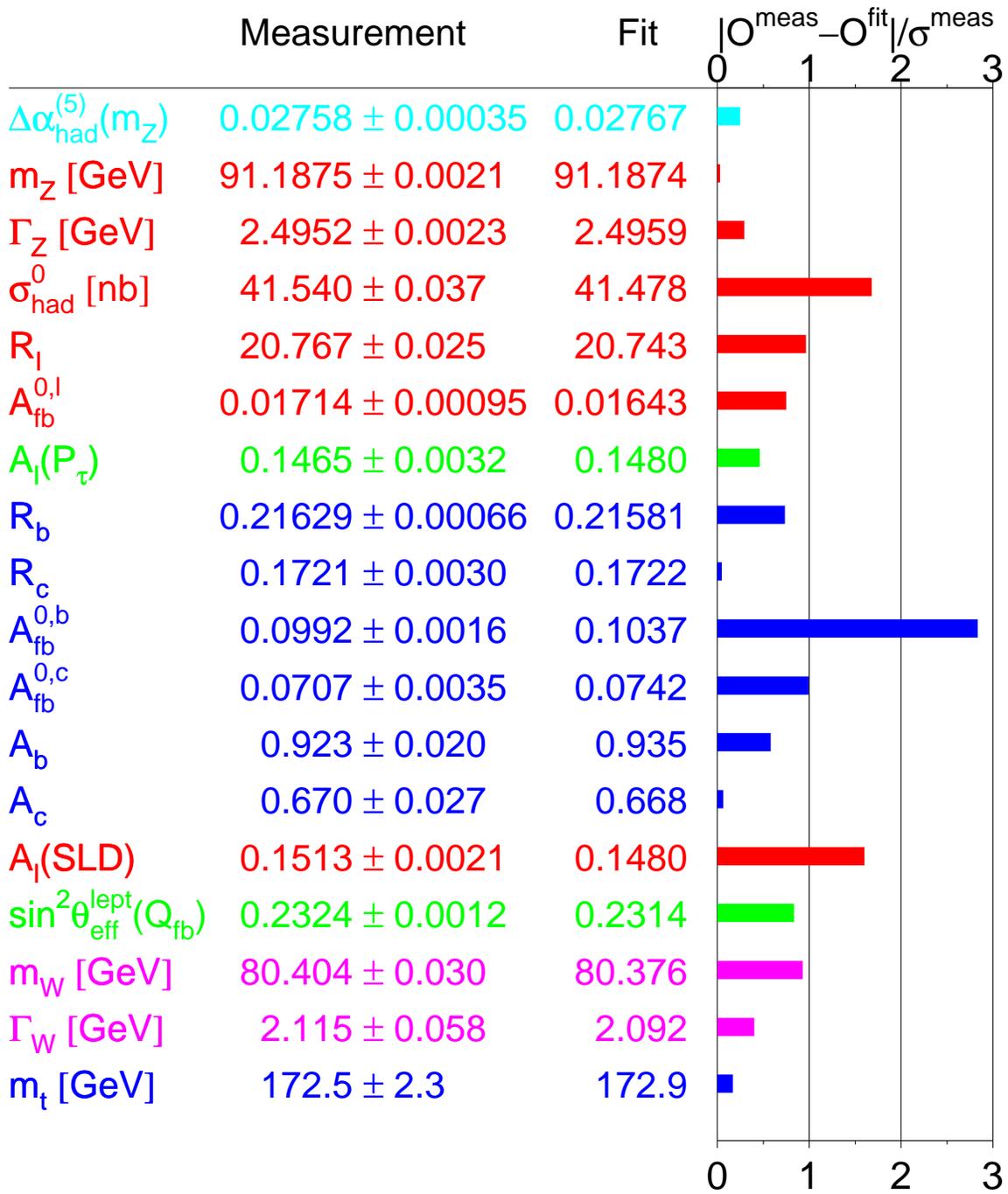}
        \end{center}
        \caption{\label{fig:ewfit}Results of a global fit to electroweak precision data. The right
          column shows the deviation of the fit from measured values in units of the standard
          deviation. From~\cite{ewwg}.} 
      \end{figure}
    
      The electroweak theory is a chiral gauge theory with gauge group
      $\mathsf{SU(2)}_W \times \mathsf{U(1)}_Y$. This symmetry is spontaneously broken down to 
      $\mathsf{U(1)}_\text{em}$ by the Higgs mechanism which generates the gauge boson and
      Higgs masses, and also  all
      fermion masses, since direct mass terms are forbidden by gauge invariance.

      The electroweak theory is extremely well tested experimentally, to the level of 0.1\%,
      which probes loop effects of the non-Abelian gauge theory. The
      results of a global electroweak fit are shown in Fig.~\ref{fig:ewfit}. There is one deviation
      of almost $3\sigma$, all other quantities agree within less than $2\sigma$.
      
      This impressive agreement is only possible due to two properties of the electroweak
      interactions: They can be tested in lepton-lepton collisions, which allow for very
      precise measurements, and they can be reliably calculated in
      perturbation  theory. QCD, on the other hand, requires hadronic processes which are
      experimentally known with less accuracy and also theoretically subject to larger
      uncertainties.


  \chapter{The Higgs Profile}
    The only missing building block of the standard model is the Higgs boson. Spontaneously broken
    electroweak symmetry, however, is a cornerstone of the standard model, and so the discovery of
    the Higgs boson and the detailed study of its interactions is a topic of prime importance
    for the LHC and also the ILC.

    The investigation of the Higgs sector can be expected to
    to give important insight also on physics beyond the standard model. Since the Higgs is
    a scalar particle, its mass is subject to quadratically divergent quantum corrections,
    and an enormous ``fine-tuning'' of the tree-level mass term is needed to keep the Higgs light 
    (this is usually
    referred to as the ``naturalness problem'' of the Higgs sector)\index{Naturalness problem}.  
    Such considerations have motivated various extensions of the standard model:
    \begin{itemize}
      \item Supersymmetry retains an elementary scalar Higgs (and actually adds four more), while
        radiative corrections with opposite signs from bosons and fermions cancel.
      \item Technicolour theories model the Higgs as a composite particle of size 
        $1/\Lambda_\text{TC}$, where $\Lambda_\text{TC} \sim 1\ \text{TeV}$ is the confinement
         scale of a new non-Abelian gauge interaction. These
        theories generically have problems with electroweak precision tests and the generation of 
        fermion  masses.
      \item A related idea regards the Higgs as a pseudo-Goldstone boson of some approximate
        global symmetry spontaneously broken at an energy scale above the electroweak scale. The
        Higgs mass is then related to the explicit breaking of this symmetry.
      \item In theories with large extra dimensions new degrees of freedom occur, and the Higgs 
        field can be identified, for instance, as the fifth component of a
        five-dimensional vector field.
    \end{itemize}
    All such ideas can be tested at the LHC and the ILC, since the unitarity
    of $WW$ scattering implies that the standard model Higgs and/or other effects related
    to electroweak symmetry breaking become manifest at energies below $\sim 1$~TeV.

    \section{Higgs Couplings and Decay}
      Suppose a resonance is found at the LHC with a mass above $114\ \text{GeV}$ and zero charge.
      How can one
      establish that it indeed is the Higgs?
      
      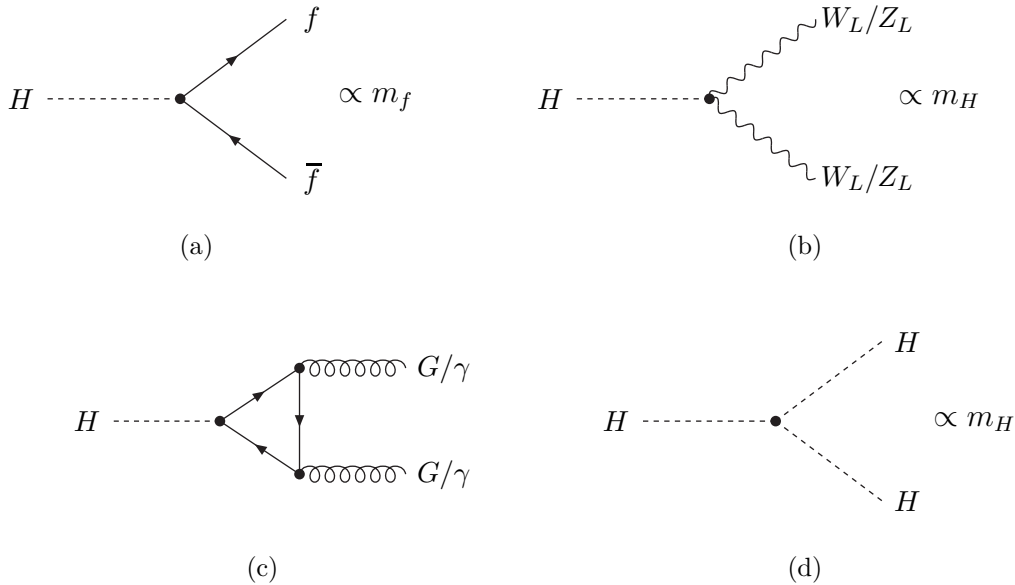
\begin{figure}[htbp]
        \begin{center}
          \subfigure[]{\label{fig:Hff}
            \begin{minipage}{160pt}
              \begin{picture}(160,80)\small
                \DashLine(20,40)(70,40){2} \Vertex(70,40){2}
                \Text(10,40)[]{$H$}
                \ArrowLine(70,40)(110,70)
                \Text(120,70)[]{$f$}
                \ArrowLine(110,10)(70,40)
                \Text(120,10)[]{$\ol{f}$}
                \normalsize
                \Text(130,40)[l]{$\displaystyle \propto m_f$}
              \end{picture}          
            \end{minipage}}
          \hspace{1cm}
          \subfigure[]{\label{fig:HWW}
            \begin{minipage}{220pt}
              \begin{picture}(200,80)\small
                \DashLine(20,40)(70,40){2} \Vertex(70,40){2}
                \Text(10,40)[]{$H$}
                \Photon(70,40)(110,70){2}{6}
                \Text(130,70)[]{$W_L/Z_L$}
                \Photon(110,10)(70,40){2}{6}
                \Text(130,10)[]{$W_L/Z_L$}
                \normalsize
                \Text(130,40)[l]{$\displaystyle \quad \propto m_H $}
              \end{picture}
            \end{minipage}}
          \subfigure[]{\label{fig:Hgg}
            \begin{minipage}{160pt}
              \begin{picture}(160,80)\small
                \DashLine(20,40)(60,40){2} \Vertex(60,40){2} 
                \Text(10,40)[]{$H$}
                \ArrowLine(60,40)(90,60) \ArrowLine(90,60)(90,20) \ArrowLine(90,20)(60,40)
                \Vertex(90,60){2} \Vertex(90,20){2}
                \Gluon(90,60)(130,60){3}{6} \Gluon(90,20)(130,20){3}{6}
                \Text(145,60)[]{$G/\gamma$} \Text(145,20)[]{$G/\gamma$}
                \normalsize 
              \end{picture}
            \end{minipage}}
          \hspace{1cm}
          \subfigure[]{\label{fig:HHH}
            \begin{minipage}{170pt}
              \begin{picture}(160,80)\small
                \DashLine(20,40)(70,40){2} \Vertex(70,40){2}
                \Text(10,40)[]{$H$}
                \DashLine(70,40)(110,70){2}
                \Text(120,70)[]{$H$}
                \DashLine(110,10)(70,40){2}
                \Text(120,10)[]{$H$}
                \normalsize
                \Text(130,40)[l]{$\displaystyle \propto m_H$}
              \end{picture}
            \end{minipage}}        
          \end{center}
        \vspace{-1cm}\caption{\label{fig:Hdecay} Higgs boson decays. Tree-level couplings are proportional to
        masses, but there also are loop-induced decays into massless particles. The cubic Higgs
        self-coupling can be probed at the ILC and possibly at the LHC.}
      \end{figure}

      The Higgs boson can be distinguished from other scalar particles as they occur, for instance, in
      supersymmetric theories, by its special couplings to standard model particles. All couplings 
      are proportional to the mass of the particle, since it is generated by the Higgs mechanism.
      Hence, the Higgs decays dominantly into the heaviest particles kinematically allowed, which
      are $t\ol{t}$ or, for a light Higgs, $b\ol{b}$ and $\tau\ol{\tau}$ pairs. It also has a strong
      coupling $\propto m_H$ to the longitudinal component of $W$ and $Z$ bosons. The tree-level
      diagrams are given in Figs.~\ref{fig:Hff} and~\ref{fig:HWW}.  In addition, there are important
      loop-induced couplings to  massless gluons and photons (see Fig.~(\ref{fig:Hgg}).
          
      The tree level decay widths in the approximation $m_H\gg m_f,M_W$ are given by
      \begin{subequations}
        \begin{align}
          \Gamma\!\left(H\to f\ol{f}\right) &= \frac{G_\text{F} m_H m_f^2}{4\pi \2} N_\text{c}\;,\\
          \Gamma\!\left(H\to Z_L Z_L\right) &=\frac{1}{2}\Gamma\!\left(H\to W_L W_L\right)=
          \frac{G_\text{F} m_H^3}{32\pi \2}\;.  
        \end{align}
      \end{subequations}
      The branching fractions of the Higgs into different decay products strongly depend on the Higgs mass, 
      as shown in Fig.~\ref{fig:BRH}. For a heavy Higgs, with \ $m_H>2 M_W$, the decay
      into a pair of $W$ bosons dominates. At the threshold the width increases by two orders of 
      magnitude, and it almost equals the Higgs mass at $m_H \sim 1\ \text{TeV}$ where the Higgs
      dynamics becomes nonperturbative.
      For a light Higgs with a mass just above the present experimental limit,
      $m_H>114$~GeV, the decay into two photons might be the best possible detection channel given the large
      QCD background for the decay into two gluons at the LHC. It is clearly an experimental challenge 
      to establish the mass dependence of the Higgs couplings, so the true discovery of the Higgs
      is likely to take several years of LHC data!

      \begin{figure}[h!t]
        \begin{center}
          \subfigure{\includegraphics[height=.49\textwidth,angle=270]{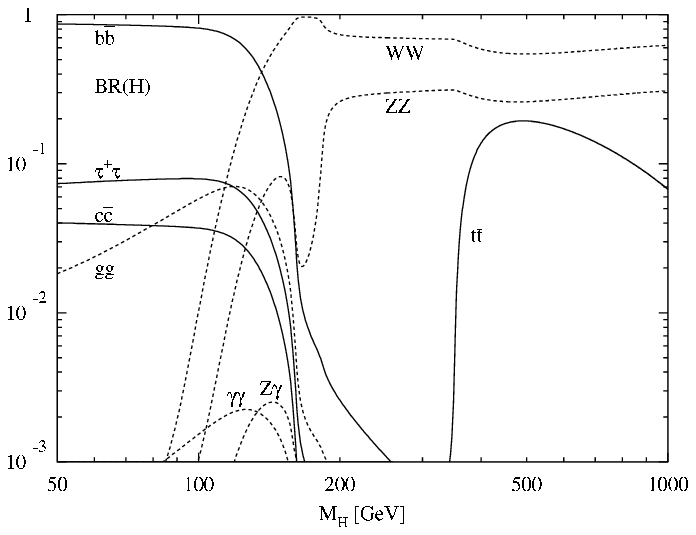}} \hfill
          \subfigure{\includegraphics[height=.495\textwidth,angle=270]{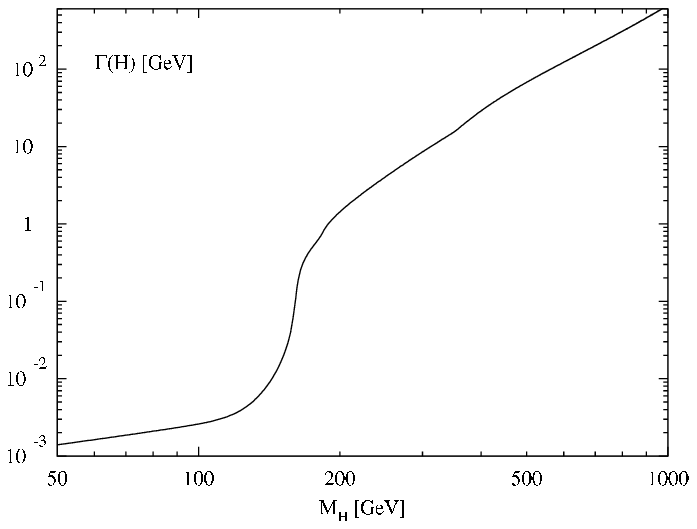}} 
        \end{center}
        \caption{\label{fig:BRH} Left: Higgs branching ratios as function of
        the Higgs mass. Right: Higgs decay width as function of the Higgs mass. It increases by 
        two orders of magnitude at the $WW$ threshold. From \cite{Spira:1997dg}.}
      \end{figure}

    \section{Higgs Mass Bounds}
      
      We now turn to the issue of the Higgs mass. Within the standard model, $m_H^2=2\lambda v^2$ is
      a free parameter which cannot be predicted. There are, however, theoretical consistency 
      arguments which yield stringent upper and lower bounds on the Higgs mass.
      
      Before we present these argument, we first recall the experimental bounds:
      \begin{itemize}
        \item The Higgs has not been seen at LEP. This gives a lower bound on the mass,
          $m_H>114$~GeV.
        \item The Higgs contributes to radiative corrections, in particular for the $\rho$
          parameter. Hence, precision measurements yield indirect constraints
          on the Higgs mass. The result of a global fit is shown in the blue-band plot,
          Fig.~\ref{fig:blueband}. The current 95\% confidence level upper bound is 
          $m_H<185$~GeV, an impressive result! One should keep in mind, however, that the loop
          corrections used to determine the Higgs mass strongly depend on the top mass as well. A
          shift of a few GeV in the top mass, well within the current uncertainties, can shift the
          Higgs mass best fit by several tens of GeV, as can be seen by comparing he plots in
          Fig.~\ref{fig:blueband}. 
      \end{itemize}

      \begin{figure}[h!t]
        \begin{center}
          \includegraphics[width=.24\textwidth]{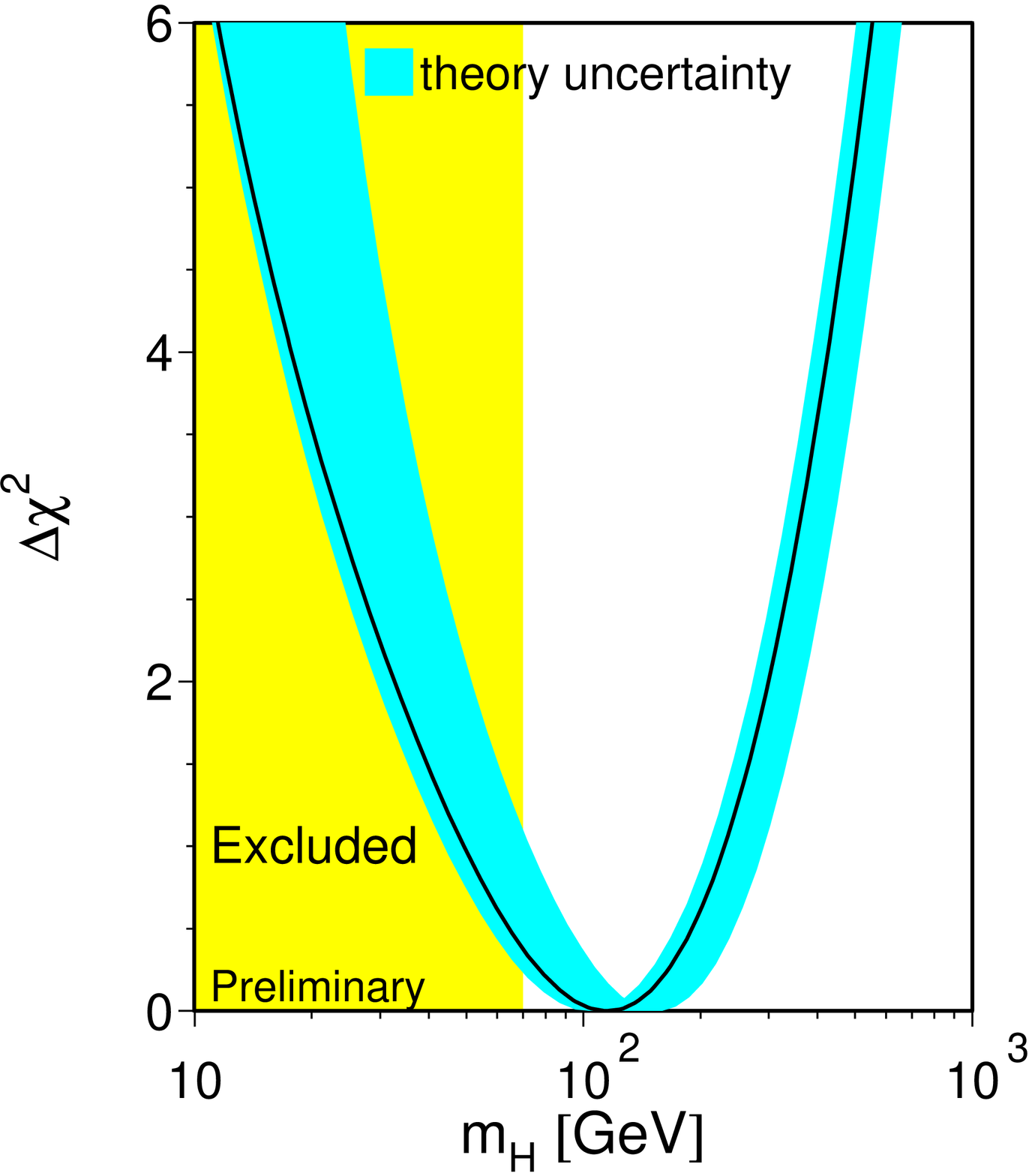} \hfill
          \includegraphics[width=.24\textwidth]{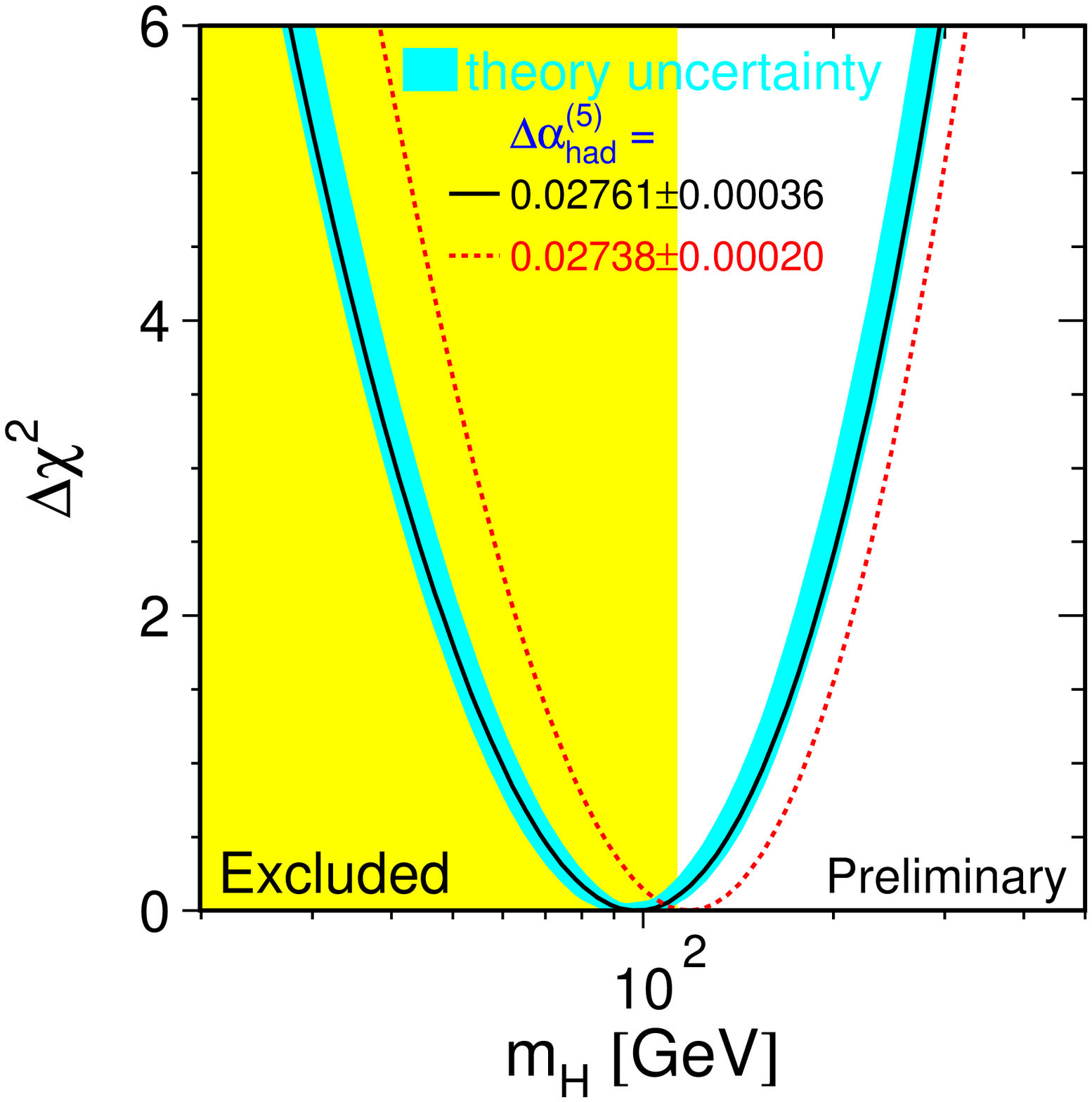} \hfill
          \includegraphics[width=.24\textwidth]{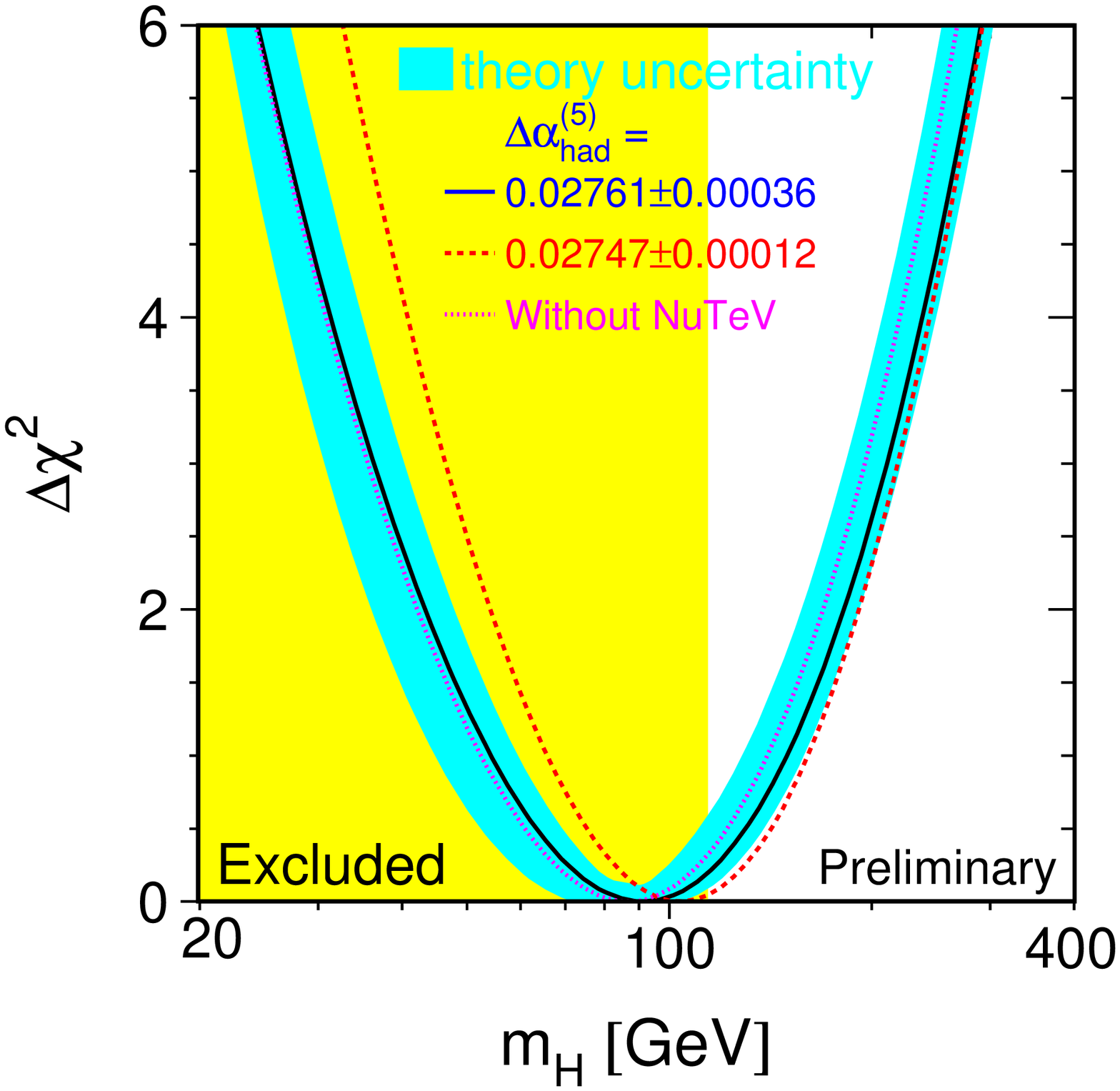} \hfill
          \includegraphics[width=.24\textwidth]{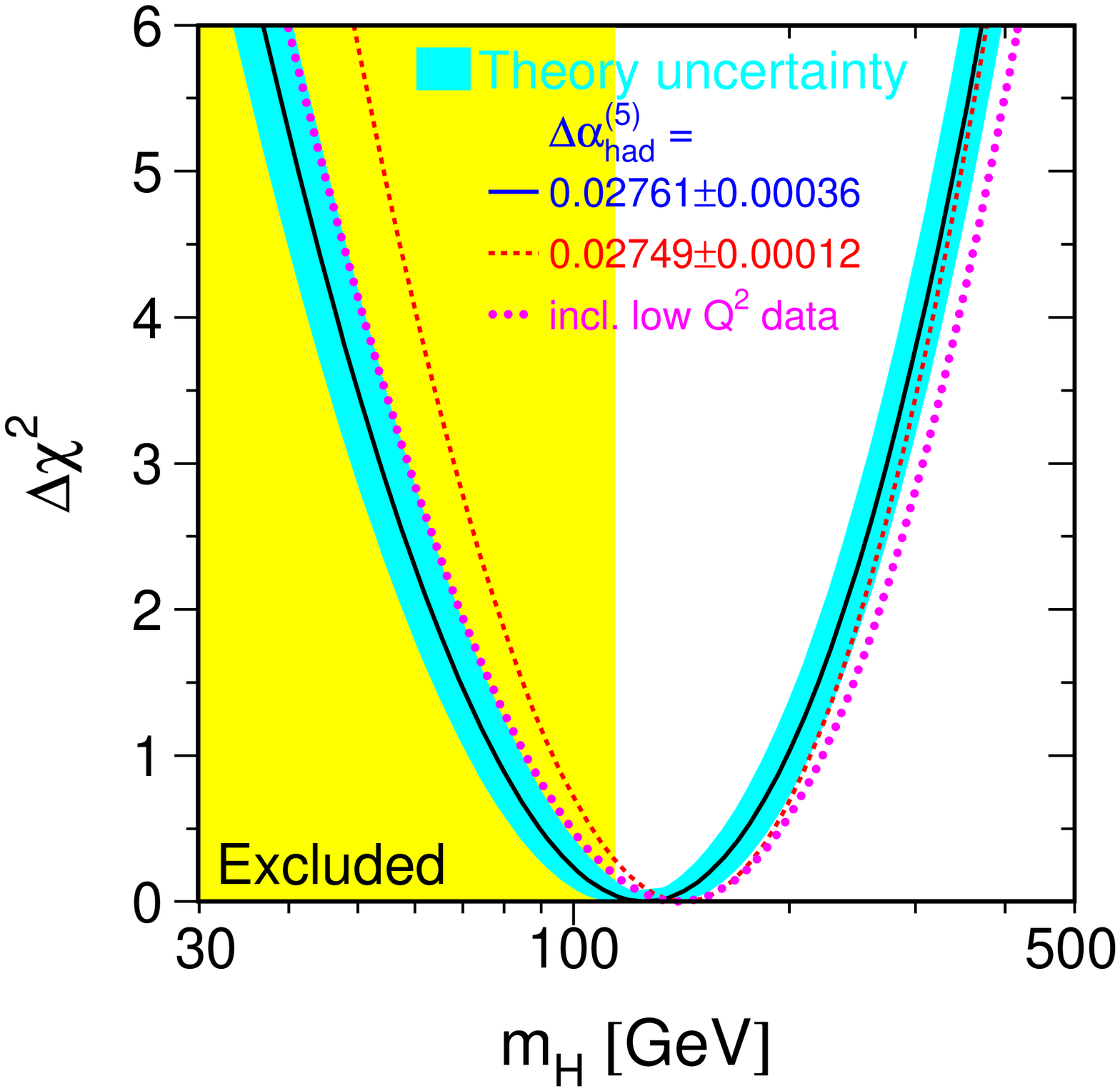}\\[5mm]
          \includegraphics[width=.6\textwidth]{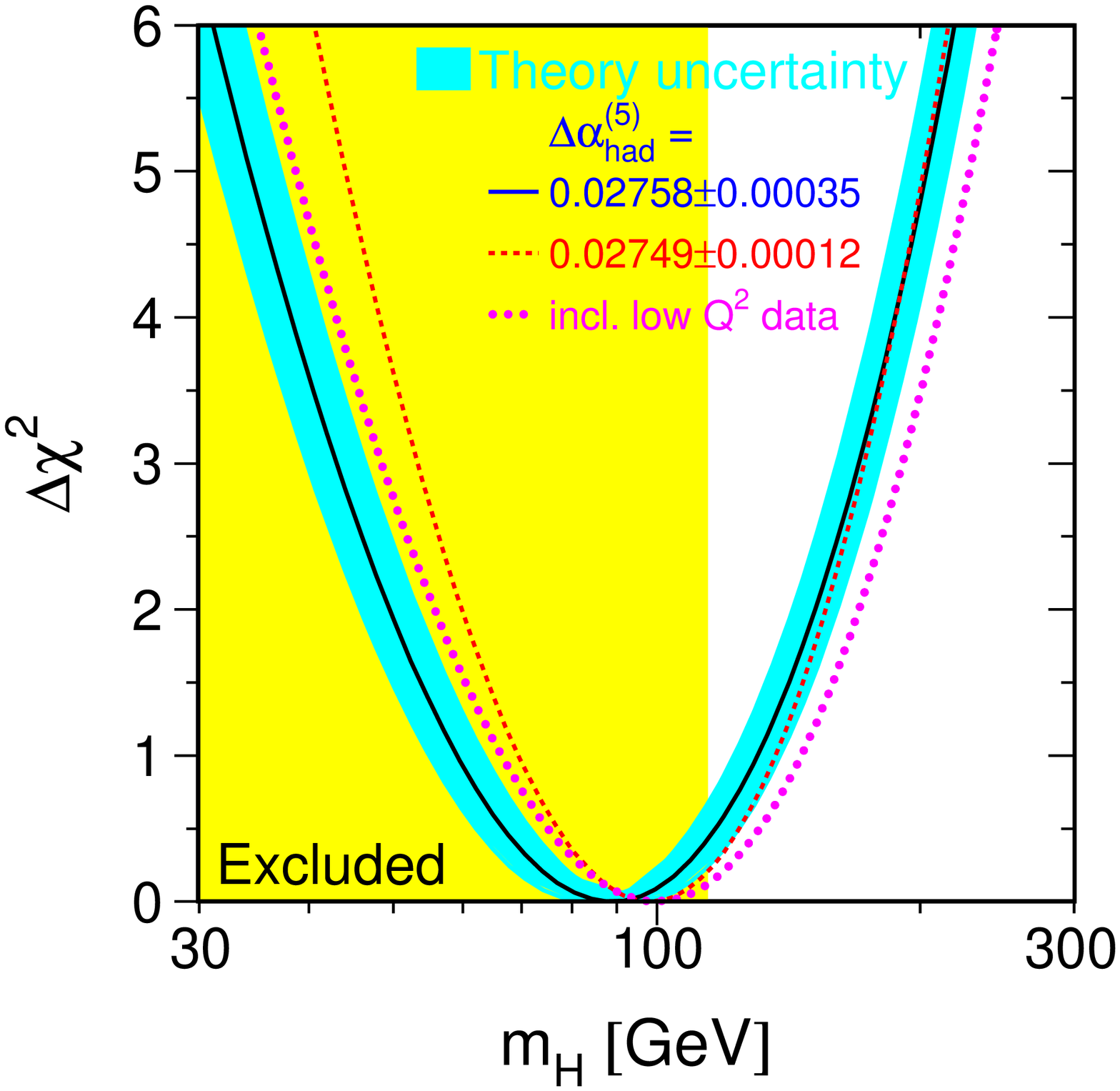}
        \end{center}
        \caption{\label{fig:blueband}The blue-band plot showing the constraints on the Higgs mass
          from precision measurements. The small plots show the same plot from winter conferences of
          different years: 1997, 2001, 2003 and 2005 (left to right). The big plot dates from winter
          2006. The best fit and the width of the parabola vary, most notable due to shifts in the
          top mass and its uncertainty. From~\cite{ewwg}.}
      \end{figure}

      Theoretical bounds on the Higgs mass arise, even in the standard model, from two
      consistency requirements: (Non-)Triviality and vacuum stability. In the minimal supersymmetric
      standard model (MSSM), on the other hand, the Higgs self-coupling is given by the gauge
      couplings, which implies the upper bound $m_H\lesssim 135$~GeV.
      
      The mass bounds in the standard model arise from the scale dependence of couplings, as explained
      in Chapter 4. Most relevant are the quartic Higgs self-coupling $\lambda$ and the top quark 
      Yukawa coupling
      $h_t$ which gives the top mass via $m_t=h_t v$. Other Yukawa couplings are
      much smaller and can be ignored. The renormalisation group equations for 
      the couplings $\lambda(\mu)$ and $h_t(\mu)$ are
      \begin{subequations}
        \begin{align}
          \mu \frac{\partial }{\partial \mu} \lambda\!\left(\mu\right) &=
          \beta_\lambda\!\left(\lambda,h_t\right) =\frac{1}{\left(4\pi\right)^2} \left(12 \lambda^2
            -12 h_t^4  +\ldots\right)\;, \\
          \mu \frac{\partial }{\partial \mu} h_t\!\left(\mu\right) &=
          \beta_\lambda\!\left(\lambda,h_t\right) = \frac{h_t}{\left(4\pi\right)^2}
          \left(\frac{9}{2}h_t^2 -8 g_\text{s}^2 +\ldots\right)\;.
        \end{align}
      \end{subequations}
      These equations imply that $h_t$ decreases with increasing $\mu$ whereas the behaviour of
      $\lambda(\mu)$ depends on the initial condition $\lambda(v)$, i.e., on the Higgs mass.

      For the standard model to be a consistent theory from the electroweak scale $v$ up to
      some high-energy cutoff $\Lambda$, one needs to satisfy the following two conditions in
      the range $v<\mu<\Lambda$:
      \begin{itemize}
        \item The triviality bound: $\lambda\!\left(\mu\right)<\infty$. If $\lambda$ would
          hit the Landau pole at some scale $\mu_\text{L} < \Lambda$, a finite value
          $\lambda(\mu_\text{L})$ would require $\lambda(v)=0$, i.e., the theory would be
          ``trivial''. 
        \item The vacuum stability bound: $\lambda\!\left(\mu\right)>0$. If $\lambda$ would become
          negative, the Higgs potential would not be bounded from below anymore, and the electroweak
          vacuum would no longer be the ground state of the theory.
      \end{itemize}
      These two requirements define allowed regions in the $m_H$-$m_t$--plane as function of the
      cutoff $\Lambda$ (see Fig.~(\ref{fig:trivstab}a)).
      For a given
      top mass, this translates into an upper and lower bound on the Higgs mass.
      For increasing $\Lambda$, the allowed
      region shrinks, and for the known top quark mass and $\Lambda\sim\Lambda_\text{GUT}\sim
      10^{16}$~GeV, the Higgs mass is constrained to lie in a narrow region, $130\text{~GeV}<m_H
      <180\text{~GeV}$ (see Fig.~(\ref{fig:trivstab}b)). 
      
      \begin{figure}[htbp]
        \begin{center}
          \subfigure[]{\includegraphics[width=.45\textwidth]{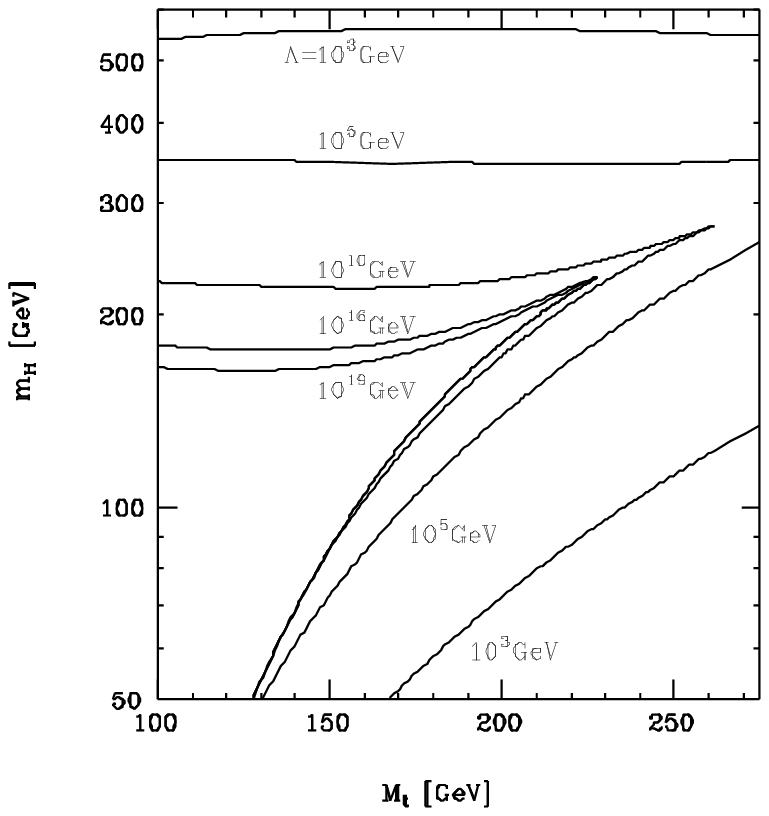}
          }~\subfigure[]{\includegraphics[width=.52\textwidth]{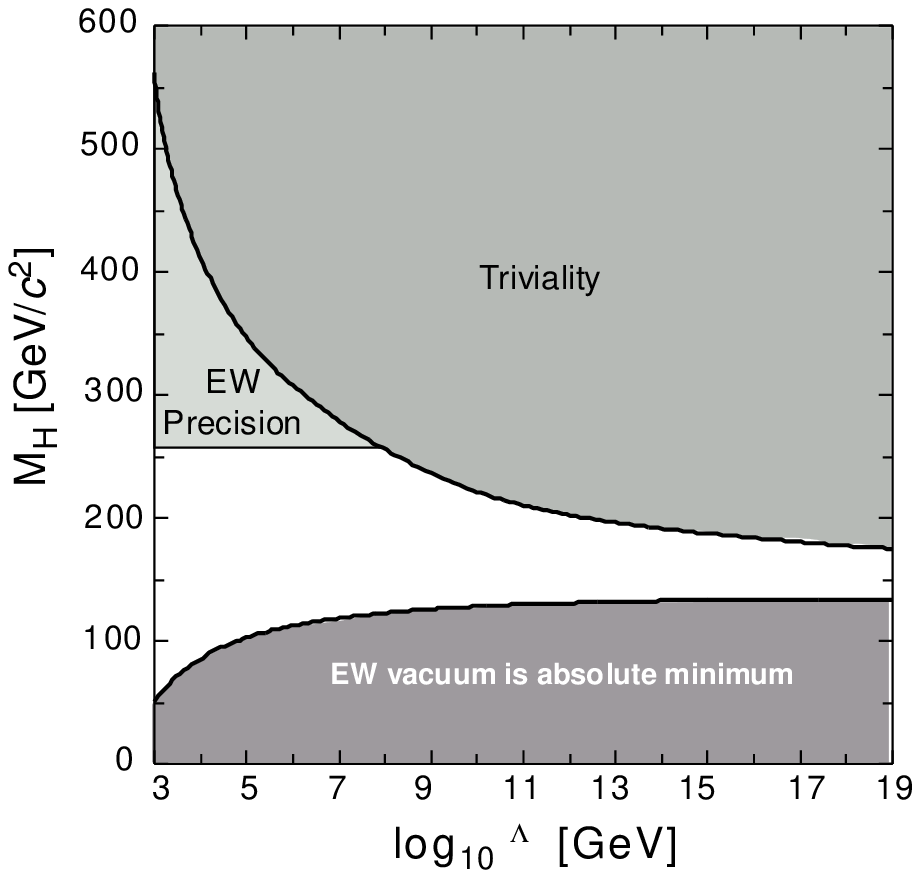}
          \begin{picture}(0,0)(0,0)
            \CBox(-110,10)(-100,17){White}{White} \Text(-105,11)[]{\large$\Lambda$}
          \end{picture}}
        \end{center}
        \vspace{-.5cm}\caption{\label{fig:trivstab}Bounds on the Higgs and top mass from triviality
          and vacuum stability. Panel~(a) shows the combined bounds for different values of
          $\Lambda$ (from~\cite{Froggatt:2003ef}). Panel~(b) gives the bounds on the Higgs mass for
          the known top mass (from~\cite{Quigg:1999xg}).} 
      \end{figure}

      The impressively narrow band of allowed Higgs masses, which one obtains from the triviality
      and vacuum stability bounds,
      assumes that the standard model is valid up to $\Lambda_\text{GUT}$, the scale of
      grand unification.
      This might seem a bold extrapolation, given the fact that our present experimental 
      knowledge ends at the electroweak scale, $\sim 10^2$~GeV. There are, however, two
      indications for such a ``desert'' between the electroweak scale and the GUT scale:
      First, the gauge couplings empirically unify at the GUT scale, especially in the 
      supersymmetric standard model, if 
      there are no new particles between $\sim 10^2$~GeV and $\Lambda_\text{GUT}$; second,
      via the seesaw mechanism, the evidence for small neutrino masses is also consistent
      with an extrapolation to $\Lambda_\text{GUT}$ without new physics at intermediate scales.


    \chapter{History and Outlook}
      Finally, instead of a summary, we shall briefly recall the history of ``The making of the
      Standard Model'' following a review by S.~Weinberg \cite{Weinberg:2004kv}. It is very
      instructive to look at this process as the interplay of some ``good ideas'' and some
      ``misunderstandings'' which often prevented progress for many years.   

      \begin{enumerate}
        \item A ``good idea'' was the quark model, proposed in 1964 independently by Gell-Mann and
          Zweig. The hypothesis that hadrons are made out of three quarks and antiquarks allowed one
          to understand their quantum numbers and mass spectrum in terms of an approximate
          $\mathsf{SU(3)}$ flavour symmetry, the ``eightfold way''. Furthermore, the deep-inelastic
          scattering experiments at SLAC in 1968 could be interpreted as elastic scattering of
          electrons off point-like partons inside the proton, and it was natural to identify these
          partons with quarks.    

          But were quarks real or just some mathematical entities? Many physicists did not believe
          in quarks since no particles with third integer charges were found despite many
          experimental searches.

         \item Another ``good idea'' was the invention of non-Abelian gauge theories by Yang and Mills
           in 1954. The local symmetry was the isospin group $\mathsf{SU(2)}$, and one hoped to
           obtain in this way a theory of strong interactions with the $\rho$-mesons as gauge
           bosons. Only several years later, after the $V-A$-structure of the weak interactions
           had been identified, Bludman, Glashow, Salam and Ward and others developed theories of 
           the weak 
           interactions with intermediate vector bosons.

           But all physical applications of non-Abelian gauge theories seemed to require massive
           vector bosons because no massless ones had been found, neither in strong nor weak 
           interactions. Such mass terms had to be inserted by hand, breaking explicitly the 
           local gauge symmetry and thereby destroying the rationale for introducing non-Abelian
           local symmetries in the first place. Furthermore, it was realized that non-Abelian gauge
           theories with mass terms would be non-renormalisable, plagued by the same divergencies as
           the four-fermion theory of weak interactions.

        \item A further ``good idea'' was spontaneous symmetry breaking: There can be symmetries of
           the Lagrangean that are not symmetries of the vacuum.  According to the Goldstone
           theorem there must be a massless spinless particle for every spontaneously broken global 
           symmetry. On the other hand, there is no experimental evidence for any massless scalar 
           with strong or weak interactions. In 1964 Higgs and Englert and Brout found a way to
           circumvent Goldstone's theorem: The theorem does not apply if the symmetry is a gauge
           symmetry as in electrodynamics or the non-Abelian Yang--Mills theory. Then the Goldstone
           boson becomes the helicity-zero part of the gauge boson, which thereby acquires a mass.   

           But again, these new developments were applied to broken symmetries in strong interactions,
           and in 1967 Weinberg still considered the chiral  
           $\mathsf{SU(2)}_\text{L}\times\mathsf{SU(2)}_\text{R}$ symmetry of strong interactions
           to be a gauge theory with the $\rho$ and $a1$ mesons as gauge bosons. In the same year,
           however, 
           he then applied the idea of spontaneous symmetry breaking to the weak interactions of
           the leptons of the first family, $(\nu_L, e_L)$ and $e_R$ (he did not believe in quarks!).
           This led to the gauge group $\mathsf{SU(2)}\times\mathsf{U(1)}$, massive $W$ and
           $Z$ bosons, a massless photon and the Higgs boson! 

      \end{enumerate}
           The next steps on the way to the Standard Model are well known: The proof by 't~Hooft
           and Veltman that non-Abelian gauge theories are renormalisable and  the discovery of
           asymptotic freedom by Gross and Wilczek and Politzer. Finally, it was realised that the
           infrared properties of non-Abelian gauge theories lead to the confinement of quarks and
           massless gluons, and the generation of hadron masses. So, by 1973 ``The making of the
           Standard Model'' was completed!

           Since 1973 many important experiments have confirmed that the Standard Model is indeed
           the correct theory of elementary particles:
       \begin{itemize}
           \item 1973: discovery of neutral currents
            
           \item 1979: discovery of the gluon

           \item 1983: discovery of the $W$ and $Z$ bosons

           \item 1975 - 2000: discovery of the third family, $\tau$, $b$, $t$ and $\nu_\tau$

           \item During the past decade impressive quantitative tests have been performed of
           the electroweak theory at LEP, SLC and Tevatron, and of QCD at LEP, HERA and Tevatron.
        \end{itemize}

       Today, there are also a number of ``good ideas'' on the market, which lead beyond the Standard Model.
       These include grand unification, dynamical symmetry breaking, supersymmetry and string theory.
       Very likely, there are again some ``misunderstandings'' among theorists, but we can soon hope for
       clarifications from the results of the LHC.

       \bigskip
       \noindent
       \hspace{1cm}\hrulefill\hspace{1cm}

       \vfill
       \noindent
       We would like to thank the participants of the school for stimulating questions and the
       organisers for arranging an enjoyable and fruitful meeting in Kitzb\"{u}hel.


  \appendix
  \chapter{Vectors, Spinors and $\boldsymbol \gamma$-Algebra}
    \label{sec:algebra-appendix}
    
    \section{Metric Conventions}
      Our spacetime metric is mostly minus,
      \begin{align}
        g_{\mu\nu}&=\diag\!\left(+,-,-,-\right)\,,
      \end{align}
      so timelike vectors $v^\mu$ have positive norm $v_\mu v^\mu>0$. The coordinate four-vector is
      $x^\mu=\left(t,\vec{x}\right)$ (with upper index), and derivatives with respect to $x^\mu$ are
      denoted by 
      \begin{align}
        \partial_\mu =\frac{\partial}{\partial x^\mu}=\left(\frac{\partial}{\partial
            t},\vec{\nabla}\right)\,. 
      \end{align}
      Greek indices $\mu,\nu,\rho,\ldots$ run from 0 to 3, purely spatial vectors are indicated by
      an vector arrow. 

    \section[$\gamma$-Matrices]{$\boldsymbol \gamma$-Matrices}
      In four dimensions, the $\gamma$-matrices are defined by their anticommutators,
      \begin{align}
        \left\{\gamma_\mu,\gamma_\nu\right\} &= 2g_{\mu\nu} \mathbbm{1}\,,\quad \mu=0,\ldots,3
      \end{align}
      In addition, $\gamma_0=\gamma_0^\dagger$ is Hermitean while the $\gamma_i=-\gamma_i^\dagger$ are
      anti-Hermitean, and all $\gamma^\mu$ are traceless. The matrix form of the $\gamma$-matrices
      is not fixed by the algebra, and there  
      are several common representations, like the Dirac and Weyl representations,
      Eqs.~(\ref{eq:gammadirac}) and~(\ref{eq:gammaweyl}), respectively. However, the following
      identities hold regardless of the representation.
      
      The product of all $\gamma$-matrices is
      \begin{align}
        \gamma^5 &= \i \gamma^0 \gamma^1 \gamma^2 \gamma^3
      \end{align}
      which is Hermitean, squares to one and anticommutes with all $\gamma$-matrices,
      \begin{align}
        \left\{ \gamma^5,\gamma^\mu\right\} &=0\,.
      \end{align}
      
      The chiral projectors $P_{\ls/\rs}$ are defined as
      \begin{align}
        P_{\ls/\rs} &=\frac{1}{2} \left(1\pm \gamma^5 \right)\,,\quad P_\ls P_\rs=P_\rs P_\ls=0\,,\quad
        P_{\ls/\rs}^2=P_{\ls/\rs} \,.
      \end{align}
      
      To evaluate Feynman diagrams like for the anomalous magnetic moment, one often needs to contract
      several $\gamma$-matrices such as
      \begin{subequations}
        \begin{align}
          \gamma^\mu \gamma_\mu &=4\\
          \gamma^\mu \gamma^\nu \gamma_\mu &= -2\gamma^\nu\\
          \gamma^\mu \gamma^\nu \gamma^\rho \gamma_\mu &= 4 g^{\nu\rho}\\
          \gamma^\mu \gamma^\nu \gamma^\rho \gamma^\sigma \gamma_\mu &= -2 \gamma^\sigma \gamma^\rho
          \gamma^\nu \quad \text{etc.}
        \end{align}
      \end{subequations}

      For a vector $v^\mu$ we sometimes use the slash $\slash[2]{v}=\gamma^\mu v_\mu$

    \section{Dirac, Weyl and Majorana Spinors}
      The solutions of the Dirac equation in momentum space are fixed by the equations
      \begin{align}
        \left(\slash[2]{p} -m \right) u^{(i)}(p)&=0 & \left(\slash[2]{p} +m \right) v^{(i)}(p)&=0\,.
      \end{align}
      Here it is convenient to choose the Weyl representation~(\ref{eq:gammaweyl}) of the Dirac
      matrices,
      \begin{align*}
        \gamma^0&=
        \begin{pmatrix}
          0 &\mathbbm{1}_2 \\ \mathbbm{1}_2 &0
        \end{pmatrix}\,,
        \quad \gamma^i=
        \begin{pmatrix}
          0 &\sigma^i\\-\sigma^i &0
        \end{pmatrix}\,,\quad \Rightarrow \quad
        \gamma^5=
        \begin{pmatrix}
          -\mathbbm{1}_2 & \\ 0&\mathbbm{1}_2
        \end{pmatrix}\,.
      \end{align*}
      
      In this basis, the spinors $u\!\left(p\right)$ and $v\!\left(p\right)$ are given by 
      \begin{align}
        u^s\!\left(p\right) &= \begin{pmatrix}\sqrt{E \mathbbm{1}_2
            +\vec{p}\cdot\vec{\sigma}\rule[10pt]{3pt}{0pt}}\, \xi^s\\ 
          \sqrt{E \mathbbm{1}_2 -\vec{p}\cdot\vec{\sigma}\rule[10pt]{3pt}{0pt}}\,\xi^s \end{pmatrix}
        \,,\quad v^s\!\left(p\right) 
        = \begin{pmatrix}\sqrt{E \mathbbm{1}_2 +\vec{p}\cdot\vec{\sigma}\rule[10pt]{3pt}{0pt}} \,\eta^s\\  -\sqrt{E
            \mathbbm{1}_2 -\vec{p}\cdot\vec{\sigma}\rule[10pt]{3pt}{0pt}}\,\eta^s \end{pmatrix}  \,.
      \end{align}
      Here $\xi$ and $\eta$ are two-component unit spinors. Choosing the momentum along the $z$-axis
      and e.g.\ $\xi=\left(1,0\right)^T$, the positive-energy spinor becomes
      \begin{align}
        u^+ &= \begin{pmatrix} \sqrt{E+p_z\rule[10pt]{3pt}{0pt}}\\0 \\
          \sqrt{E-p_z\rule[10pt]{3pt}{0pt}}\\0\end{pmatrix}\,, 
      \end{align}
      which has spin $+\frac{1}{2}$ along the $z$-axis. For $\xi=\left(0,1\right)^T$, the spin is
      reversed, and similar for $\eta$ and the negative energy spinors.

      The spinors considered so far are called Dirac spinors: They are restricted only by the Dirac
      equation and have four degrees of freedom (particle and antiparticle, spin up and spin
      down). There are two restricted classes of spinors, Weyl and Majorana spinors, which only have
      two degrees of freedom.

      Weyl or chiral spinors are subject to the constraint
      \begin{align}
        P_\ls \psi_\ls &= \psi_\ls \quad \text{or} \quad P_\rs \psi_\rs = \psi_\rs 
      \end{align}
      and correspond to purely left- or right-handed fermions. In the language of $u$'s and $v$'s,
      chiral spinors correspond to sums $u \pm\gamma^5 v$.Chiral spinors can have a kinetic 
      term, but no usual mass term, since 
      \begin{align}
        \overline{\left(\psi_\ls\right)}=\overline{P_\ls \psi_\ls} = \left(P_\ls\psi_\ls\right)^\dagger \gamma^0 =
        \psi_\ls^\dagger P_\ls\gamma^0= \psib_\ls P_\rs\,
      \end{align}
      and hence
      \begin{align}
        \overline{\psi_\ls}\, \psi_\ls =\psib_\ls \underbrace{P_\rs P_\ls}_{=0} \psi_\ls=0\,.
      \end{align}
      However, there is the possibility of a Majorana mass term via the charge conjugate spinor
      $\psi^\cs$: 
      \begin{align}
        \psi^\cs =C\psib^T \quad \text{with the charge conjugation matrix}\quad C=\i\gamma^0 \gamma^2\,.
      \end{align}
      $\psi^\cs$ is of opposite chirality to $\psi$, so it can be used to build a bilinear
      $\psib^\cs \psi$ for a mass term. However, this term violates all symmetries under which
      $\psi$ is charged, so it is only acceptable for complete singlets, like right-handed
      neutrinos.

  \printindex


\begin{thebibliography}{9}
    \bibitem{Weinberg:2004kv}
      S.~Weinberg,
      ``The making of the standard model,''
      Eur.\ Phys.\ J.\ C {\bf 34} (2004) 5
      {\ttfamily [arXiv:hep-ph/0401010]}
 
    \bibitem{Nachtmann:1990ta}
      O.~Nachtmann, ``Elementary Particle Physics: Concepts And Phenomena,''\hspace*{\fill}\linebreak Springer~1990

    \bibitem{peskin}
      M.~E.~Peskin and D.~V.~Schroeder,
      ``An Introduction to quantum field theory,'' Perseus Books~1995
 
    \bibitem{ellis}
      J.~Ellis, these proceedings

    \bibitem{ecker}
      G.~Ecker, these proceedings

    \bibitem{fleischer}
      R.~Fleischer, these proceedings

    \bibitem{lindner}
      M.~Lindner, these proceedings

    \bibitem{gravi}
      S.~Weinberg,
      ``Gravitation and Cosmology'' (Wiley, New York, 1972), pp. 61-63

    \bibitem{thooft}
      G.~'t~Hooft, M.~Veltman, ``Diagrammar'', CERN report 73-9 (1973)

    \bibitem{kinoshita}
      T.~Kinoshita, Nucl.\ Phys.\ Proc.\ Suppl.\  {\bf 157} (2006) 101

    \bibitem{ewwg}
      LEP Electroweak Working Group,~~{\ttfamily http://lepewwg.web.cern.ch}

    \bibitem{Spira:1997dg}
      M.~Spira,
      Fortsch.\ Phys.\  {\bf 46} (1998) 203
      {\ttfamily [arXiv:hep-ph/9705337]} 

      \bibitem{Froggatt:2003ef}
       C.~D.~Froggatt,
       Surveys High Energ.\ Phys.\  {\bf 18} (2003) 77
       {\ttfamily [arXiv:hep-ph/0307138]}
       
     \bibitem{Quigg:1999xg}
       C.~Quigg,
       Acta Phys.\ Polon.\ B {\bf 30} (1999) 2145
       {\ttfamily [arXiv:hep-ph/9905369]}

  \end{thebibliography}
\end{document}